\documentclass[fleqn,10pt]{wlscirep}
\usepackage[utf8]{inputenc}
\usepackage[T1]{fontenc}
\usepackage{array}
\usepackage{booktabs}
\usepackage{pdflscape}
\usepackage{afterpage}
\usepackage{caption}

\usepackage{setspace}
\usepackage{amsmath, amssymb}
\usepackage{rotating}
\usepackage{graphicx}
\usepackage{authblk}
\usepackage{longtable}

\usepackage{setspace}

\title{Accelerating unrest at Campi Flegrei signals a critical transition within the next decade}

\author[1,2,*]{Davide Zaccagnino}
\author[1,$\dag$]{Didier Sornette}
\author[3]{Antonio Giovanni Iaccarino}
\author[3,4]{Matteo Picozzi}

\affil[1]{Institute of Risk Analysis, Prediction and Management (Risks-X), Academy for Advanced Interdisciplinary Studies, Southern University of Science and Technology (SUSTech), Shenzhen, 518055, Guangdong, China}
\affil[2]{Istituto Nazionale di Geofisica e Vulcanologia (INGV), Rome, 00143, Italy}  
\affil[3]{Department of Physics, Federico II University, Via Cintia, 80126, Napoli, Italy}
\affil[4]{Istituto Nazionale di Oceanografia e di Geofisica Sperimentale - OGS Borgo Grotta Gigante 42/C, 34010 - Sgonico, Italy}

\affil[*]{zaccagnino@sustech.edu.cn}
\affil[$\dag$]{dsornette@ethz.ch}
 

\begin{abstract}
Campi Flegrei, a large caldera in southern Italy, is among the most hazardous volcanic systems on Earth, directly threatening over one million people. Since 2005, it has entered a phase of accelerating uplift accompanied by intensified seismicity, raising the key question of whether this evolution will culminate in eruption, a bradyseismic peak, or another regime change.
Here, we show that the acceleration of seismicity and geodetic deformation is better described by a regularised finite-time singularity than by exponential growth, implying not just a better empirical representation but a different underlying process with potentially dire consequences for the system's subsequent evolution. Independent analyses converge on a critical time $t_c \approx$ 2030--2034, with uplift projected to reach about 4 metres by the early 2030s. Geochemical and statistical evidence indicates that deep magmatic volatile input drives this evolution by progressively pressurising the crust.
Although no evidence of imminent eruption is found, the system appears to be approaching a critical mechanical threshold whose outcome remains uncertain, requiring sustained high-resolution monitoring and continuously updated forecasts.
\end{abstract}

\begin{document}

\maketitle
\thispagestyle{empty}

\section*{INTRODUCTION}

Large caldera systems rank among the most dynamic and hazardous environments on Earth, hosting long-lived magmatic reservoirs capable of generating high-impact eruptions \cite{rampino1992volcanic, rampino2002supereruptions, self2006effects, miller2008supervolcanoes, meredith2026high}. Campi Flegrei exemplifies these systems and provides a unique natural laboratory to investigate the dynamics of caldera unrest. Since the mid-2000s, its evolution has been marked by a sustained acceleration of key observables over decadal timescales, with both ground deformation and seismicity increasing markedly \cite{kilburn2023, iaccarino2025}. This behaviour indicates that the system has entered a new dynamical regime characterized by strengthened feedbacks and progressive damage accumulation in the shallow crust.

More generally, caldera unrest is expressed through departures of geophysical and geochemical signals from background conditions, including seismicity, ground deformation, and gas emissions. Despite decades of monitoring and modeling, interpreting these signals remains a major challenge. Similar surface patterns may arise from fundamentally different subsurface processes, including magma intrusion, volatile exsolution or resorption, hydrothermal pressurization, and permeability changes \cite{acocella2015, allison2021highly, keller2026volatile, reid2004massive}. This non-uniqueness severely limits the reliability of eruption forecasting. Indeed, most unrest episodes do not culminate in large-scale failure, while some significant eruptions have been preceded by only weak or ambiguous precursory signals \cite{sparks2003forecasting, dempsey2020automatic, gomez2025there}.
This ambiguity reflects the intrinsically nonlinear and coupled nature of caldera dynamics, where unrest arises from the interaction between magmatic and hydrothermal processes mediated by feedbacks among fluid migration, stress redistribution, and progressive rock damage \cite{chiodini2016}. Within this framework, sustained acceleration of key observables emerges as a natural expression of the system's evolution toward increasingly unstable conditions.

At Campi Flegrei, this behavior is particularly pronounced. Since the mid-2000s, both ground deformation and seismicity have exhibited a marked and persistent acceleration over decadal timescales \cite{bevilacqua2022}, indicating a transition to a new dynamical regime characterized by intensified feedbacks and progressive damage accumulation in the shallow crust. This acceleration is not merely descriptive but carries essential information about the underlying processes and their trajectory. The central question is to identify the underlying process revealed by the functional form of this acceleration, and to infer from it the likely subsequent evolution of the system.

Two broad classes of behavior can account for accelerating signals. The first is exponential growth, 
\begin{equation}
    x(t) \sim e^{\gamma t},
\end{equation}
which arises when the rate of change of an observable is proportional to its current value ($\dot{x} = \gamma x$ with a constant growth rate $\gamma$). Exponential growth already signals an unstable and unsustainable regime, implying that the system must eventually transition to a different state. In the language of dynamical systems, this behavior is the generic signature of a loss of stability of an attractor or fixed point. Specifically, when linearizing the dynamics around a previously stable state, the evolution of small perturbations is governed by the Jacobian operator. Stability requires that all its eigenvalues have negative real parts; when the real part of the leading eigenvalue of the Jacobian crosses zero and becomes positive, the attractor loses stability. The system then exhibits exponential growth of perturbations at a rate $\gamma$ set by the positive real part of this dominant eigenvalue. Such exponential amplification reflects the onset of an instability but does not, by itself, encode information about the timing of the subsequent regime transition.

The second class corresponds to super-exponential acceleration consistent with a finite-time singularity (FTS), which can be expressed as 
\cite{voight1988method,voight1989relation,sornette2004,lei2025log,lei2025unified}
\begin{equation}
    x(t) \sim (t_c - t)^{-\beta}~,  \quad {\rm with}~~\beta>0~.
    \label{thwrtbgq}
\end{equation}
This behavior reflects nonlinear positive feedbacks in which the growth rate $\gamma$ increases itself with the state variable as $\gamma \sim x^\delta$, with $\delta=1/\beta$. A key feature of this formulation is the presence of a critical time $t_c$, which emerges from the dynamics itself and is determined by the initial conditions and the nonlinear positive feedbacks. Expression (\ref{thwrtbgq}) is valid for $t < t_c$, where $t_c$ should be interpreted as the expected time of a regime transition. Such a transition may correspond, for instance, to an eruption, which fundamentally reorganizes the system by releasing accumulated stress and resetting the underlying structure and nonlinear feedbacks. Beyond this point, the assumptions underlying Expression (\ref{thwrtbgq}) no longer hold, and the dynamics must be described by a different regime.

Discriminating between these two classes of acceleration is therefore of central importance. An exponential trajectory indicates an ongoing destabilization without predictive power regarding its endpoint, whereas a finite-time singularity, if supported by the data, implies the existence of a finite horizon for a qualitative transition \cite{lei2025unified}. Importantly, the question is not whether natural systems exhibit true mathematical singularities, but whether their transient dynamics are better captured by functional forms that encode a finite lifetime and thus inform the timing of a potential regime shift.

The goal of the present study is to analyze the ongoing acceleration phase of Campi Flegrei using physics-based models and advanced statistical inference in order to constrain its functional form and extract as much information as possible about future scenarios. Specifically, we test whether the observed dynamics are more consistent with exponential growth or with a finite-time singularity, estimate the associated parameters (including a range of possible critical times $t_c$), and assess the robustness of these estimates under realistic sources of uncertainty. By doing so, we aim to provide quantitative guidance on the mid-term evolution of the system and to inform risk assessment in a context where decision-making must contend with deep physical uncertainty.

\begin{figure}[!htb]
	\centering
	\includegraphics[width=\linewidth]{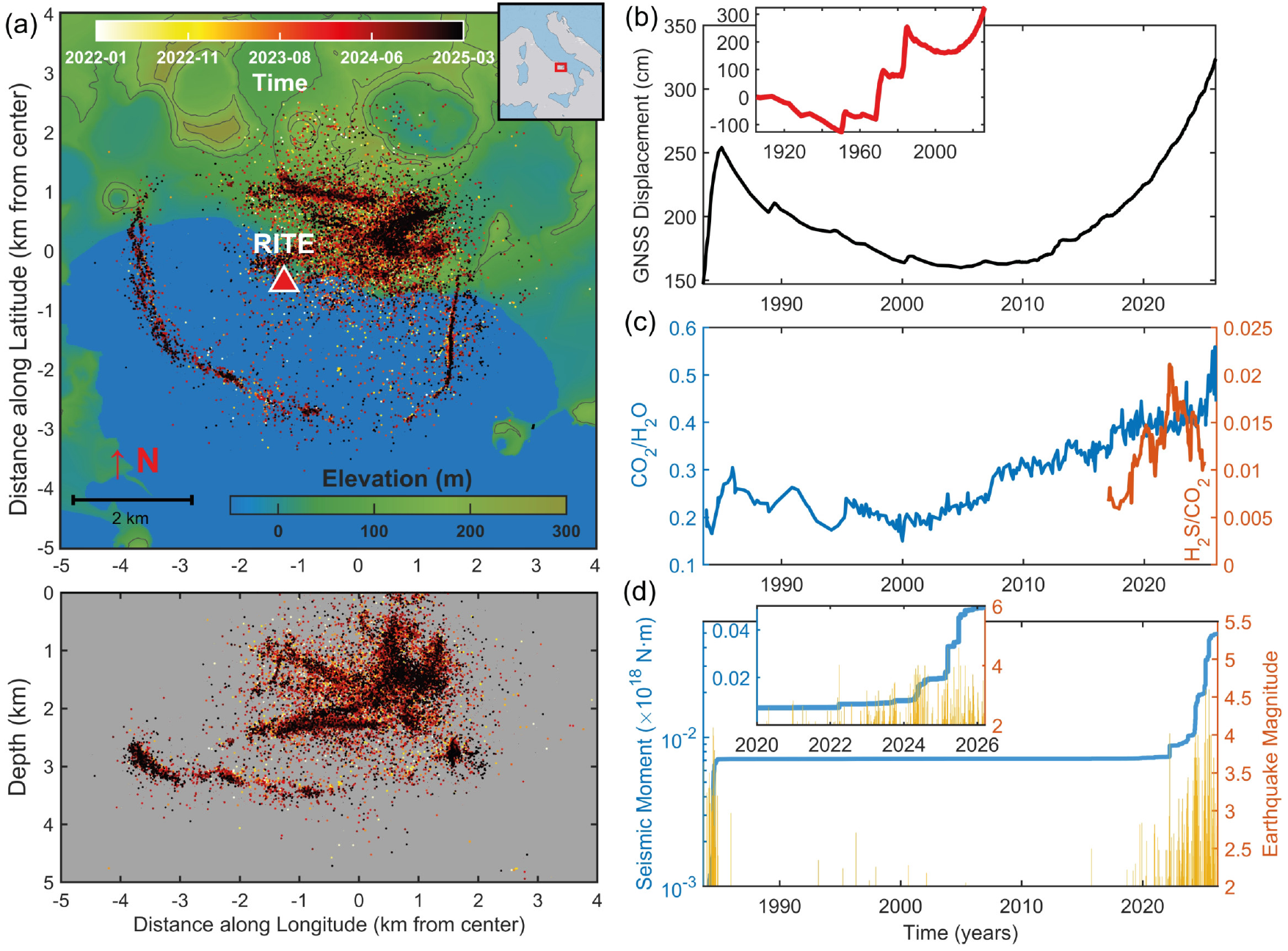}
	\caption{Recent evolution of seismic and volcanic activity at Campi Flegrei, Italy. (a) Spatial and temporal organization of seismicity from the high-quality, relocated machine-learning-augmented catalog of \cite{tan2025clearer}. Most seismicity occurs within a thin brittle cap above shallow magma reservoirs, confined to depths of 2--3 km and distributed through a widely fractured volume. Topographic data are from \cite{vilardo_2024_11190448}. A minority of events distributes along ring faults that mark the boundaries of the volcanic complex undergoing the strongest bradyseismic process. (b) Vertical displacement recorded at GNSS station RITE (near the center of the deforming region) from 1983 to 2026. In the inset, the long-term trend since 1900 inferred from repeated sea-level referenced measurements in the same area \cite{del2010unrest}. (c) Trends in the CO$_2$/H$_2$O (blue) and H$_2$S/CO$_2$ (red) ratios from measurements reported in \cite{vesuviano2024bollettini}. 
		(d) Earthquake magnitudes (orange bars) and cumulative seismic moment over time (blue line).}
	\label{fig1}
\end{figure}

\section*{RESULTS}
\subsection*{Campi Flegrei: system description and evidence for a finite-time singularity regime}

The Campi Flegrei caldera, an almost 15-km-wide volcanic depression nestled immediately west of Naples, is among the most intensively monitored volcanic systems on Earth \cite{cardellini2017monitoring, sabbarese2020continuous, bianco2022permanent}. Unlike the iconic cone of Vesuvius, Campi Flegrei is a sprawling, restless crater floor, home to over 75,000 people in the town of Pozzuoli alone, with about one million living within or along its margins \cite{carlino2021brief}. Its name is from ancient Greek, meaning ``burning fields'': fumaroles vent steam and sulfurous gases, hot springs bubble, and the ground itself has been rising and falling for millennia in a phenomenon known as bradyseism \cite{cannatelli2020ground}. During the most dramatic recent episode, between 1982 and 1984, the caldera floor surged upward by almost 3 meters in three decades, forcing the evacuation of 40,000 inhabitants from Pozzuoli's historic centre \cite{ricci2013volcanic}.

Campi Flegrei is classified as a supervolcano, having produced Europe's largest known eruption 39,000 years ago, a blast that blanketed the Mediterranean in ash and likely triggered a volcanic winter \cite{scarpati1993neapolitan, deino2004age, gianchiglia2025fine}. While an eruption of that magnitude is unlikely in the near future \cite{natale2025magma, caricchi2026scenario}, the ongoing unrest raises urgent questions about the caldera's mid-term evolution \cite{iervolino2024seismic, rolandi20251538, giudicepietro2025burst, tan2025clearer, caricchi2026scenario}. The caldera has undergone recurrent bradyseismic episodes over recent decades and is among the most intensively monitored volcanic systems worldwide \cite{bevilacqua2022}. Since 2005, cumulative uplift has exceeded 1.6 metres, accompanied by a marked intensification of shallow seismicity (Figure~\ref{fig1}b, d). 
The trend of the temperature at Campi Flegrei fumaroles has also increased over time since 2005 reaching about 260$^\circ$C nowadays \cite{buono2023discriminating}, supporting the evidence of a destabilised  system.
This ongoing unrest is not an isolated episode but part of a broader, cumulative process affecting the shallow crust. Both \cite{bevilacqua2022} and \cite{kilburn2023} have shown that deformation and seismicity exhibit long-term acceleration since 2005, with seismicity increasing faster than deformation and, in some intervals, surpassing exponential trends. \cite{iaccarino2025} further demonstrated that seismicity is strongly controlled by deformation rate, with a transition around 2022 from predominantly elastic to inelastic behaviour, evidence that the system is entering a regime of distributed damage and irreversible strain.

The physical driver of this evolution is increasingly tied to deep fluid input. Geochemical studies document a progressive increase in the magmatic contribution to the hydrothermal system, evidenced by rising CO$_2$ fluxes, changing sulfur speciation, and isotopic signatures consistent with deeper fluid sources \cite{chiodini2016, caliro2025}. These observations indicate sustained magmatic volatile supply without requiring shallow magma intrusion or implying imminent eruption. Rather, they highlight a long-term trajectory toward more energetic and pressurized crustal conditions. This picture is reinforced by evidence for progressive stress field evolution and the role of crustal heterogeneity in modulating deformation and seismicity \cite{danesi2024, tramelli2024}.

Our analysis demonstrates that the ongoing unrest at Campi Flegrei, which entered its current accelerating phase since 2005--2008, follows a trajectory that is systematically better described by a finite-time singularity than by an exponential growth. The  finite-time singularity model provides a more compelling description than the exponential alternative, as shown by the Akaike Information Criterion (AIC) differences for both the GNSS vertical displacement at RITE and the cumulative Benioff strain. This constitutes strong statistical evidence that the observed acceleration reflects a singular underlying dynamics rather than simple exponential growth (Supplementary Tables S1--S2). This result is confirmed to be robust over most of the tested analysis windows spanning start dates from 2000 to 2025. As a support of this result, the inverse-time linearity diagnostic, which tests whether the observable scales as $1/(t_c - t)$ as expected for singular dynamics, shows coefficients of determination $R^2 > 0.95$ for optimal fitting windows. Moreover, the singularity exponent $\beta$ stabilises around $1.4$--$1.5$ for GNSS and $0.6$--$0.7$ for Benioff strain during the emergence of the singular regime, consistent with values reported for accelerating precursory sequences in other volcanic systems \cite{cornelius1995graphical, bell2013limits, hao2017accelerating}. This convergence of independent statistical evidence supports the idea that the current unrest at Campi Flegrei reflects a dynamical regime of positive feedback in which the rate of deformation and seismic energy release is proportional to a power of the already accumulated quantity, a hallmark of systems approaching a critical mechanical threshold and a swift change of regime.

\subsection*{Estimation of the critical time and its uncertainty}
\begin{figure}[!htb]
	\centering
	\includegraphics[width=\linewidth]{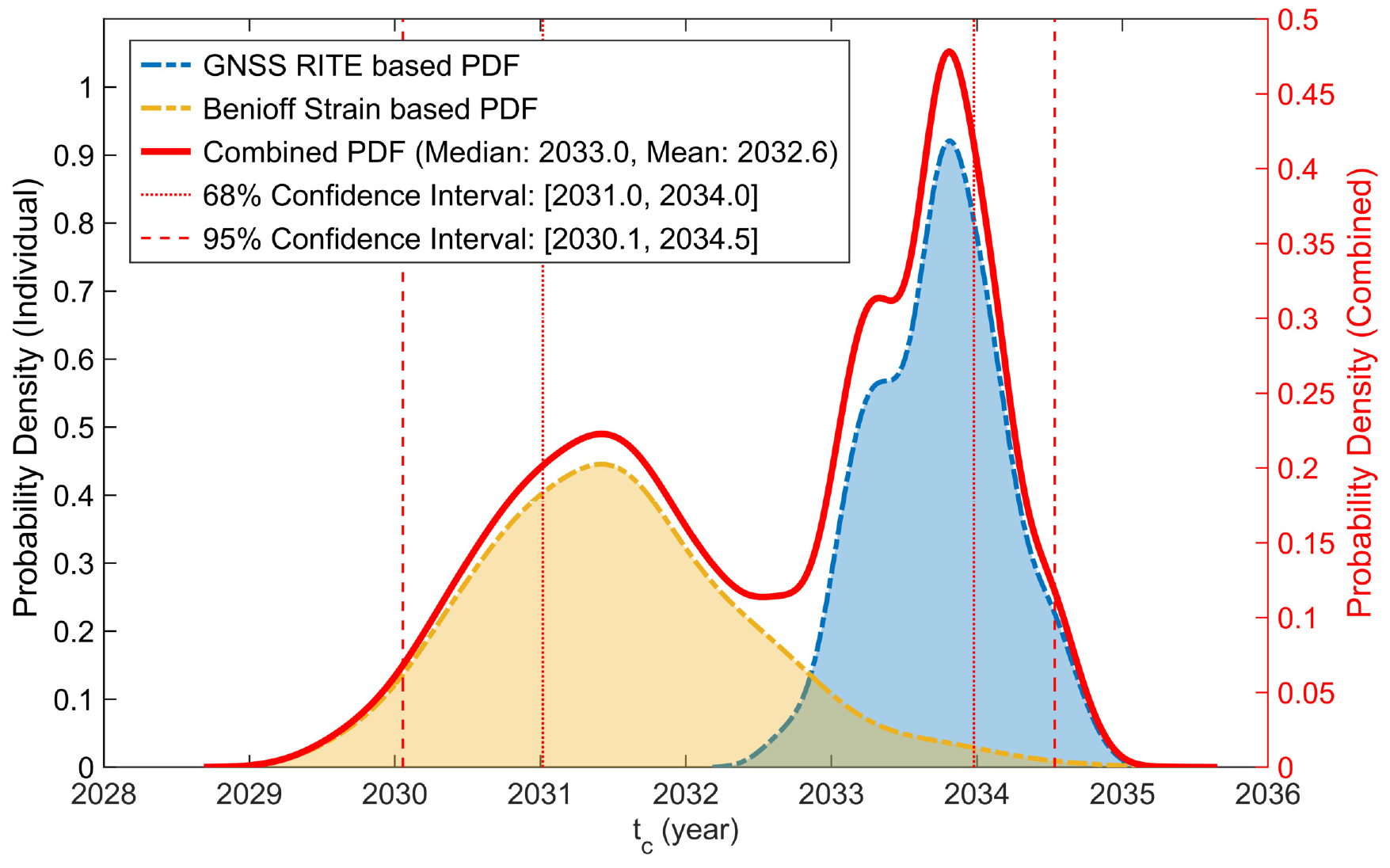}
	\caption{Probability density of the regularized finite-time singularity time ($t_c$), defined as the time at which a change of regime is predicted from the current accelerating upscaling trend in seismic and volcanic activity. The estimate based on geodetic data from GNSS station RITE (2000--2026) is shown in light blue, while the output obtained by analyzing the Benioff strain computed from seismicity is in orange color. The first estimate yields $t_c^g = 2033.83 \pm ^{0.32}_{0.58}$ ([2032.917, 2034.598]), and the second yields $t_c^s = 2031.54 \pm ^{0.86}_{0.96}$ (95\% CI: [2029.931, 2033.615]). The two estimates are compatible within 2$\sigma$, supporting the conclusion that a change of regime will occur within the next decade. The best combined estimate of $t_c \approx 2033.0$ (95\% CI: $[2030.1, 2034.5]$, red vertical dashed lines), obtained via direct quantile estimation from the combined probability density (red line).}
	\label{fig2}
\end{figure}

The critical time $t_c$ estimated independently from geodetic deformation and seismic datasets converges on a consistent horizon. The best geodetic estimate suggests the change of regime will occur in 2033 or 2034, being more likely, given the available information, in autumn 2033 ($t_c^g \approx 2033.83$ with 95\% confidence interval $[2032.92, 2034.60]$). The seismic estimate gives a highest probability in 2031--2032 but with a wider uncertainty ($t_c^s \approx 2031.54$, 95\% CI: $[2029.93, 2033.62]$). The compatibility of these independent estimates within $2\sigma$ strongly reinforces the robustness of the critical time inference. By combining the evidence via direct quantile estimation from joint probability density, the best estimate of the critical time is $t_c \approx 2033.0$ (95\% CI: $[2030.1, 2034.5]$; Figure~\ref{fig2}). Therefore, based on the available information, a change of regime is likely expected between 2030 and 2035.

\begin{figure}[!htb]
	\centering
	\includegraphics[width=\linewidth]{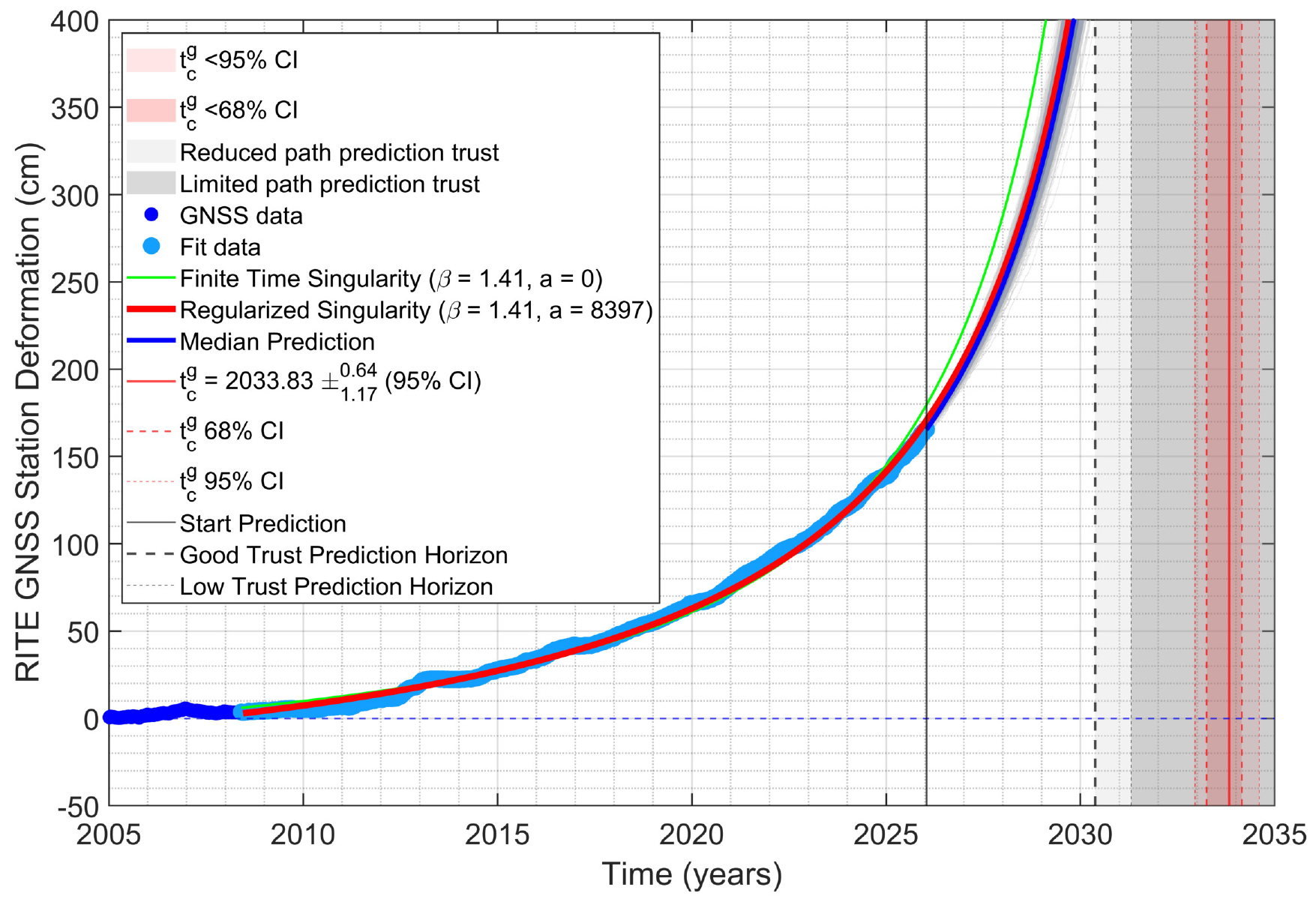}
	\caption{Forecast of vertical soil displacement (in centimeters) at GNSS station RITE. Blue dots represent past data with zero offset set at January 2005. Light blue dots indicate the optimal data range for the forecast based on the Lagrange multiplier methodology (see Methods). The green line shows the singular model used to estimate the scaling exponent of the power law diverging trend. The red line is the regularized finite-time singularity optimized over the entire data range, while the blue line represents the same model optimized to match the final one-year trajectory of the station. Thin gray lines correspond to 100 stochastic simulations using the best-tracking regularized finite-time singularity with Brownian noise, where fluctuations are calibrated from the residuals of the light blue data relative to the best-fitting model. The vertical red solid line marks the estimated $t_c$ (geodetic estimate), with 68\% and 95\% confidence intervals shown by the dark and light reddish shaded regions, respectively, bounded by vertical red dashed lines. Gray zones indicate time horizons of reduced or limited confidence in the GNSS displacement forecast, where the estimated standard deviation of the station position reaches 25\% (light gray) and 50\% (dark gray) of the median predicted position.}
	\label{fig3}
\end{figure}

Barring a change of regime mediated by processes not currently expressed in the observables, the deformation of the caldera is projected to reach approximately 4 metres of cumulative uplift relative to the 2005 post-deflation minimum by the early 2030, compared with the present value of roughly 1.6--1.7 metres (Figure~\ref{fig3}). This amplitude would substantially exceed the metre-scale uplifts of previous bradyseismic crises in the twentieth century, underscoring the exceptional nature of the current unrest phase, but still lower than the peak forewarning the 1538 Monte Nuovo century eruption \cite{rolandi20251538}.

However, the approach to a finite-time singularity imposes fundamental limits on predictability. In singular dynamical regimes, small uncertainties in the present state and in model parameters are amplified as the critical time is approached. It is a manifestation of the same positive feedback that drives the acceleration. Stochastic forward simulations show that the geodetic deformation can be predicted with good degree of belief until 2030 (Figure~\ref{fig3}). Beyond this horizon, the forecast of precise displacement values becomes qualitative rather than quantitative. Indeed, the limited trust horizon, where uncertainty reaches 50\% of the median prediction, follows shortly thereafter. These horizons are not artifacts of model inadequacy but intrinsic consequences of the singular dynamics itself, reflecting the limits on long-term predictability in systems governed by positive feedback \cite{bree2013prediction}.

\section*{DISCUSSIONS}
\subsection*{Physical dynamics and interpretation of the unrest}

\begin{figure}[!htb]
	\centering
	\includegraphics[width=\linewidth]{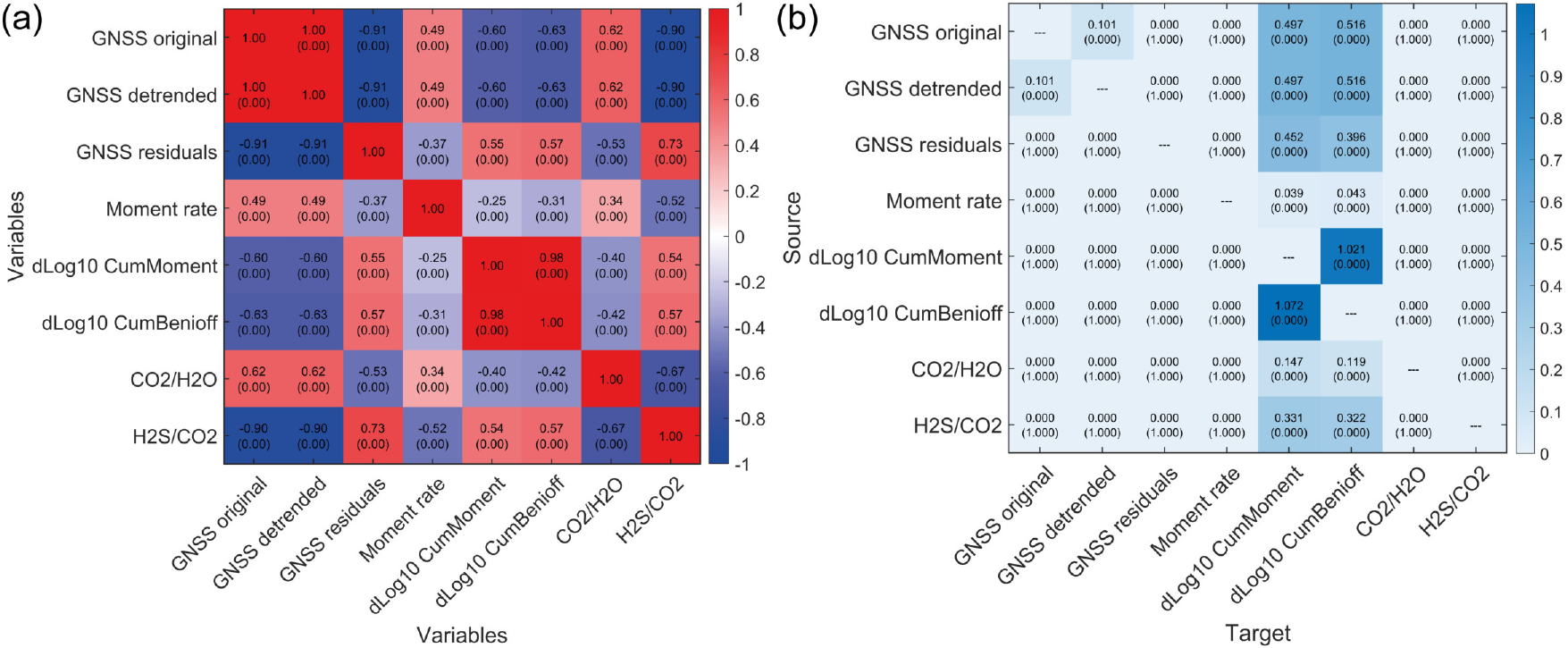}
	\caption{Dependence between physical quantities describing seismic-volcanic activity at Campi Flegrei. This dependence is quantified using (a) Spearman's rank correlation coefficient to assess monotonic similarities between time series, and (b) transfer entropy to infer directional causality. The analysis shows that, as expected, crustal deformation is positively correlated with seismicity (expressed as cumulative seismic moment and Benioff strain), with deformation acting as the primary driver of seismic release. Furthermore, a causal relationship is identified between the increasing H$_2$S/CO$_2$ ratio and both Benioff strain rate and seismic moment rate. This finding points to a progressive involvement of deeper, magmatic-derived fluids in the seismogenic process, as H$_2$S enrichment relative to CO$_2$ typically indicates input of fluids from undegassed magma at depth. In contrast, the CO$_2$/H$_2$O ratio shows a weaker and less evident causal signal, consistent with its stronger contribution from shallow hydrothermal sources and atmospheric contamination.}
	\label{fig4}
\end{figure}

The statistical evidence for a finite-time singularity raises the question of which physical processes drive the observed acceleration and what they imply for the system's future evolution. The dependence structure among the monitored variables -- including geodetic deformation, cumulative Benioff strain, and geochemical indicators such as the H$_2$S/CO$_2$ and CO$_2$/H$_2$O ratios measured at fumaroles -- provides key insight, revealing a coherent causal architecture linking magmatic processes to surface deformation and seismicity (Figure~\ref{fig4}).

At a first level, Spearman correlation analysis confirms a strong positive monotonic association between GNSS displacement and seismic variables, consistent with the established understanding that crustal stressing loads faults and triggers seismicity \cite{hainzl2026deformation}. Transfer entropy analysis refines this result by resolving the directionality of the coupling: significant information flow is detected from GNSS deformation to both cumulative Benioff strain and seismic moment rate ($T_{X \to Y} \geq 0.5$, $p < 0.001$), while the reverse direction is statistically insignificant. This asymmetry indicates that deformation acts as the primary driver of seismic release, with seismicity emerging as a consequence of stress accumulation rather than as an independent forcing agent. In parallel, increasingly intense and frequent bursts of seismicity are observed, consistent with \cite{giudicepietro2025burst} (see Supplementary materials). This behaviour is consistent with the damage mechanics framework developed by Kilburn and co-workers \cite{kilburn2017, kilburn2023}, in which progressive pressurisation of a magmatic source fractures the overlying crust, producing accelerating deformation and seismicity as coupled manifestations of the same underlying process.

Beneath this deformation-driven coupling, the geochemical signals provide evidence for a deeper driver of the unrest. 
Spearman correlation shows that GNSS is weakly correlated to CO$_2$/H$_2$O ratio, while the signal is more robust for the H$_2$S/CO$_2$ ratio. Transfer entropy analysis reveals a significant causal relationship between the fumarolic ratio H$_2$S/CO$_2$ and both the Benioff strain rate and seismic moment rate ($p < 0.001$), whereas the CO$_2$/H$_2$O ratio exhibits a weaker and less consistent signal. An increasing H$_2$S/CO$_2$ ratio reflects a growing contribution of volatiles sourced from deeper, relatively undegassed magma, as H$_2$S preferentially partitions into the magmatic vapour phase during early degassing stages \cite{chiodini2016, caliro2025}. In contrast, CO$_2$/H$_2$O is more strongly influenced by shallow hydrothermal processes and meteoric contamination, reducing its sensitivity to deep magmatic input. The causal link from H$_2$S/CO$_2$ to seismic energy release therefore indicates that accelerating seismicity is not solely a passive response to crustal stressing, but is actively promoted by the influx of deep magmatic volatiles. These fluids increase pore pressure, reduce effective normal stress on faults, and chemically weaken fracture surfaces through stress-corrosion processes, facilitating brittle failure at lower differential stresses \cite{sibson1994crustal}. The convergence of geophysical and geochemical observations supports an integrated magmatic-hydrothermal system in which sustained volatile supply drives both pressurisation and progressive damage accumulation in the shallow crust, consistent with \cite{astort2024tracking, amstutz2025volcano, giacomuzzi2025causal, rapagnani2025coupled, vanorio2025recurrence}. This interpretation is further supported by mutual information analysis (see Supplementary materials).

Spectral coherence analysis (see Supplementary materials) provides additional support, showing moderate coherence ($C_{XY} \sim 0.7$--$0.9$) between geodetic and seismic variables at periods longer than one year, and strong coherence between the H$_2$S/CO$_2$ ratio and seismic moment rate at periods of 2--4 months. These results reinforce the interpretation of a dynamically coupled system linking deep volatile input, deformation, and seismic release.

No log-periodic oscillations or discrete scaling signatures, of the type observed in other systems approaching failure \cite{lei2025unified, lei2025log}, are detected. Such signatures typically reflect large-scale coherent destabilisation and hierarchical organisation of damage preceding catastrophic failure. Their absence suggests that, although the system is accelerating toward a critical mechanical threshold, it has not yet entered a phase of system-wide coordination. Damage within the crust remains spatially distributed rather than organised into a fully connected failure structure.

This interpretation is consistent with the historical evolution of the system. The current uplift differs from previous bradyseismic episodes, which were characterised by shorter-duration, near-linear uplift phases followed by relatively abrupt reversals (Figure~\ref{fig1}B). In contrast, the present phase has sustained a power law acceleration for nearly two decades, consistent with cumulative damage resulting from multiple unrest episodes since the mid-twentieth century. Rather than reflecting a single self-limiting intrusion, the system appears to retain memory of past deformation, with each episode progressively weakening the crust and reducing its mechanical integrity.

In this framework, the estimated critical time $t_c \approx 2033$ does not necessarily correspond to a specific event, but instead marks the horizon at which the current accelerating trajectory, if maintained, is expected to culminate in a change of regime and transition to a new dynamical phase.

\subsection*{Conclusions and perspectives}

Taken together, our results depict Campi Flegrei as a system evolving toward a critical mechanical threshold under the sustained influence of deep magmatic volatile input, with deformation and seismicity emerging as coupled surface expressions of progressive crustal damage. The convergence of independent geodetic and seismic estimates toward a critical time in the range 2030--2034, the identified causal link between deep volatile supply and seismic energy release, and the clear departure from the dynamics of past bradyseismic episodes collectively indicate that the caldera has entered a new phase of heightened unrest and increasing hazard.

At the same time, several lines of evidence call for a cautious interpretation. The absence of log-periodic signatures, the spatially distributed character of seismicity, and the comparison with historical behaviour suggest that the system has not yet reached the level of large-scale mechanical coordination often associated with imminent eruptive transitions \cite{lei2025log}. In this context, the finite-time singularity trajectory provides a robust quantitative framework for characterising the accelerating dynamics and constraining the timing of a potential regime shift, but it does not uniquely determine the nature of the forthcoming transition, which remains intrinsically non-deterministic at this stage.

As the system evolves toward the inferred critical horizon, real-time monitoring combined with regularly updated forecasts will be essential. Such an approach is required both to support risk management in a densely populated and highly vulnerable area and to improve our understanding of predictability limits in complex volcanic systems governed by nonlinear and feedback-driven processes.


\section*{METHODS}
    \subsection*{Observational data and derived quantities}
	The analysis integrates two geophysical time series spanning the interval from January 2000 to January 2026, covering approximately 26 years of unrest at Campi Flegrei caldera. 
	Let $x(t)$ denote a scalar precursory observable defined over time $t$. The first observable consists of continuous Global Navigation Satellite System (GNSS) measurements of vertical displacement recorded at station RITE (14.126$^\circ$E, 40.823$^\circ$N), situated within the central sector of the caldera and recognised as the most sensitive station to the ongoing uplift. 
	We only select data from RITE following \cite{bevilacqua2024accelerating, iaccarino2025} since the deformation recorded at this site can be considered representative of the vertical displacements observed at all other GNSS stations. Indeed, the vertical displacements at other stations are lower than those recorded at the RITE station, consistent with the observed bell-shaped deformation pattern centered in the town of Pozzuoli.
	Geodetic time series from 2000 to 2026 and both the geochemical time-series have been digitally extracted from monthly reports published by Osservatorio Vesuviano \cite{vesuviano2024bollettini} using OpenCV library \cite{bradski2000opencv}. Deformation data from 1905 to 2000 (as showed in Figure \ref{fig1}b but not used in the analysis) refer to the Serapeo (14.120$^\circ$E, 40.826$^\circ$N, 500 m from RITE) water level \cite{del2010unrest}.
	The second observable is derived from the homogenized seismic catalogue for Italy, HORUS \cite{lolli2020homogenized}, from which we extract events located within a radius $R = 15$ km of the caldera center (14.14$^\circ$E, 40.83$^\circ$N) capturing local seismicity directly associated with the volcanic system while excluding tectonic seismic activity unrelated to the unrest process.
	
	For each earthquake in the filtered catalogue, characterised by its moment magnitude $M_w$, we compute the seismic moment $M_0$ (expressed in N$\cdot$m) through the standard empirical relation~\cite{hanks1979moment}
	\begin{equation}
		M_0 = 10^{1.5 M_w + 9.1}.
	\end{equation}
	The cumulative Benioff strain over time is then constructed as the chronological sum of the square roots of the individual seismic moments~\cite{benioff1951global} as 
	\begin{equation}
		x_{\text{s}}(t) = \sum_{t_i \leq t} \sqrt{M_0(t_i)}. 
	\end{equation}
	This scalar quantity, with physical dimensions of N$^{1/2} \cdot$ m$^{1/2}$, serves as a proxy for the cumulative brittle strain released by the seismogenic volume. In volcanic contexts like the one we investigate, accelerating rates of cumulative Benioff strain have been empirically associated with impending eruptive activity and are interpreted as a manifestation of progressive rock failure within the caldera~\cite{voight1988method, main1999applicability}.
	
	For the geodetic displacement time series $x_{\text{g}}(t)$, we remove a transient decaying component observed during the earliest portion of the record. This initial phase, characterised by a decreasing vertical position, is attributed to post-deflation relaxation following the 1982--1984 bradyseismic crisis. We model this transient using an exponential decay function of the form $x_{\text{decay}}(t) = p_1 e^{-p_2 (t - t_0)} + p_3$, where $t_0$ denotes the earliest observation time whose parameters are optimized via a standard fit procedure. The detrended displacement is then defined as $x(t) = x_{\text{g}}(t) - x_{\text{decay}}(t)$. This detrended quantity, together with the cumulative Benioff strain, constitutes the observational basis for the critical-time analysis.
	
	\subsection*{Forecasting models and finite-time singularities}
	\subsubsection*{The Voight model}
	The accelerating dynamics of both geophysical observables indicates that the system is evolving toward instability, but the precise functional form governing this acceleration is not known a priori. 
	
	A general framework for describing such an acceleration is provided by the empirical relationship proposed by Voight \cite{voight1989relation}, which states that the rate of change of the precursory observable $\dot{x}(t) = dx/dt$ is proportional to a power of the observable itself, according to the equation 
	\begin{equation}
		\dot{x}(t) = \gamma \, [x(t)]^\alpha,
		\label{eq:voight_rate}
	\end{equation}
	where $\gamma$ is a rate constant and the exponent $\alpha$ determines the nature of the acceleration. By integrating equation~\ref{eq:voight_rate}, different functional forms are obtained depending on the value of $\alpha$, corresponding to physically distinct processes:
	\begin{itemize}
		\item 	The case  $\alpha > 1$ corresponds to the presence of positive feedback, which induces a finite-time singularity -- a divergence of the observable at a critical time $t_c$. Setting the exponent parametrisation $\alpha = 1 + 1/\beta$ with $\beta > 0$, the solution of (\ref{eq:voight_rate}) takes the power law form
		\begin{equation}
			x(t) = A (t_c - t)^{-\beta} .
			\label{eq:singularity_solution}
		\end{equation}
		Here, $A > 0$ is an amplitude parameter determined by initial conditions and the rate constant $\gamma$, $\beta=1/(\alpha -1)$ is the positive exponent characterising the nature of the acceleration, $C$ is an offset representing a background level, and $t_c$ is the critical time at which the observable would diverge in the absence of saturation or regularising mechanisms. $t_c$ is called a ``movable singularity'' determined by the initial conditions.
		\item For $\alpha = 1$, the solution of (\ref{eq:voight_rate}) reads	
		\begin{equation}
			x_{\text{exp}}(t) = A e^{\gamma (t - t_0)} ~,
			\label{eq:exponential_solution}
		\end{equation}
		corresponding to an exponential growth.
		\item For $\alpha < 1$, the solution becomes
		\begin{equation}
			x_{\text{pl}}(t) = A (t + t_0)^{1 \over 1-\alpha}
			\label{eq:PL}
		\end{equation}
		corresponding to a slower power law growth as a function of time.
	\end{itemize}
	The physical interpretations of these three functional forms differ fundamentally. The singular model (equation~\ref{eq:singularity_solution}) describes a system governed by positive feedback that is intrinsically unsustainable, and therefore expected to culminate in a terminal transition at a finite critical time $t_c$. In contrast, the exponential model (equation~\ref{eq:exponential_solution}) and the power-law model (equation~\ref{eq:PL}) also represent accelerating dynamics (for $\alpha>0$), but do not impose a finite-time bound: their growth can, in principle, continue indefinitely without requiring a transition. Comparing these models thus provides a direct means of discriminating between an acceleration driven by a finite-time runaway process and one reflecting a more gradual, unbounded instability.
	
	\subsubsection*{Generalised regularised singularity model}
	The pure power law singularity $x(t) \sim A (t_c - t)^{-\beta}$ provides an excellent description of an intermediate asymptotic regime, capturing a substantial portion of the dynamics, before a crossover to other functional forms ensures finiteness and removes the apparent divergence as $t \to t_c^-$.
 In physical systems, the growth described by such a power law is naturally tempered as $t$ approaches $t_c$, with finite-size effects, dissipative processes, and negative feedback mechanisms progressively intervening to arrest the acceleration and ensure a bounded evolution, leading to so-called \textquotedblleft ghost singularities'' \cite{smug2018predicting}. In volcanic systems, such regularisation can arise from the finite volume of the magma reservoir, the transition from quasi-elastic to ductile or brittle failure of the overlying rock, and the limited availability of strain energy for seismic release.
 	
To account for the onset of nonlinear saturation effects while preserving the interpretation of the negative term as the leading analytic feedback counteracting the singular growth, we consider the generalisation of equation (\ref{eq:voight_rate})
\begin{equation}
\label{btggrb}
    \frac{dx}{dt} = \gamma x^\alpha - a x^m,
    \qquad
    m = \lfloor \alpha \rfloor + 1,
    \qquad
    \gamma>0,\; a\ge 0,\; \alpha>1.
\end{equation}
Here, \(m\) is the smallest integer strictly larger than \(\alpha\), so that the regularising term is the first analytic negative feedback term compatible with the leading nonlinearity \(x^\alpha\).

For \(a=0\), the solution is given by expression (\ref{eq:singularity_solution})
with $\beta = \frac{1}{\alpha-1}$, $A = \big[\gamma(\alpha-1)\big]^{-\beta}$ and $ t_c = t_0 + \frac{x_0^{\,1-\alpha}}{\gamma(\alpha-1)}$.

We seek a perturbative solution of the form $x(t)=x_{\mathrm{pl}}(t)+a\,x_1(t)+\mathcal{O}(a^2)$, considering $a$ as a small parameter.
Substituting into (\ref{btggrb})
and expanding to first order in \(a\) gives $\dot{x}_{\mathrm{pl}} + a \dot{x}_1  =
    \gamma\left(x_{\mathrm{pl}}^\alpha + a \alpha x_{\mathrm{pl}}^{\alpha-1}x_1\right)   - a x_{\mathrm{pl}}^m   + \mathcal{O}(a^2)$.
Using \(\dot{x}_{\mathrm{pl}}=\gamma x_{\mathrm{pl}}^\alpha\), the \(\mathcal{O}(a)\) equation is
\begin{equation}
    \dot{x}_1 - \alpha \gamma x_{\mathrm{pl}}^{\alpha-1}x_1 = -x_{\mathrm{pl}}^m.
\end{equation}
Introducing  $\Delta = t_c-t$, we finally obtain the reduced equation
\begin{equation}
    \frac{d}{d\Delta}\!\left(
        \Delta^{\frac{\alpha}{\alpha-1}}x_1
    \right)
    =
    \big[\gamma(\alpha-1)\big]^{-\frac{m}{\alpha-1}}
    \Delta^{\frac{\alpha-m}{\alpha-1}}.
\end{equation}

The solution takes the general form 
\begin{equation}
    x(t)
    \approx
    \frac{A}{(t_c-t)^\beta}
    -
    a\,B\,(t_c-t)^{-\eta},
\end{equation}
with
\begin{equation}
    \beta=\frac{1}{\alpha-1},
    \qquad
    \eta=\frac{m-\alpha+1}{\alpha-1},
    \qquad
    B=\frac{\alpha-1}{m-2\alpha+1}\big[\gamma(\alpha-1)\big]^{-\frac{m}{\alpha-1}},
\end{equation}
in the non-resonant case \(2\alpha-m-1 \neq 0\), and the logarithmic replacement
\begin{equation}
    x(t)
    \approx
    \frac{A}{(t_c-t)^\beta}
    - a\, {\widetilde{B} \over (t_c-t)^{\beta+1}}
    \ln\!\left(\frac{t_c-t_0}{t_c-t}\right)
\end{equation}
in the resonant case \(2\alpha-m-1=0\). We include an additive baseline \(C\) to account for a pre-existing background level of the variable.

Given the strong positive correlation between the exponent \(\beta\) and the regularisation parameter \(a\) (see Supplementary materials), we adopt a sequential estimation strategy. First, the unregularised model is fitted with \(\beta\) free to vary, returning an optimal exponent \(\beta^*\). This value is then fixed in the regularised model, and the remaining parameters \((t_c, A, C, a)\) are estimated via nonlinear least squares over a joint grid in \((t_c, a)\). For each candidate pair \((t_c, a)\), the linear parameters \((A, C)\) are determined analytically as the solution to the normal equations. Because the resonant logarithmic form makes the fit considerably more robust against the partial degeneracy between \(\beta\) and \(a\) that would otherwise destabilise the estimation, we adopt it as our default model; the fitted exponent \(\beta^*\) should accordingly be interpreted as an effective exponent within this regularisation scheme rather than as a strict determination of the underlying nonlinearity index \(\alpha\).
	
	\subsubsection*{Prediction trust horizons}
The regularised formulation, together with the amplification of stochastic fluctuations as the system approaches the critical regime, naturally limits the predictive reliability of the model. To quantify this limitation, we introduce two trust horizons based on the relative growth of prediction uncertainty compared to the expected signal.

Let \(\bar{x}(t)\) denote the median trajectory obtained from an ensemble of stochastic realisations of the regularised model, and let \([x_{16}(t), x_{84}(t)]\) and \([x_{2.5}(t), x_{97.5}(t)]\) denote the corresponding 68\% and 95\% prediction intervals. As \(t \to t_c\), these intervals broaden rapidly due to the combined effects of uncertainty in the inferred critical time \(t_c\) and the intrinsic amplification of fluctuations by the accelerating dynamics.

We define the \textquotedblleft good trust horizon'' \(t_h^{(1)}\) as the earliest time at which the width of the 95\% prediction interval exceeds 25\% of the magnitude of the median prediction, i.e.,
\[
x_{97.5}(t) - x_{2.5}(t) \ge 0.25\, |\bar{x}(t)|.
\]
Similarly, the \textquotedblleft limited trust horizon'' \(t_h^{(2)}\) is defined using the more permissive threshold of 50\%.

For \(t < t_h^{(1)}\), the forecast remains quantitatively reliable. In the interval \(t_h^{(1)} < t < t_h^{(2)}\), uncertainty becomes a significant fraction of the signal and forecasts should be interpreted with caution. For \(t > t_h^{(2)}\), uncertainty dominates the signal, and predictions are best regarded as qualitative indicators of the evolving dynamics rather than precise estimates.

	\subsection*{Diagnostic tests}
	For the specific case $\beta = 1$, the singular model reduces to $x(t) = A/(t_c - t) + C$. This formulation suggests a simple and powerful diagnostic: if the system follows singular dynamics, then plotting the observable $x(t)$ minus the background constant $C$ against the inverse transformed time $1/(t_c - t)$ returns a linear relationship \cite{voight1988method,voight1989relation,cornelius1995graphical}. For a candidate $t_c$, the coefficient of determination $R^2_{\text{lin}}$ from a linear regression of $x(t)$ on $1/(t_c - t)$ quantifies the degree to which the data support a singular acceleration toward that critical time. In our analysis, this linearising test always return excellent $R^2$ values (exceeding 0.99 for optimal windows; Supplementary Tables), providing strong empirical support for the singular ansatz.
	
	However, the key assessment and physical interpretation of seismic-volcanic activity evolution rests on understanding whether positive feedback mechanisms are at work or not. Therefore, we employ specific tests to answer this question. 
	
	As a preliminary step, we identify the time window where the primary driver is optimally detected and less affected by second order effects (e.g., the initial post 1983-1984 crisis deflating relaxation). We define candidate starting times $t_s$ for the analysis. For each of them, the parameters of the singularity model $(t_c, A, \beta, C)$ are estimated through a two-stage optimisation procedure. In the first step, an initial estimate of $t_c$ is obtained by minimizing the residuals. In the second stage, the full four-parameter nonlinear model is fitted using the Levenberg--Marquardt algorithm \cite{marquardt1963algorithm}. The exponential model is fitted analogously over the accelerating portion of the time series, defined as the same interval.
	
	Model comparison is evaluated over the common time interval up to present time $t \in [t_s, t_{\text{end}}]$ employing the Root-mean-square error defined as $\text{RMSE} = \sqrt{\frac{1}{n}\sum_{i=1}^n (x_i^{\text{obs}} - x_i^{\text{pred}})^2}$, where $n$ is the number of data points, and the Akaike Information Criterion \cite{akaike2003new}, where $\text{AIC} = n \ln(\text{RSS}/n) + 2k$, with three degrees of freedom, $k=3$, for the exponential and the power law singularity models. Lower values of AIC indicate superior model performance after penalising for parameter complexity. A difference $|\Delta\text{AIC}| > 3$ is usually considered substantial evidence in favour of the model with the lower AIC.
	
	For each parameter, uncertainties are quantified using a residual bootstrap procedure. For each bootstrap iteration, a synthetic dataset is generated by adding randomly resampled residuals with replacement to the best-fitting model prediction. The full estimation procedure is then repeated, leading to a bootstrap distribution for each parameter. 
	
	\subsubsection*{Sensitivity analysis and regime identification}
	A fundamental challenge in applying singularity models to precursory phenomena is the potential dependence of the estimated critical time $t_c$ on the choice of the analysis starting time $t_s$. To investigate this sensitivity and to identify the emergence of a coherent singular regime in geodetic data, we systematically varied $t_s$ from 2000 to 2015 (2025 for the Benioff strain) in relatively small increments (6-12 months). For each $t_s$, we applied the complete estimation procedure described above.
	
	The quality of the singularity fit, as measured by the RMSE, exhibits a systematic dependence on the time span $\Delta t = t_{\text{end}} - t_s$. As the system approaches a critical point, the singular behaviour is expected to become progressively more pronounced, leading to an improvement in fit quality relative to the background trend. To isolate this effect, we modeled the expected RMSE as a function of the time span. For the GNSS data, a linear decreasing trend is observed; while for the Benioff strain data, the RMSE exhibits exponentially decreasing residuals. 
	The residuals from these fitted trends quantify the deviation of each analysis window from the expected fit quality. This procedure is conceptually identical to the method of Lagrange multipliers for removing a systematic trend \cite{demosdiSor19}, allowing us to identify the time span at which the singular fit performs optimally relative to the expectation based on data density alone. The starting time $t_s^*$ with the minimum (most negative) residual identifies the analysis window for which the singular regime is most clearly expressed and must therefore being considered as more informative than other windows. We employ a specific weighting procedure for this goal. 
	
	\subsubsection*{Weighted sensitivity estimates}
	The sensitivity analysis provides an ensemble of critical time estimates $\{t_c(t_s)\}$ corresponding to different choices of the starting time $t_s$, each accompanied by an uncertainty $\sigma_{t_c}(t_s)$ and a fit quality metric $\text{RMSE}(t_s)$. To synthesise these multiple estimates into a single robust forecast, we construct a weighted combination that accounts for both the formal parameter uncertainty and the degree to which each analysis window captures the singular regime.
	
	We defined a total normalised uncertainty for each starting time $t_s$ as the quadrature sum of two components
	\begin{equation}
		U(t_s) = \sqrt{ \lambda_\sigma^2 \, \tilde{\sigma}_{t_c}^2(t_s) + \lambda_{\text{RMSE}}^2 \, \widetilde{\Delta\text{RMSE}}^2(t_s) },
	\end{equation}
	where $\tilde{\sigma}_{t_c}(t_s) = \sigma_{t_c}(t_s) / \max_{t_s'} \sigma_{t_c}(t_s')$ is the normalised bootstrap uncertainty, and $\widetilde{\Delta\text{RMSE}}(t_s)$ quantifies the normalised deviation of the RMSE residual from its minimum value. The coefficients are set such that they reflect the reasonable choice that regime quality should dominate over formal fitting uncertainty in determining estimate reliability. See Supplementary Material for details.  
	The composite weight assigned to the estimate from starting time $t_s$ is then 
	\begin{equation}
		w(t_s) \propto 1 / U(t_s). 
	\end{equation}
	The weighted mean critical time is computed as 
	\begin{equation}
		\bar{t}_c = \sum_{t_s} w(t_s) t_c(t_s),
	\end{equation}
	and confidence intervals were obtained via weighted bootstrap resampling with $N_{\text{boot}} = 1000$, where $ \sum_{t_s} w(t_s)=1$.
	
	\subsection*{Dependence and causality analysis}
	To investigate the physical coupling between geodetic deformation, seismicity, and geochemical emissions at Campi Flegrei, we perform  dependence and causality analyses on the standardised time series over their common temporal interval.
	
	As a first measure of dependence between physical parameters, we employ the Spearman correlation coefficient $\rho$ \cite{spearman1961proof}, which quantifies the strength and direction of monotonic association between two variables $X$ and $Y$ without assuming linearity or normality. Values of $\rho$ near $+1$ indicate strong positive monotonic association, values near $-1$ strong negative association, and values near $0$ the absence of monotonic association. 
    
    Mutual information $I(X; Y)$ provides a more general measure of statistical dependence, without assuming any specific functional form. For two discrete random variables $X$ and $Y$ with joint probability mass function $p(x,y)$ and marginal distributions $p(x)$ and $p(y)$, the mutual information is defined as $I(X; Y) = \sum_{x \in \mathcal{X}} \sum_{y \in \mathcal{Y}} p(x,y) \log_2 [p(x,y) / (p(x) \, p(y))]$ \cite{schreiber2000measuring}. The quantity $I(X;Y)$ is non-negative, and it equals zero if and only if $X$ and $Y$ are independent. It measures the reduction in uncertainty about $X$ (in bits) gained by knowledge of $Y$. We discretise each variable into bins with the same probability and assess statistical significance via permutation testing with 100 random shuffles of the pairing between variables. A significant mutual information between GNSS and seismic variables confirms that the two data streams share information beyond what is captured by linear correlation alone. 
    
    While mutual information captures contemporaneous statistical dependence, it does not reveal the direction of information flow.
    Transfer entropy $T_{X \to Y}$ provides a directed measure of predictive information transfer between dynamical variables \cite{schreiber2000measuring}. For a time lag of one step, the transfer entropy from a source variable $X$ to a target variable $Y$ is defined as $T_{X \to Y} = \sum p(y_{t+1}, y_t, x_t) \log_2 [p(y_{t+1} \mid y_t, x_t) / p(y_{t+1} \mid y_t)]$, where $p(y_{t+1} \mid y_t)$ is the conditional probability of the future state of $Y$ given its own past, and $p(y_{t+1} \mid y_t, x_t)$ is the corresponding conditional probability when the past of $X$ is additionally known. A positive value of $T_{X \to Y}$ indicates that knowledge of $X$'s history reduces the uncertainty in predicting $Y$'s future beyond what is achievable using $Y$'s history alone. We assessed the statistical significance of the transfer entropy following the same procedure we employed for mutual information. 

    As a support in the causality analysis, we estimated the Granger causality as a complementary linear time-domain test for predictive influence~\cite{granger1969investigating}. 

    Spectral coherence $C_{XY}(f)$ measures the frequency-domain linear correlation between two signals $X$ and $Y$, defined as 
    $C_{XY}(f) = |P_{xy}(f)|^2 / [P_{xx}(f) P_{yy}(f)]$,
    where $P_{XY}(f)$ is the cross-spectral density and $P_{XX}(f)$, $P_{YY}(f)$ are the power spectral densities \cite{koopmans1995spectral}. Coherence ranges from 0 (no linear relationship) to 1 (perfect linear relationship at frequency $f$). High coherence at specific frequency bands would indicate that the two variables share common periodicities. 
	
\bibliography{Biblio_Flegrei}

\begin{thebibliography}{10}
\urlstyle{rm}
\expandafter\ifx\csname url\endcsname\relax
  \def\url#1{\texttt{#1}}\fi
\expandafter\ifx\csname urlprefix\endcsname\relax\def\urlprefix{URL }\fi
\expandafter\ifx\csname doiprefix\endcsname\relax\def\doiprefix{DOI: }\fi
\providecommand{\bibinfo}[2]{#2}
\providecommand{\eprint}[2][]{\url{#2}}

\bibitem{rampino1992volcanic}
\bibinfo{author}{Rampino, M.~R.} \& \bibinfo{author}{Self, S.}
\newblock \bibinfo{journal}{\bibinfo{title}{Volcanic winter and accelerated
  glaciation following the {Toba} super-eruption}}.
\newblock {\emph{\JournalTitle{Nature}}} \textbf{\bibinfo{volume}{359}},
  \bibinfo{pages}{50--52} (\bibinfo{year}{1992}).

\bibitem{rampino2002supereruptions}
\bibinfo{author}{Rampino, M.~R.}
\newblock \bibinfo{journal}{\bibinfo{title}{Supereruptions as a threat to
  civilizations on {Earth-like} planets}}.
\newblock {\emph{\JournalTitle{Icarus}}} \textbf{\bibinfo{volume}{156}},
  \bibinfo{pages}{562--569} (\bibinfo{year}{2002}).

\bibitem{self2006effects}
\bibinfo{author}{Self, S.}
\newblock \bibinfo{journal}{\bibinfo{title}{The effects and consequences of
  very large explosive volcanic eruptions}}.
\newblock {\emph{\JournalTitle{Philosophical Transactions of the Royal Society
  A: Mathematical, Physical and Engineering Sciences}}}
  \textbf{\bibinfo{volume}{364}}, \bibinfo{pages}{2073--2097}
  (\bibinfo{year}{2006}).

\bibitem{miller2008supervolcanoes}
\bibinfo{author}{Miller, C.~F.} \& \bibinfo{author}{Wark, D.~A.}
\newblock \bibinfo{journal}{\bibinfo{title}{Supervolcanoes and their explosive
  supereruptions}}.
\newblock {\emph{\JournalTitle{Elements}}} \textbf{\bibinfo{volume}{4}},
  \bibinfo{pages}{11--15} (\bibinfo{year}{2008}).

\bibitem{meredith2026high}
\bibinfo{author}{Meredith, E.~S.}, \bibinfo{author}{Handley, H.},
  \bibinfo{author}{Jenkins, S.~F.}, \bibinfo{author}{Chim, M.~M.} \&
  \bibinfo{author}{Gregg, C.}
\newblock \bibinfo{journal}{\bibinfo{title}{High-impact low-probability events:
  Exposure to potential large-magnitude explosive volcanic eruptions}}.
\newblock {\emph{\JournalTitle{Anthropocene}}} \bibinfo{pages}{100542}
  (\bibinfo{year}{2026}).

\bibitem{kilburn2023}
\bibinfo{author}{Kilburn, C.}, \bibinfo{author}{Carlino, S.},
  \bibinfo{author}{Danesi, S.} \emph{et~al.}
\newblock \bibinfo{journal}{\bibinfo{title}{Potential for rupture before
  eruption at {Campi Flegrei caldera, Southern Italy}}}.
\newblock {\emph{\JournalTitle{Commun Earth Environ}}}
  \textbf{\bibinfo{volume}{4}}, \bibinfo{pages}{190},
  \doiprefix\url{10.1038/s43247-023-00842-1} (\bibinfo{year}{2023}).

\bibitem{iaccarino2025}
\bibinfo{author}{Iaccarino, A.}, \bibinfo{author}{Picozzi, M.},
  \bibinfo{author}{De~Landro, G.} \& \bibinfo{author}{Spallarossa, D.}
\newblock \bibinfo{journal}{\bibinfo{title}{Preparatory phase of major
  earthquakes during {Campi Flegrei unrest (2020-2024)}}}.
\newblock {\emph{\JournalTitle{Journal of Geophysical Research: Solid Earth}}}
  \textbf{\bibinfo{volume}{130}}, \bibinfo{pages}{e2025JB031777},
  \doiprefix\url{10.1029/2025JB031777} (\bibinfo{year}{2025}).

\bibitem{acocella2015}
\bibinfo{author}{Acocella, V.}, \bibinfo{author}{Di~Lorenzo, R.},
  \bibinfo{author}{Newhall, C.} \& \bibinfo{author}{Scandone, R.}
\newblock \bibinfo{journal}{\bibinfo{title}{An overview of recent (1988 to
  2014) caldera unrest: Knowledge and perspectives}}.
\newblock {\emph{\JournalTitle{Rev. Geophys.}}} \textbf{\bibinfo{volume}{53}},
  \bibinfo{pages}{896--955}, \doiprefix\url{10.1002/2015RG000492}
  (\bibinfo{year}{2015}).

\bibitem{allison2021highly}
\bibinfo{author}{Allison, C.~M.}, \bibinfo{author}{Roggensack, K.} \&
  \bibinfo{author}{Clarke, A.~B.}
\newblock \bibinfo{journal}{\bibinfo{title}{Highly explosive basaltic eruptions
  driven by {CO2} exsolution}}.
\newblock {\emph{\JournalTitle{Nature communications}}}
  \textbf{\bibinfo{volume}{12}}, \bibinfo{pages}{217} (\bibinfo{year}{2021}).

\bibitem{keller2026volatile}
\bibinfo{author}{Keller, F.}, \bibinfo{author}{Townsend, M.},
  \bibinfo{author}{Troch, J.} \& \bibinfo{author}{Huber, C.}
\newblock \bibinfo{journal}{\bibinfo{title}{Volatile resorption expedites
  eruption onset in large silicic systems}}.
\newblock {\emph{\JournalTitle{Nature Communications}}}
  (\bibinfo{year}{2026}).

\bibitem{reid2004massive}
\bibinfo{author}{Reid, M.~E.}
\newblock \bibinfo{journal}{\bibinfo{title}{Massive collapse of volcano
  edifices triggered by hydrothermal pressurization}}.
\newblock {\emph{\JournalTitle{Geology}}} \textbf{\bibinfo{volume}{32}},
  \bibinfo{pages}{373--376} (\bibinfo{year}{2004}).

\bibitem{sparks2003forecasting}
\bibinfo{author}{Sparks, R. S.~J.}
\newblock \bibinfo{journal}{\bibinfo{title}{Forecasting volcanic eruptions}}.
\newblock {\emph{\JournalTitle{Earth and Planetary Science Letters}}}
  \textbf{\bibinfo{volume}{210}}, \bibinfo{pages}{1--15}
  (\bibinfo{year}{2003}).

\bibitem{dempsey2020automatic}
\bibinfo{author}{Dempsey, D.}, \bibinfo{author}{Cronin, S.~J.},
  \bibinfo{author}{Mei, S.} \& \bibinfo{author}{Kempa-Liehr, A.~W.}
\newblock \bibinfo{journal}{\bibinfo{title}{Automatic precursor recognition and
  real-time forecasting of sudden explosive volcanic eruptions at {Whakaari,
  New Zealand}}}.
\newblock {\emph{\JournalTitle{Nature communications}}}
  \textbf{\bibinfo{volume}{11}}, \bibinfo{pages}{3562} (\bibinfo{year}{2020}).

\bibitem{gomez2025there}
\bibinfo{author}{Gomez-Patron, A.} \emph{et~al.}
\newblock \bibinfo{journal}{\bibinfo{title}{Are there thermal precursors to
  eruptions detectable by {ASTER? Evaluating 22 years of global medium
  resolution satellite thermal observations at 200+ volcanoes}}}.
\newblock {\emph{\JournalTitle{Journal of Geophysical Research: Solid Earth}}}
  \textbf{\bibinfo{volume}{130}}, \bibinfo{pages}{e2024JB030427}
  (\bibinfo{year}{2025}).

\bibitem{chiodini2016}
\bibinfo{author}{Chiodini, G.}, \bibinfo{author}{Paonita, A.},
  \bibinfo{author}{Aiuppa, A.} \emph{et~al.}
\newblock \bibinfo{journal}{\bibinfo{title}{Magmas near the critical degassing
  pressure drive volcanic unrest towards a critical state}}.
\newblock {\emph{\JournalTitle{Nat Commun}}} \textbf{\bibinfo{volume}{7}},
  \bibinfo{pages}{13712}, \doiprefix\url{10.1038/ncomms13712}
  (\bibinfo{year}{2016}).

\bibitem{bevilacqua2022}
\bibinfo{author}{Bevilacqua, A.}, \bibinfo{author}{De~Martino, P.},
  \bibinfo{author}{Giudicepietro, F.} \emph{et~al.}
\newblock \bibinfo{journal}{\bibinfo{title}{Data analysis of the unsteadily
  accelerating {GPS and seismic records at Campi Flegrei caldera from 2000 to
  2020}}}.
\newblock {\emph{\JournalTitle{Sci Rep}}} \textbf{\bibinfo{volume}{12}},
  \bibinfo{pages}{19175}, \doiprefix\url{10.1038/s41598-022-23628-5}
  (\bibinfo{year}{2022}).

\bibitem{voight1988method}
\bibinfo{author}{Voight, B.}
\newblock \bibinfo{journal}{\bibinfo{title}{A method for prediction of volcanic
  eruptions}}.
\newblock {\emph{\JournalTitle{Nature}}} \textbf{\bibinfo{volume}{332}},
  \bibinfo{pages}{125--130} (\bibinfo{year}{1988}).

\bibitem{voight1989relation}
\bibinfo{author}{Voight, B.}
\newblock \bibinfo{journal}{\bibinfo{title}{A relation to describe
  rate-dependent material failure}}.
\newblock {\emph{\JournalTitle{Science}}} \textbf{\bibinfo{volume}{243}},
  \bibinfo{pages}{200--203} (\bibinfo{year}{1989}).

\bibitem{sornette2004}
\bibinfo{author}{Sornette, D.}
\newblock \emph{\bibinfo{title}{{Critical Phenomena in Natural Sciences: Chaos,
  Fractals, Self-organization and Disorder: Concepts and Tools}}}.
\newblock Springer Series in Synergetics (\bibinfo{publisher}{Springer},
  \bibinfo{address}{Berlin Heidelberg}, \bibinfo{year}{2004}).

\bibitem{lei2025log}
\bibinfo{author}{Lei, Q.} \& \bibinfo{author}{Sornette, D.}
\newblock \bibinfo{journal}{\bibinfo{title}{Log-periodic signatures prior to
  volcanic eruptions: evidence from 34 events}}.
\newblock {\emph{\JournalTitle{Earth and Planetary Science Letters}}}
  \textbf{\bibinfo{volume}{666}}, \bibinfo{pages}{119496}
  (\bibinfo{year}{2025}).

\bibitem{lei2025unified}
\bibinfo{author}{Lei, Q.} \& \bibinfo{author}{Sornette, D.}
\newblock \bibinfo{journal}{\bibinfo{title}{Unified failure model for
  landslides, rockbursts, glaciers, and volcanoes}}.
\newblock {\emph{\JournalTitle{Communications Earth \& Environment}}}
  \textbf{\bibinfo{volume}{6}}, \bibinfo{pages}{390} (\bibinfo{year}{2025}).

\bibitem{tan2025clearer}
\bibinfo{author}{Tan, X.} \emph{et~al.}
\newblock \bibinfo{journal}{\bibinfo{title}{A clearer view of the current phase
  of unrest at {Campi Flegrei caldera}}}.
\newblock {\emph{\JournalTitle{Science}}} \textbf{\bibinfo{volume}{390}},
  \bibinfo{pages}{70--75} (\bibinfo{year}{2025}).

\bibitem{vilardo_2024_11190448}
\bibinfo{author}{Vilardo, G.}, \bibinfo{author}{Bellucci~Sessa, E.} \&
  \bibinfo{author}{Sansivero, F.}
\newblock \bibinfo{title}{Campi {Flegrei topographic 3D elevation Map
  (Italy)}}, \doiprefix\url{10.5281/zenodo.11190448} (\bibinfo{year}{2024}).

\bibitem{del2010unrest}
\bibinfo{author}{Del~Gaudio, C.}, \bibinfo{author}{Aquino, I.},
  \bibinfo{author}{Ricciardi, G.}, \bibinfo{author}{Ricco, C.} \&
  \bibinfo{author}{Scandone, R.}
\newblock \bibinfo{journal}{\bibinfo{title}{Unrest episodes at {Campi Flegrei:
  A reconstruction of vertical ground movements during 1905--2009}}}.
\newblock {\emph{\JournalTitle{Journal of Volcanology and Geothermal
  Research}}} \textbf{\bibinfo{volume}{195}}, \bibinfo{pages}{48--56}
  (\bibinfo{year}{2010}).

\bibitem{vesuviano2024bollettini}
\bibinfo{author}{Vesuviano, I.-O.}
\newblock \bibinfo{journal}{\bibinfo{title}{Bollettini di {Sorveglianza dei
  Vulcani Campani}}}.
\newblock {\emph{\JournalTitle{Osservatorio Vesuviano. Istituto Nazionale Di
  Geofisica e Vulcanologia (INGV)}}}  (\bibinfo{year}{2024}).

\bibitem{cardellini2017monitoring}
\bibinfo{author}{Cardellini, C.} \emph{et~al.}
\newblock \bibinfo{journal}{\bibinfo{title}{Monitoring diffuse volcanic
  degassing during volcanic unrests: the case of {Campi Flegrei (Italy)}}}.
\newblock {\emph{\JournalTitle{Scientific Reports}}}
  \textbf{\bibinfo{volume}{7}}, \bibinfo{pages}{6757} (\bibinfo{year}{2017}).

\bibitem{sabbarese2020continuous}
\bibinfo{author}{Sabbarese, C.} \emph{et~al.}
\newblock \bibinfo{journal}{\bibinfo{title}{Continuous radon monitoring during
  seven years of volcanic unrest at {Campi Flegrei caldera (Italy)}}}.
\newblock {\emph{\JournalTitle{Scientific Reports}}}
  \textbf{\bibinfo{volume}{10}}, \bibinfo{pages}{9551} (\bibinfo{year}{2020}).

\bibitem{bianco2022permanent}
\bibinfo{author}{Bianco, F.} \emph{et~al.}
\newblock \bibinfo{title}{The permanent monitoring system of the {Campi Flegrei
  caldera, Italy}}.
\newblock In \emph{\bibinfo{booktitle}{Campi Flegrei: A restless Caldera in a
  densely populated area}}, \bibinfo{pages}{219--237}
  (\bibinfo{publisher}{Springer}, \bibinfo{year}{2022}).

\bibitem{carlino2021brief}
\bibinfo{author}{Carlino, S.}
\newblock \bibinfo{journal}{\bibinfo{title}{Brief history of volcanic risk in
  the {Neapolitan area (Campania, southern Italy)}: a critical review}}.
\newblock {\emph{\JournalTitle{Natural Hazards and Earth System Sciences}}}
  \textbf{\bibinfo{volume}{21}}, \bibinfo{pages}{3097--3112}
  (\bibinfo{year}{2021}).

\bibitem{cannatelli2020ground}
\bibinfo{author}{Cannatelli, C.}, \bibinfo{author}{Spera, F.~J.},
  \bibinfo{author}{Bodnar, R.~J.}, \bibinfo{author}{Lima, A.} \&
  \bibinfo{author}{De~Vivo, B.}
\newblock \bibinfo{journal}{\bibinfo{title}{Ground movement (bradyseism) in the
  {Campi Flegrei} volcanic area: a review}}.
\newblock {\emph{\JournalTitle{Vesuvius, Campi Flegrei, and campanian
  volcanism}}} \bibinfo{pages}{407--433} (\bibinfo{year}{2020}).

\bibitem{ricci2013volcanic}
\bibinfo{author}{Ricci, T.}, \bibinfo{author}{Barberi, F.},
  \bibinfo{author}{Davis, M.}, \bibinfo{author}{Isaia, R.} \&
  \bibinfo{author}{Nave, R.}
\newblock \bibinfo{journal}{\bibinfo{title}{Volcanic risk perception in the
  {Campi Flegrei area}}}.
\newblock {\emph{\JournalTitle{Journal of Volcanology and Geothermal
  Research}}} \textbf{\bibinfo{volume}{254}}, \bibinfo{pages}{118--130}
  (\bibinfo{year}{2013}).

\bibitem{scarpati1993neapolitan}
\bibinfo{author}{Scarpati, C.}, \bibinfo{author}{Cole, P.} \&
  \bibinfo{author}{Perrotta, A.}
\newblock \bibinfo{journal}{\bibinfo{title}{The {Neapolitan Yellow Tuff large
  volume multiphase eruption from Campi Flegrei, southern Italy}}}.
\newblock {\emph{\JournalTitle{Bulletin of Volcanology}}}
  \textbf{\bibinfo{volume}{55}}, \bibinfo{pages}{343--356}
  (\bibinfo{year}{1993}).

\bibitem{deino2004age}
\bibinfo{author}{Deino, A.~L.}, \bibinfo{author}{Orsi, G.},
  \bibinfo{author}{de~Vita, S.} \& \bibinfo{author}{Piochi, M.}
\newblock \bibinfo{journal}{\bibinfo{title}{The age of the {Neapolitan Yellow
  Tuff caldera-forming eruption (Campi Flegrei caldera - Italy) assessed by
  40Ar/39Ar} dating method}}.
\newblock {\emph{\JournalTitle{Journal of Volcanology and Geothermal
  Research}}} \textbf{\bibinfo{volume}{133}}, \bibinfo{pages}{157--170}
  (\bibinfo{year}{2004}).

\bibitem{gianchiglia2025fine}
\bibinfo{author}{Gianchiglia, F.} \emph{et~al.}
\newblock \bibinfo{journal}{\bibinfo{title}{Fine ash from the {Campanian
  Ignimbrite super-eruption, 40 ka, southern Italy}: implications for dispersal
  mechanisms and health hazard}}.
\newblock {\emph{\JournalTitle{Scientific Reports}}}
  \textbf{\bibinfo{volume}{15}}, \bibinfo{pages}{20039} (\bibinfo{year}{2025}).

\bibitem{natale2025magma}
\bibinfo{author}{Natale, J.} \& \bibinfo{author}{Vitale, S.}
\newblock \bibinfo{journal}{\bibinfo{title}{Magma chamber failure and dyke
  injection threshold for magma-driven unrest at {Campi Flegrei caldera}}}.
\newblock {\emph{\JournalTitle{Nature Communications}}}
  \textbf{\bibinfo{volume}{16}}, \bibinfo{pages}{7658} (\bibinfo{year}{2025}).

\bibitem{caricchi2026scenario}
\bibinfo{author}{Caricchi, L.}, \bibinfo{author}{Lormand, C.},
  \bibinfo{author}{Carlino, S.}, \bibinfo{author}{Pivetta, T.} \&
  \bibinfo{author}{Simpson, G.}
\newblock \bibinfo{journal}{\bibinfo{title}{Scenario-based forecast of the
  evolution of 75 years of unrest at {Campi Flegrei caldera (Italy)}}}.
\newblock {\emph{\JournalTitle{Communications Earth \& Environment}}}
  \textbf{\bibinfo{volume}{7}}, \bibinfo{pages}{37} (\bibinfo{year}{2026}).

\bibitem{iervolino2024seismic}
\bibinfo{author}{Iervolino, I.} \emph{et~al.}
\newblock \bibinfo{journal}{\bibinfo{title}{Seismic risk mitigation at {Campi
  Flegrei} in volcanic unrest}}.
\newblock {\emph{\JournalTitle{Nature communications}}}
  \textbf{\bibinfo{volume}{15}}, \bibinfo{pages}{10474} (\bibinfo{year}{2024}).

\bibitem{rolandi20251538}
\bibinfo{author}{Rolandi, G.}, \bibinfo{author}{Troise, C.},
  \bibinfo{author}{Sacchi, M.}, \bibinfo{author}{Di~Lascio, M.} \&
  \bibinfo{author}{De~Natale, G.}
\newblock \bibinfo{journal}{\bibinfo{title}{The 1538 eruption at the {Campi
  Flegrei resurgent caldera: implications for future unrest and eruptive
  scenarios}}}.
\newblock {\emph{\JournalTitle{Natural Hazards and Earth System Sciences}}}
  \textbf{\bibinfo{volume}{25}}, \bibinfo{pages}{3421--3453}
  (\bibinfo{year}{2025}).

\bibitem{giudicepietro2025burst}
\bibinfo{author}{Giudicepietro, F.} \emph{et~al.}
\newblock \bibinfo{journal}{\bibinfo{title}{Burst-like swarms in the {Campi
  Flegrei caldera accelerating unrest from 2021 to 2024}}}.
\newblock {\emph{\JournalTitle{Nature Communications}}}
  \textbf{\bibinfo{volume}{16}}, \bibinfo{pages}{1548} (\bibinfo{year}{2025}).

\bibitem{buono2023discriminating}
\bibinfo{author}{Buono, G.}, \bibinfo{author}{Caliro, S.},
  \bibinfo{author}{Paonita, A.}, \bibinfo{author}{Pappalardo, L.} \&
  \bibinfo{author}{Chiodini, G.}
\newblock \bibinfo{journal}{\bibinfo{title}{Discriminating carbon dioxide
  sources during volcanic unrest: The case of {Campi Flegrei caldera
  (Italy)}}}.
\newblock {\emph{\JournalTitle{Geology}}} \textbf{\bibinfo{volume}{51}},
  \bibinfo{pages}{397--401} (\bibinfo{year}{2023}).

\bibitem{caliro2025}
\bibinfo{author}{Caliro, S.}, \bibinfo{author}{Chiodini, G.},
  \bibinfo{author}{Avino, R.} \emph{et~al.}
\newblock \bibinfo{journal}{\bibinfo{title}{Escalation of caldera unrest
  indicated by increasing emission of isotopically light sulfur}}.
\newblock {\emph{\JournalTitle{Nat. Geosci.}}} \textbf{\bibinfo{volume}{18}},
  \bibinfo{pages}{167--174}, \doiprefix\url{10.1038/s41561-024-01632-w}
  (\bibinfo{year}{2025}).

\bibitem{danesi2024}
\bibinfo{author}{Danesi, S.}, \bibinfo{author}{Pino, N.},
  \bibinfo{author}{Carlino, S.} \& \bibinfo{author}{Kilburn, C.}
\newblock \bibinfo{journal}{\bibinfo{title}{Evolution in unrest processes at
  {Campi Flegrei} caldera as inferred from local seismicity}}.
\newblock {\emph{\JournalTitle{Earth and Planetary Science Letters}}}
  \textbf{\bibinfo{volume}{626}}, \bibinfo{pages}{118530},
  \doiprefix\url{10.1016/j.epsl.2023.118530} (\bibinfo{year}{2024}).

\bibitem{tramelli2024}
\bibinfo{author}{Tramelli, A.}, \bibinfo{author}{Convertito, V.} \&
  \bibinfo{author}{Godano, C.}
\newblock \bibinfo{journal}{\bibinfo{title}{b value enlightens different
  rheological behaviour in {Campi Flegrei caldera}}}.
\newblock {\emph{\JournalTitle{Commun Earth Environ}}}
  \textbf{\bibinfo{volume}{5}}, \bibinfo{pages}{275},
  \doiprefix\url{10.1038/s43247-024-01447-y} (\bibinfo{year}{2024}).

\bibitem{cornelius1995graphical}
\bibinfo{author}{Cornelius, R.~R.} \& \bibinfo{author}{Voight, B.}
\newblock \bibinfo{journal}{\bibinfo{title}{Graphical and {PC-software analysis
  of volcano eruption precursors according to the Materials Failure Forecast
  Method (FFM)}}}.
\newblock {\emph{\JournalTitle{Journal of Volcanology and Geothermal
  Research}}} \textbf{\bibinfo{volume}{64}}, \bibinfo{pages}{295--320}
  (\bibinfo{year}{1995}).

\bibitem{bell2013limits}
\bibinfo{author}{Bell, A.~F.}, \bibinfo{author}{Naylor, M.} \&
  \bibinfo{author}{Main, I.~G.}
\newblock \bibinfo{journal}{\bibinfo{title}{The limits of predictability of
  volcanic eruptions from accelerating rates of earthquakes}}.
\newblock {\emph{\JournalTitle{Geophysical Journal International}}}
  \textbf{\bibinfo{volume}{194}}, \bibinfo{pages}{1541--1553}
  (\bibinfo{year}{2013}).

\bibitem{hao2017accelerating}
\bibinfo{author}{Hao, S.}, \bibinfo{author}{Yang, H.} \&
  \bibinfo{author}{Elsworth, D.}
\newblock \bibinfo{journal}{\bibinfo{title}{An accelerating precursor to
  predict ``time-to-failure'' in creep and volcanic eruptions}}.
\newblock {\emph{\JournalTitle{Journal of Volcanology and Geothermal
  Research}}} \textbf{\bibinfo{volume}{343}}, \bibinfo{pages}{252--262}
  (\bibinfo{year}{2017}).

\bibitem{bree2013prediction}
\bibinfo{author}{Br{\'e}e, D.~S.}, \bibinfo{author}{Challet, D.} \&
  \bibinfo{author}{Peirano, P.~P.}
\newblock \bibinfo{journal}{\bibinfo{title}{Prediction accuracy and sloppiness
  of log-periodic functions}}.
\newblock {\emph{\JournalTitle{Quantitative Finance}}}
  \textbf{\bibinfo{volume}{13}}, \bibinfo{pages}{275--280}
  (\bibinfo{year}{2013}).

\bibitem{hainzl2026deformation}
\bibinfo{author}{Hainzl, S.}, \bibinfo{author}{Dahm, T.} \&
  \bibinfo{author}{Tramelli, A.}
\newblock \bibinfo{journal}{\bibinfo{title}{A deformation-driven earthquake
  interaction model for seismicity at {Campi Flegrei}}}.
\newblock {\emph{\JournalTitle{Communications Earth \& Environment}}}
  (\bibinfo{year}{2026}).

\bibitem{kilburn2017}
\bibinfo{author}{Kilburn, C.}, \bibinfo{author}{De~Natale, G.} \&
  \bibinfo{author}{Carlino, S.}
\newblock \bibinfo{journal}{\bibinfo{title}{Progressive approach to eruption at
  {Campi Flegrei caldera in southern Italy}}}.
\newblock {\emph{\JournalTitle{Nat Commun}}} \textbf{\bibinfo{volume}{8}},
  \bibinfo{pages}{15312}, \doiprefix\url{10.1038/ncomms15312}
  (\bibinfo{year}{2017}).

\bibitem{sibson1994crustal}
\bibinfo{author}{Sibson, R.~H.}
\newblock \bibinfo{journal}{\bibinfo{title}{Crustal stress, faulting and fluid
  flow}}.
\newblock {\emph{\JournalTitle{Geological Society, London, Special
  Publications}}} \textbf{\bibinfo{volume}{78}}, \bibinfo{pages}{69--84}
  (\bibinfo{year}{1994}).

\bibitem{astort2024tracking}
\bibinfo{author}{Astort, A.} \emph{et~al.}
\newblock \bibinfo{journal}{\bibinfo{title}{Tracking the 2007--2023
  magma-driven unrest at {Campi Flegrei caldera (Italy)}}}.
\newblock {\emph{\JournalTitle{Communications Earth \& Environment}}}
  \textbf{\bibinfo{volume}{5}}, \bibinfo{pages}{506} (\bibinfo{year}{2024}).

\bibitem{amstutz2025volcano}
\bibinfo{author}{Amstutz, F.~M.} \emph{et~al.}
\newblock \bibinfo{journal}{\bibinfo{title}{Volcano-tectonic controls on
  magmatic evolution at {Campi Flegrei, Italy: insights from thermodynamic
  modelling}}}.
\newblock {\emph{\JournalTitle{Journal of Petrology}}}
  \textbf{\bibinfo{volume}{66}}, \bibinfo{pages}{egaf068}
  (\bibinfo{year}{2025}).

\bibitem{giacomuzzi2025causal}
\bibinfo{author}{Giacomuzzi, G.}, \bibinfo{author}{Fonzetti, R.},
  \bibinfo{author}{Govoni, A.}, \bibinfo{author}{De~Gori, P.} \&
  \bibinfo{author}{Chiarabba, C.}
\newblock \bibinfo{journal}{\bibinfo{title}{Causal processes of shallow and
  deep seismicity at {Campi Flegrei caldera}}}.
\newblock {\emph{\JournalTitle{Communications Earth \& Environment}}}
  \textbf{\bibinfo{volume}{6}}, \bibinfo{pages}{70} (\bibinfo{year}{2025}).

\bibitem{rapagnani2025coupled}
\bibinfo{author}{Rapagnani, G.} \emph{et~al.}
\newblock \bibinfo{journal}{\bibinfo{title}{Coupled earthquakes and resonance
  processes during the uplift of {Campi Flegrei caldera}}}.
\newblock {\emph{\JournalTitle{Communications Earth \& Environment}}}
  \textbf{\bibinfo{volume}{6}}, \bibinfo{pages}{607} (\bibinfo{year}{2025}).

\bibitem{vanorio2025recurrence}
\bibinfo{author}{Vanorio, T.}, \bibinfo{author}{Geremia, D.},
  \bibinfo{author}{De~Landro, G.} \& \bibinfo{author}{Guo, T.}
\newblock \bibinfo{journal}{\bibinfo{title}{The recurrence of geophysical
  manifestations at the {Campi Flegrei} caldera}}.
\newblock {\emph{\JournalTitle{Science Advances}}}
  \textbf{\bibinfo{volume}{11}}, \bibinfo{pages}{eadt2067}
  (\bibinfo{year}{2025}).

\bibitem{bevilacqua2024accelerating}
\bibinfo{author}{Bevilacqua, A.} \emph{et~al.}
\newblock \bibinfo{journal}{\bibinfo{title}{Accelerating upper crustal
  deformation and seismicity of {Campi Flegrei caldera (Italy), during the
  2000--2023 unrest}}}.
\newblock {\emph{\JournalTitle{Communications Earth \& Environment}}}
  \textbf{\bibinfo{volume}{5}}, \bibinfo{pages}{742} (\bibinfo{year}{2024}).

\bibitem{bradski2000opencv}
\bibinfo{author}{Bradski, G.}
\newblock \bibinfo{journal}{\bibinfo{title}{The opencv library.}}
\newblock {\emph{\JournalTitle{Dr. Dobb's Journal: Software Tools for the
  Professional Programmer}}} \textbf{\bibinfo{volume}{25}},
  \bibinfo{pages}{120--123} (\bibinfo{year}{2000}).

\bibitem{lolli2020homogenized}
\bibinfo{author}{Lolli, B.}, \bibinfo{author}{Randazzo, D.},
  \bibinfo{author}{Vannucci, G.} \& \bibinfo{author}{Gasperini, P.}
\newblock \bibinfo{journal}{\bibinfo{title}{The homogenized instrumental
  seismic catalog {(HORUS) of Italy from 1960 to present}}}.
\newblock {\emph{\JournalTitle{Seismological Society of America}}}
  \textbf{\bibinfo{volume}{91}}, \bibinfo{pages}{3208--3222}
  (\bibinfo{year}{2020}).

\bibitem{hanks1979moment}
\bibinfo{author}{Hanks, T.~C.} \& \bibinfo{author}{Kanamori, H.}
\newblock \bibinfo{journal}{\bibinfo{title}{A moment magnitude scale}}.
\newblock {\emph{\JournalTitle{Journal of Geophysical Research: Solid Earth}}}
  \textbf{\bibinfo{volume}{84}}, \bibinfo{pages}{2348--2350}
  (\bibinfo{year}{1979}).

\bibitem{benioff1951global}
\bibinfo{author}{Benioff, H.}
\newblock \bibinfo{journal}{\bibinfo{title}{Global strain accumulation and
  release as revealed by great earthquakes}}.
\newblock {\emph{\JournalTitle{Geological Society of America Bulletin}}}
  \textbf{\bibinfo{volume}{62}}, \bibinfo{pages}{331--338}
  (\bibinfo{year}{1951}).

\bibitem{main1999applicability}
\bibinfo{author}{Main, I.~G.}
\newblock \bibinfo{journal}{\bibinfo{title}{Applicability of time-to-failure
  analysis to accelerated strain before earthquakes and volcanic eruptions}}.
\newblock {\emph{\JournalTitle{Geophysical Journal International}}}
  \textbf{\bibinfo{volume}{139}}, \bibinfo{pages}{F1--F6}
  (\bibinfo{year}{1999}).

\bibitem{smug2018predicting}
\bibinfo{author}{Smug, D.}, \bibinfo{author}{Ashwin, P.} \&
  \bibinfo{author}{Sornette, D.}
\newblock \bibinfo{journal}{\bibinfo{title}{Predicting financial market crashes
  using ghost singularities}}.
\newblock {\emph{\JournalTitle{Plos one}}} \textbf{\bibinfo{volume}{13}},
  \bibinfo{pages}{e0195265} (\bibinfo{year}{2018}).

\bibitem{marquardt1963algorithm}
\bibinfo{author}{Marquardt, D.~W.}
\newblock \bibinfo{journal}{\bibinfo{title}{An algorithm for least-squares
  estimation of nonlinear parameters}}.
\newblock {\emph{\JournalTitle{Journal of the society for Industrial and
  Applied Mathematics}}} \textbf{\bibinfo{volume}{11}},
  \bibinfo{pages}{431--441} (\bibinfo{year}{1963}).

\bibitem{akaike2003new}
\bibinfo{author}{Akaike, H.}
\newblock \bibinfo{journal}{\bibinfo{title}{A new look at the statistical model
  identification}}.
\newblock {\emph{\JournalTitle{IEEE transactions on automatic control}}}
  \textbf{\bibinfo{volume}{19}}, \bibinfo{pages}{716--723}
  (\bibinfo{year}{2003}).

\bibitem{demosdiSor19}
\bibinfo{author}{Demos, G.} \& \bibinfo{author}{Sornette, D.}
\newblock \bibinfo{journal}{\bibinfo{title}{Comparing nested data sets and
  objectively determining financial bubbles' inceptions}}.
\newblock {\emph{\JournalTitle{Physica A: Statistical Mechanics and its
  Applications}}} \textbf{\bibinfo{volume}{524}}, \bibinfo{pages}{661--675}
  (\bibinfo{year}{2019}).

\bibitem{spearman1961proof}
\bibinfo{author}{Spearman, C.}
\newblock \bibinfo{journal}{\bibinfo{title}{The proof and measurement of
  association between two things}}.
\newblock {\emph{\JournalTitle{The American Journal of Psychology}}}
  \textbf{\bibinfo{volume}{15}}, \bibinfo{pages}{72--101}
  (\bibinfo{year}{1904}).

\bibitem{schreiber2000measuring}
\bibinfo{author}{Schreiber, T.}
\newblock \bibinfo{journal}{\bibinfo{title}{Measuring information transfer}}.
\newblock {\emph{\JournalTitle{Physical Review Letters}}}
  \textbf{\bibinfo{volume}{85}}, \bibinfo{pages}{461} (\bibinfo{year}{2000}).

\bibitem{granger1969investigating}
\bibinfo{author}{Granger, C.~W.}
\newblock \bibinfo{journal}{\bibinfo{title}{Investigating causal relations by
  econometric models and cross-spectral methods}}.
\newblock {\emph{\JournalTitle{Econometrica: Journal of the Econometric
  Society}}} \bibinfo{pages}{424--438} (\bibinfo{year}{1969}).

\bibitem{koopmans1995spectral}
\bibinfo{author}{Koopmans, L.~H.}
\newblock \emph{\bibinfo{title}{The spectral analysis of time series}}
  (\bibinfo{publisher}{Elsevier}, \bibinfo{year}{1995}).

\end{thebibliography}

\newpage 
\section*{SUPPLEMENTARY MATERIAL}
\subsection*{Derivation of the solution of the generalised regularised singularity model}

We present the derivation of the solution of the equation including the leading analytic feedback counteracting the finite-time singularity, which reads
\begin{equation}
	\frac{dx}{dt} = \gamma x^\alpha - a x^m,
	\qquad
	m = \lfloor \alpha \rfloor + 1,
	\qquad
	\gamma>0,\; a\ge 0,\; \alpha>1.
\end{equation}
Here, \(m\) is the smallest integer strictly larger than \(\alpha\), so that the regularising term is the first analytic negative feedback term compatible with the leading nonlinearity \(x^\alpha\).

For \(a=0\), the dynamics reduces to
\begin{equation}
	\frac{dx}{dt} = \gamma x^\alpha,
\end{equation}
whose solution is the pure power law
\begin{equation}
	x_{\mathrm{pl}}(t)
	=
	\big[\gamma(\alpha-1)(t_c-t)\big]^{-\frac{1}{\alpha-1}}
	=
	\frac{A}{(t_c-t)^\beta},
\end{equation}
with
\begin{equation}
	\beta = \frac{1}{\alpha-1},
	\qquad
	A = \big[\gamma(\alpha-1)\big]^{-\beta},
\end{equation}
and
\begin{equation}
	t_c = t_0 + \frac{x_0^{\,1-\alpha}}{\gamma(\alpha-1)},
	\qquad x(t_0)=x_0.
\end{equation}

\paragraph{First-order correction in \(a\).}

We seek a perturbative solution of the form
\begin{equation}
	x(t)=x_{\mathrm{pl}}(t)+a\,x_1(t)+\mathcal{O}(a^2).
\end{equation}
Substituting into
\begin{equation}
	\frac{dx}{dt}=\gamma x^\alpha-a x^m
\end{equation}
and expanding to first order in \(a\) gives
\begin{equation}
	\dot{x}_{\mathrm{pl}} + a \dot{x}_1
	=
	\gamma\left(x_{\mathrm{pl}}^\alpha + a \alpha x_{\mathrm{pl}}^{\alpha-1}x_1\right)
	- a x_{\mathrm{pl}}^m
	+ \mathcal{O}(a^2).
\end{equation}
Using \(\dot{x}_{\mathrm{pl}}=\gamma x_{\mathrm{pl}}^\alpha\), the \(\mathcal{O}(a)\) equation is
\begin{equation}
	\dot{x}_1 - \alpha \gamma x_{\mathrm{pl}}^{\alpha-1}x_1 = -x_{\mathrm{pl}}^m.
\end{equation}

Introducing
\begin{equation}
	\Delta = t_c-t,
\end{equation}
the pure power-law solution satisfies
\begin{equation}
	x_{\mathrm{pl}}(\Delta)=\big[\gamma(\alpha-1)\Delta\big]^{-\frac{1}{\alpha-1}},
\end{equation}
hence
\begin{equation}
	\gamma x_{\mathrm{pl}}^{\alpha-1}=\frac{1}{(\alpha-1)\Delta}.
\end{equation}
Since \(d/dt=-d/d\Delta\), the equation for \(x_1\) becomes
\begin{equation}
	\frac{d x_1}{d\Delta}
	+
	\frac{\alpha}{(\alpha-1)\Delta}x_1
	=
	x_{\mathrm{pl}}^m
	=
	\big[\gamma(\alpha-1)\big]^{-\frac{m}{\alpha-1}}
	\Delta^{-\frac{m}{\alpha-1}}.
\end{equation}
The integrating factor is
\begin{equation}
	\mu(\Delta)=\Delta^{\frac{\alpha}{\alpha-1}},
\end{equation}
which we use to transform the left-hand side into an exact derivative to obtain
\begin{equation}
	\frac{d}{d\Delta}\!\left(
	\Delta^{\frac{\alpha}{\alpha-1}}x_1
	\right)
	=
	\big[\gamma(\alpha-1)\big]^{-\frac{m}{\alpha-1}}
	\Delta^{\frac{\alpha-m}{\alpha-1}}.
\end{equation}

\paragraph{Generic non-resonant case: \(2\alpha-m-1\neq 0\).}

When \(2\alpha-m-1\neq 0\), integration yields
\begin{equation}
	x_1(\Delta)
	=
	-\frac{\alpha-1}{m-2\alpha+1}
	\big[\gamma(\alpha-1)\big]^{-\frac{m}{\alpha-1}}
	\Delta^{-\frac{m-\alpha+1}{\alpha-1}}
	+ C\,\Delta^{-\frac{\alpha}{\alpha-1}}.
\end{equation}
The homogeneous term \(C\,\Delta^{-\frac{\alpha}{\alpha-1}}\) exhibits the same time dependence as the first-order variation of the leading power law solution induced by a small shift of the critical time \(t_c\); it therefore does not represent an independent physical correction and can be absorbed into a redefinition (renormalisation) of \(t_c\). Keeping only the genuinely new regularising contribution, we obtain
\begin{equation}
	x(t)
	=
	\big[\gamma(\alpha-1)(t_c-t)\big]^{-\frac{1}{\alpha-1}}
	-
	a\,
	\frac{\alpha-1}{m-2\alpha+1}
	\big[\gamma(\alpha-1)\big]^{-\frac{m}{\alpha-1}}
	(t_c-t)^{-\frac{m-\alpha+1}{\alpha-1}}
	+
	\mathcal{O}(a^2).
\end{equation}
Equivalently, in terms of \(\beta=1/(\alpha-1)\) and \(A=[\gamma(\alpha-1)]^{-\beta}\),
\begin{equation}
	x(t)
	=
	\frac{A}{(t_c-t)^\beta}
	-
	a\,
	\frac{\alpha-1}{m-2\alpha+1}
	A^m\,
	(t_c-t)^{-\beta(m-\alpha+1)}
	+
	\mathcal{O}(a^2).
\end{equation}

\paragraph{Resonant case: \(2\alpha-m-1=0\).}

When
\begin{equation}
	2\alpha-m-1=0,
\end{equation}
the integral becomes logarithmic. This occurs for the two values
\begin{equation}
	\alpha=\frac32
	\quad\text{and}\quad
	\alpha=2,
\end{equation}
since \(m=\lfloor\alpha\rfloor+1\). In that case,
\begin{equation}
	x_1(\Delta)
	=
	-\big[\gamma(\alpha-1)\big]^{-\frac{m}{\alpha-1}}
	\Delta^{-\frac{\alpha}{\alpha-1}}
	\ln\!\left(\frac{\Delta_0}{\Delta}\right)
	+ C\,\Delta^{-\frac{\alpha}{\alpha-1}},
\end{equation}
where \(\Delta_0=t_c-t_0\). Absorbing again the homogeneous term into \(t_c\), the first-order corrected solution becomes
\begin{equation}
	x(t)
	=
	\big[\gamma(\alpha-1)(t_c-t)\big]^{-\frac{1}{\alpha-1}}
	-
	a\,
	\big[\gamma(\alpha-1)\big]^{-\frac{m}{\alpha-1}}
	(t_c-t)^{-\frac{\alpha}{\alpha-1}}
	\ln\!\left(\frac{t_c-t_0}{t_c-t}\right)
	+
	\mathcal{O}(a^2).
\end{equation}

\paragraph{Special case \(\alpha=2\).}

For \(\alpha=2\), one has \(m=3\), and the above formula reduces to
\begin{equation}
	x(t)
	=
	\frac{1}{\gamma(t_c-t)}
	-
	\frac{a}{\gamma^3}
	\frac{\ln\!\left(\dfrac{t_c-t_0}{t_c-t}\right)}{(t_c-t)^2}
	+
	\mathcal{O}(a^2),
\end{equation}
which is the direct generalisation of the logarithmic first-order correction obtained previously.

\paragraph{Resulting empirical regularised singularity model.}

The perturbative result suggests the following empirical form:
\begin{equation}
	x(t)
	\approx
	\frac{A}{(t_c-t)^\beta}
	-
	a\,B\,(t_c-t)^{-\eta},
\end{equation}
with
\begin{equation}
	\beta=\frac{1}{\alpha-1},
	\qquad
	\eta=\frac{m-\alpha+1}{\alpha-1},
	\qquad
	B=\frac{\alpha-1}{m-2\alpha+1}\big[\gamma(\alpha-1)\big]^{-\frac{m}{\alpha-1}},
\end{equation}
in the non-resonant case, and the logarithmic replacement
\begin{equation}
	x(t)
	\approx
	\frac{A}{(t_c-t)^\beta}
	-
	a\,\widetilde{B}\,
	(t_c-t)^{-\frac{\alpha}{\alpha-1}}
	\ln\!\left(\frac{t_c-t_0}{t_c-t}\right)
\end{equation}
in the resonant case \(2\alpha-m-1=0\).

\newpage 
\section*{Detailed results of geodetic, seismic and multiparametric data analyses}

\begin{figure}[ht]
	\centering
	\includegraphics[width=\linewidth]{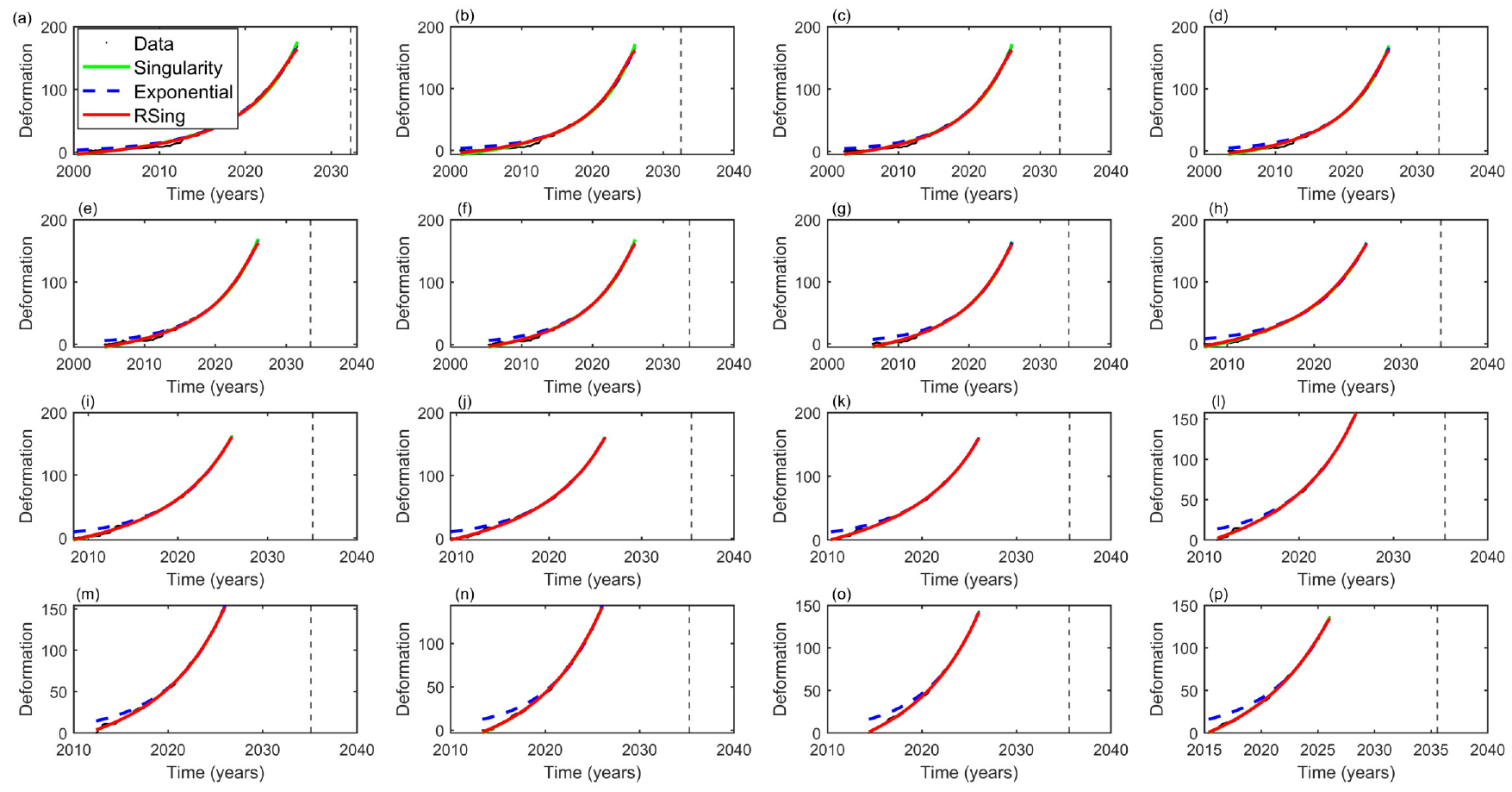}
	\caption{Vertical displacement recorded at GNSS station RITE for analysis windows with start dates ranging from 2000 to 2015 (panels a-p). The dashed blue line represents the best-fitting exponential model, the solid green line the finite-time singularity model, and the solid red line the regularized singularity model. The vertical dashed line in each panel marks the estimated critical time $t_c$ from the regularized model.}
	\label{fig1s}
\end{figure}

\begin{figure}[ht]
	\centering
	\includegraphics[width=\linewidth]{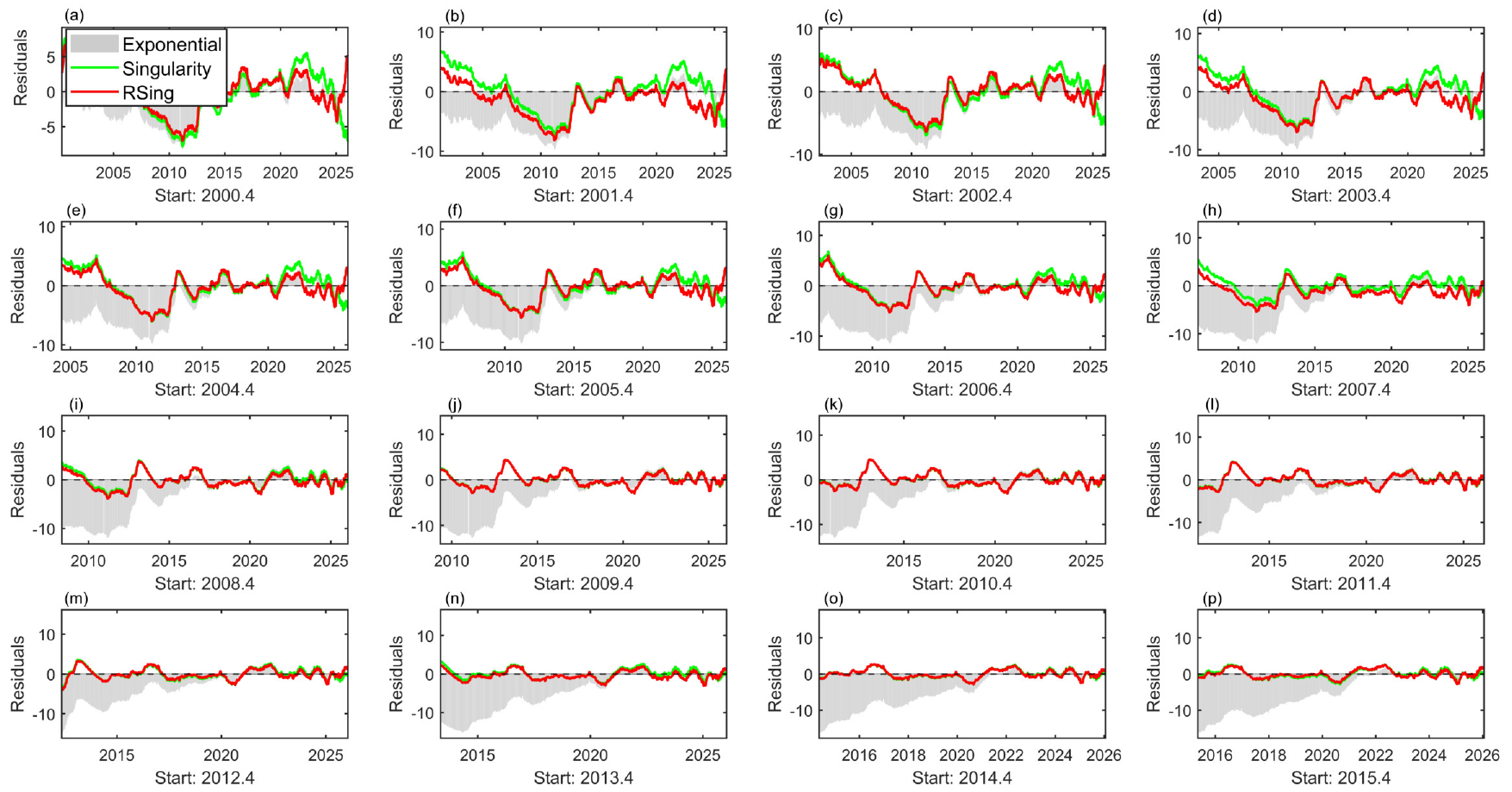}
	\caption{Residuals of the vertical displacement recorded at GNSS station RITE for different fitting functions and different analysis windows with start dates ranging from 2000 to 2015 (panels a-p). The shadowed gray area represents the residuals of the exponential model, the solid green line the finite-time singularity residuals, and the solid red line the regularized singularity residuals.}
	\label{fig2s}
\end{figure}

\begin{figure}[ht]
	\centering
	\includegraphics[width=\linewidth]{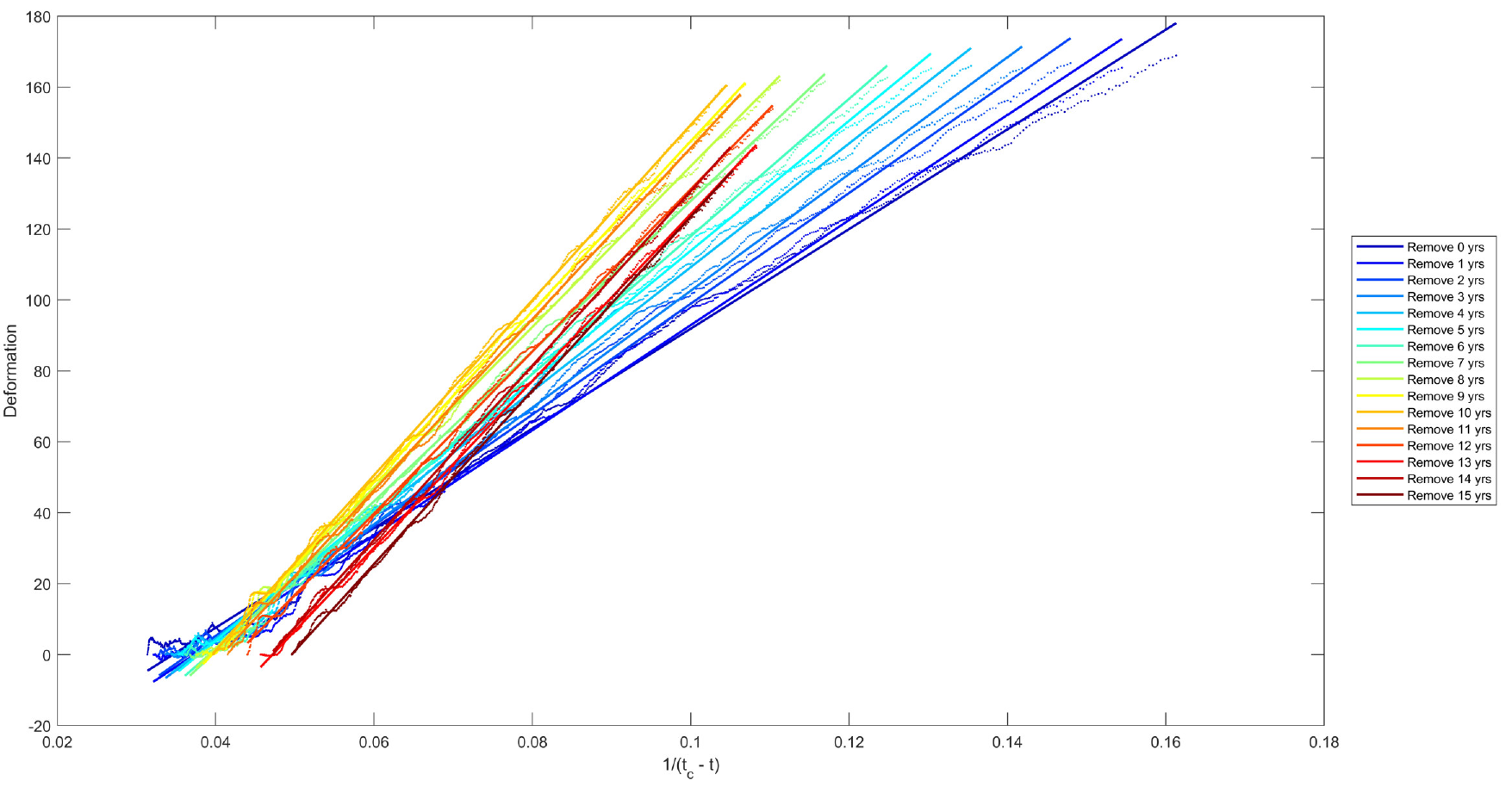}
	\caption{Inverse-time linearity diagnostic for GNSS vertical displacement at station RITE. The observable is plotted against the transformed time coordinate $1/(t_c - t)$ for analysis windows with start dates ranging from 2000 to 2015. Colours progress from blue (earliest start dates) through light blue, yellow, and orange to red (latest start dates). The linear relationship confirms that the acceleration follows a finite-time singular trajectory, with the slope evolving systematically as the singular regime emerges.}
	\label{fig3s}
\end{figure}	

\begin{figure}[ht]
	\centering
	\includegraphics[width=\linewidth]{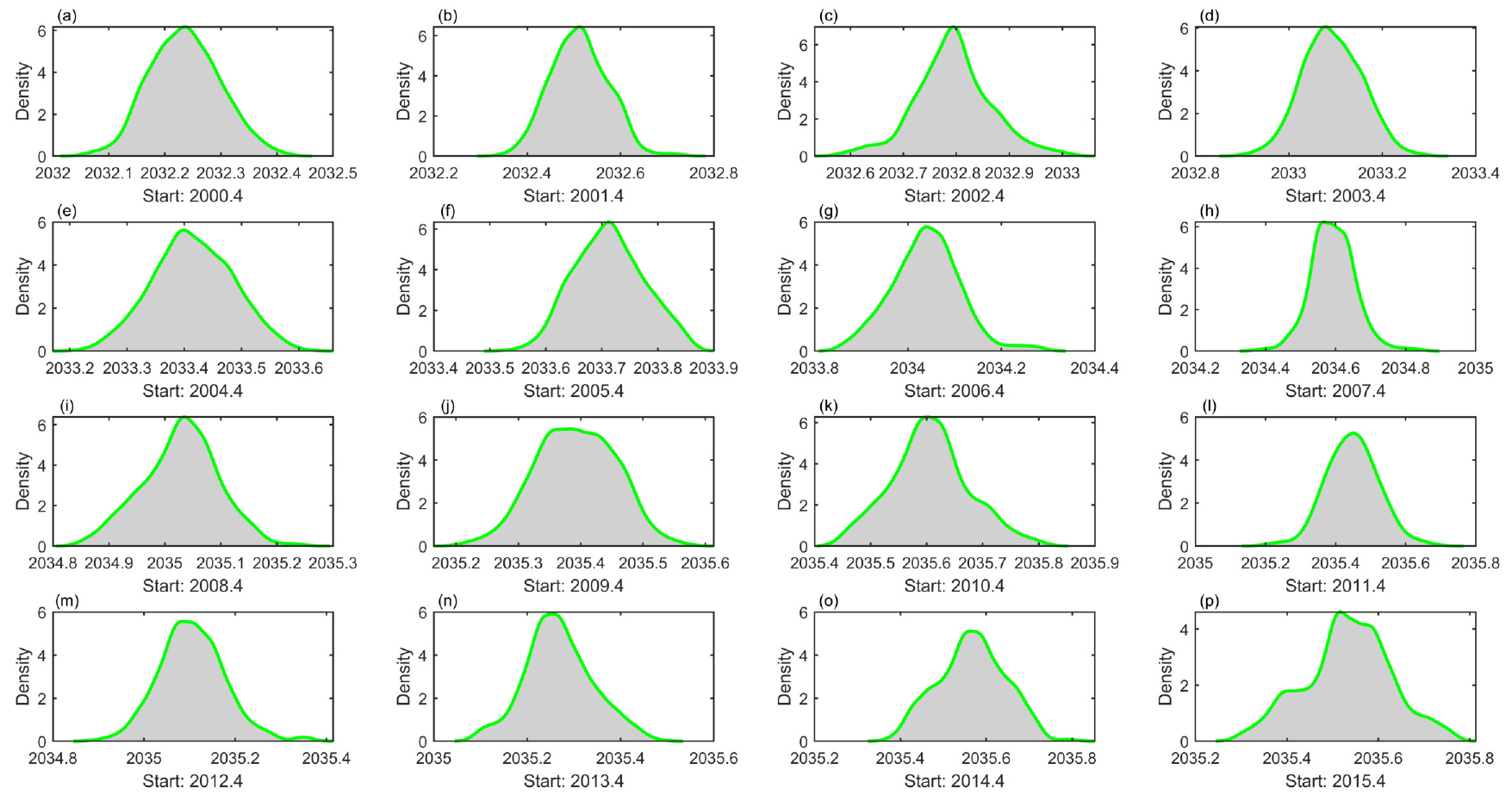}
	\caption{Probability density of the critical time $t_c$ estimated from the finite-time singularity model for GNSS vertical displacement at station RITE. Panels a-p correspond to analysis windows with start dates progressing from 2000 to 2015. For detailed average values with uncertainty, see Table 1.}
	\label{fig4s}
\end{figure}

\begin{figure}[ht]
	\centering
	\includegraphics[width=\linewidth]{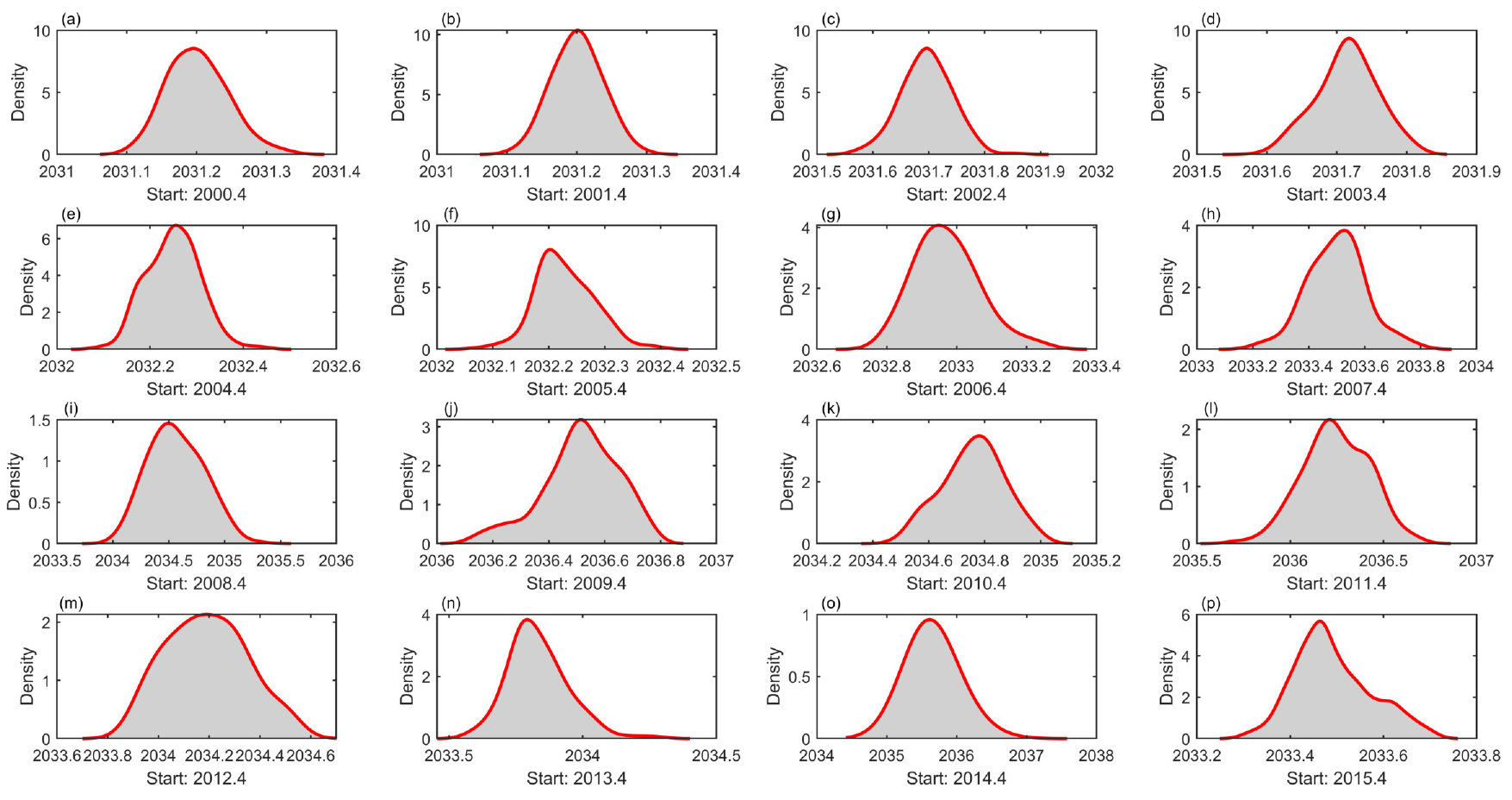}
	\caption{Probability density of the critical time $t_c$ estimated from the finite-time regularised singularity model for GNSS vertical displacement at station RITE. Panels a-p correspond to analysis windows with start dates progressing from 2000 to 2015. For detailed average values with uncertainty, see Table 3.}
	\label{fig5}
\end{figure}

\begin{figure}[ht]
	\centering
	\includegraphics[width=\linewidth]{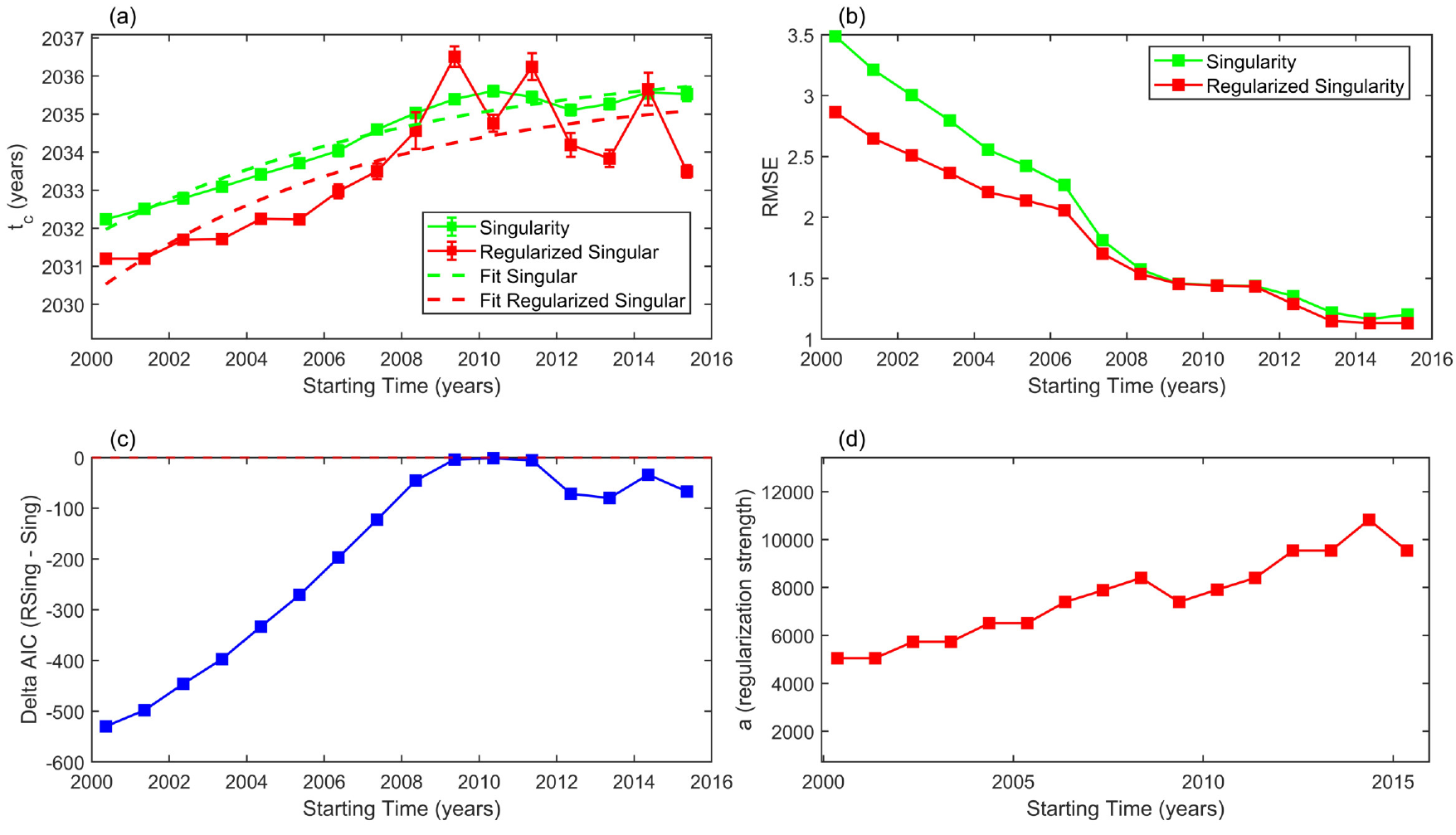}
	\caption{Comparison of the finite-time singularity and regularized singularity models for GNSS vertical displacement at station RITE. (a) Estimated critical time $t_c$ as a function of analysis start date for both models. The regularized model yields $t_c$ values systematically closer to the present, with progressive stabilisation observed after 2008 and compatibility within uncertainty. (b) Root-mean-square error (RMSE) versus start date. Both models show the expected decreasing trend as the number of data points declines, with the regularized model consistently achieving lower RMSE. (c) Difference in Akaike Information Criterion ($\Delta$AIC = AIC${\text{RSing}} - \text{AIC}{\text{Sing}}$). Systematically negative values indicate a statistical preference for the regularized model. (d) Regularization parameter $a$ as a function of start date. The magnitude of $a$ reflects the strength of the saturation term; for an intuitive visualization of its physical role, see Supplementary Figure~\ref{fig13}.}
	\label{fig6}
\end{figure}

\begin{figure}[ht]
	\centering
	\includegraphics[width=\linewidth]{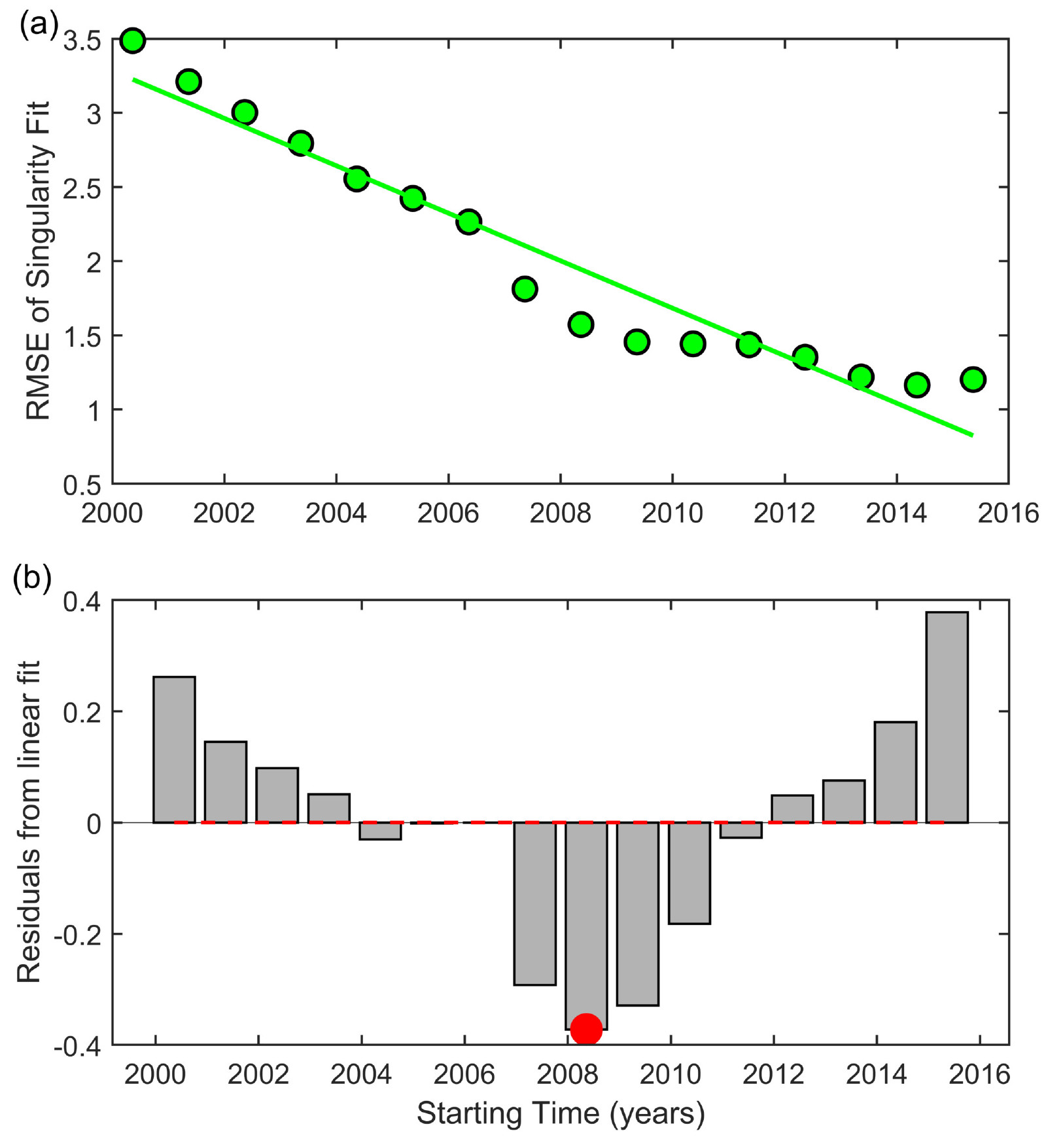}
	\caption{Identification of the optimal analysis window for the finite-time singularity model applied to GNSS vertical displacement at station RITE. (a) Root-mean-square error (RMSE) as a function of analysis start date, with a linear fit capturing the systematic trend. (b) Residuals of the RMSE with respect to the linear fit. The minimum residual occurs at 2008 (red circle), indicating that the 2008--2026 window most faithfully reflects singular dynamics. This residual profile determines 75\% of the weighting scheme used to compute the final geodetic critical time estimate $t_c^g$.}
	\label{fig7}
\end{figure}	

\begin{figure}[ht]
	\centering
	\includegraphics[width=\linewidth]{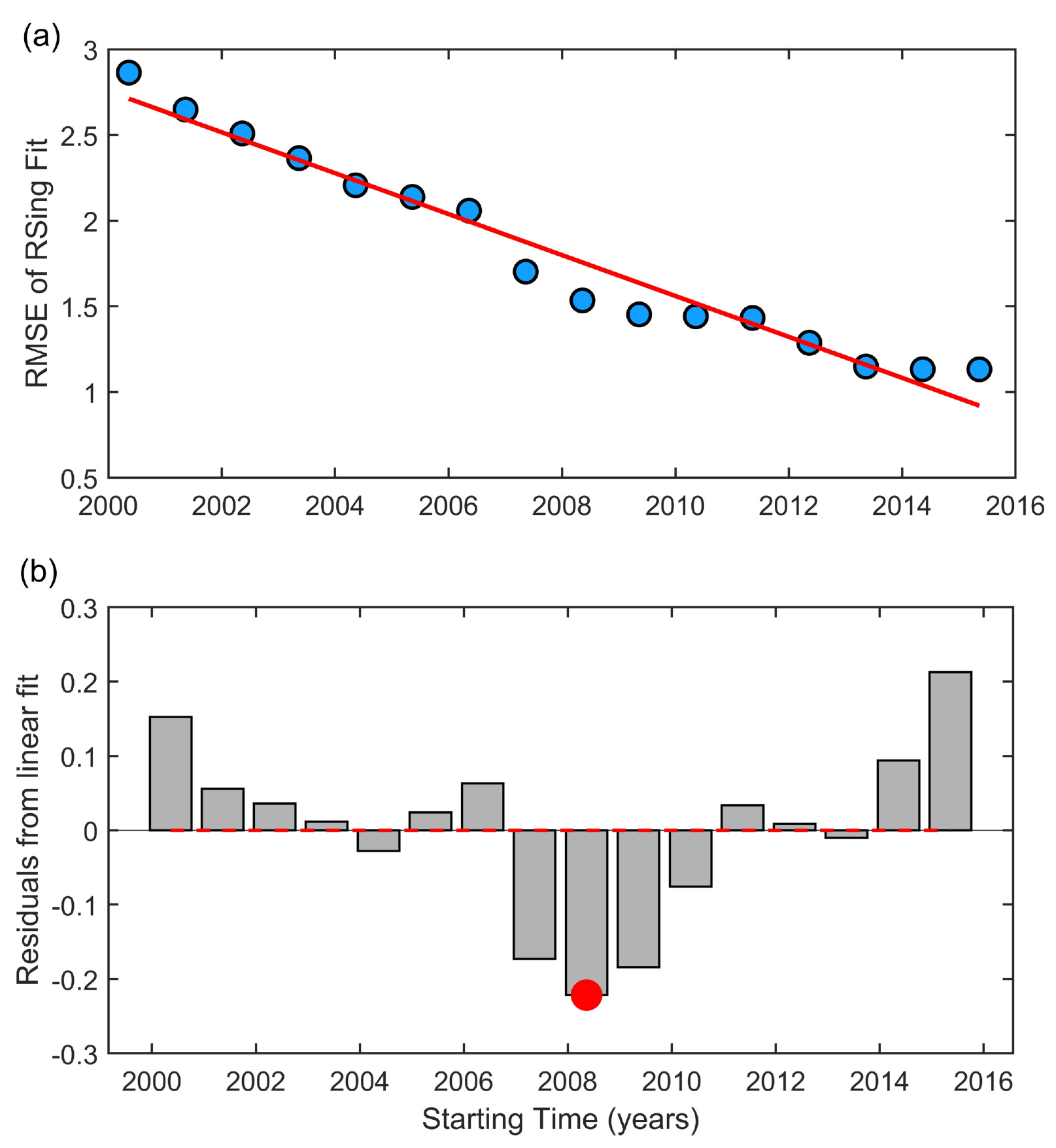}
	\caption{Identification of the optimal analysis window for the regularised finite-time singularity model applied to GNSS vertical displacement at station RITE. (a) Root-mean-square error (RMSE) as a function of analysis start date, with a linear fit capturing the systematic trend. (b) Residuals of the RMSE with respect to the linear fit. The minimum residual occurs at 2008 (red circle), indicating that the 2008--2026 window most faithfully reflects singular dynamics. This residual profile determines 75\% of the weighting scheme used to compute the final geodetic critical time estimate $t_c^g$.}
	\label{fig8}
\end{figure}

\begin{figure}[ht]
	\centering
	\includegraphics[width=\linewidth]{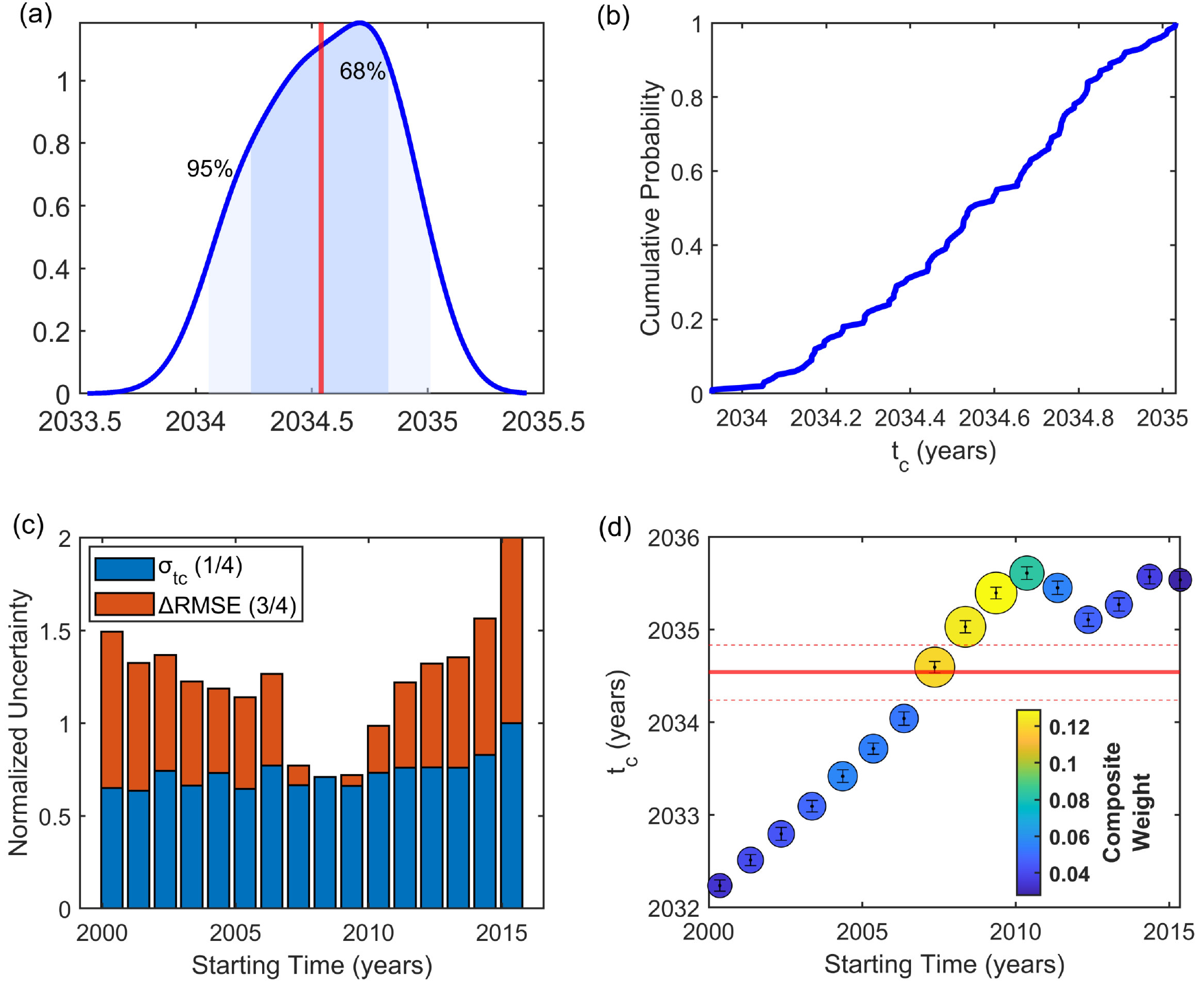}
	\caption{Weighting procedure for the combined estimation of the geodetic critical time $t_c$ from the finite-time singularity model. (a) Probability density function of $t_c$ obtained via weighted bootstrap resampling. (b) Cumulative distribution function corresponding to the density in panel a. The weights are defined as the inverse of the total normalized uncertainty, which combines the bootstrap uncertainty of $t_c$ (25\% contribution) and the RMSE residuals from Supplementary Figure~\ref{fig7}b (75\% contribution). (c) Individual $t_c$ estimates plotted against analysis start date. The size and colour of each point encode the composite weight, with the colour bar indicating the relative contribution of each window to the final probability density shown in panel a.}
	\label{fig9}
\end{figure}

\begin{figure}[ht]
	\centering
	\includegraphics[width=\linewidth]{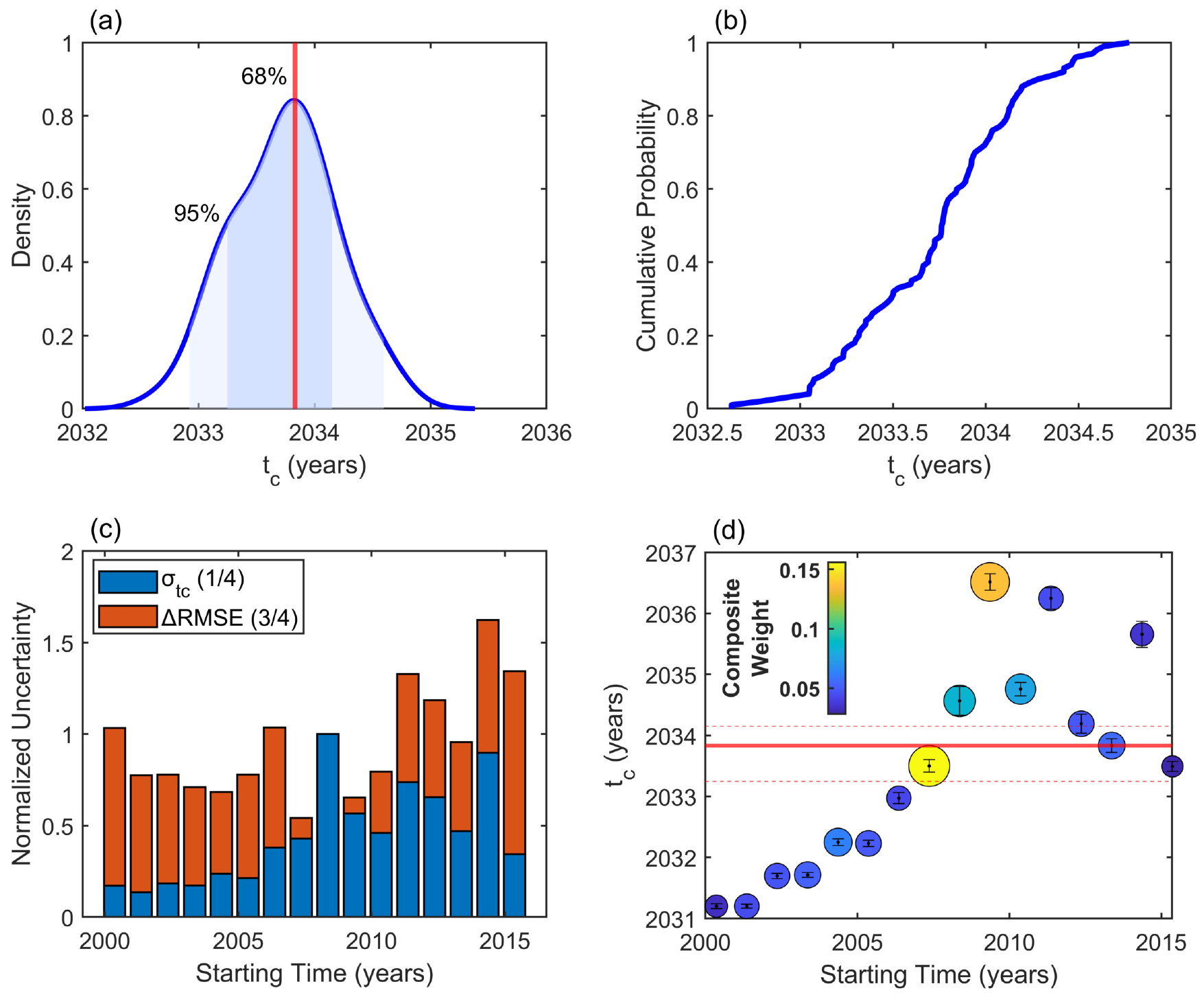}
	\caption{Weighting procedure for the combined estimation of the geodetic critical time $t_c$ from the finite-time regularised singularity model. (a) Probability density function of $t_c$ obtained via weighted bootstrap resampling. (b) Cumulative distribution function corresponding to the density in panel a. The weights are defined as the inverse of the total normalized uncertainty, which combines the bootstrap uncertainty of $t_c$ (25\% contribution) and the RMSE residuals from Supplementary Figure~\ref{fig8}b (75\% contribution). (c) Individual $t_c$ estimates plotted against analysis start date. The size and colour of each point encode the composite weight, with the colour bar indicating the relative contribution of each window to the final probability density shown in panel a.}
	\label{fig10}
\end{figure}

\begin{figure}[ht]
	\centering
	\includegraphics[width=\linewidth]{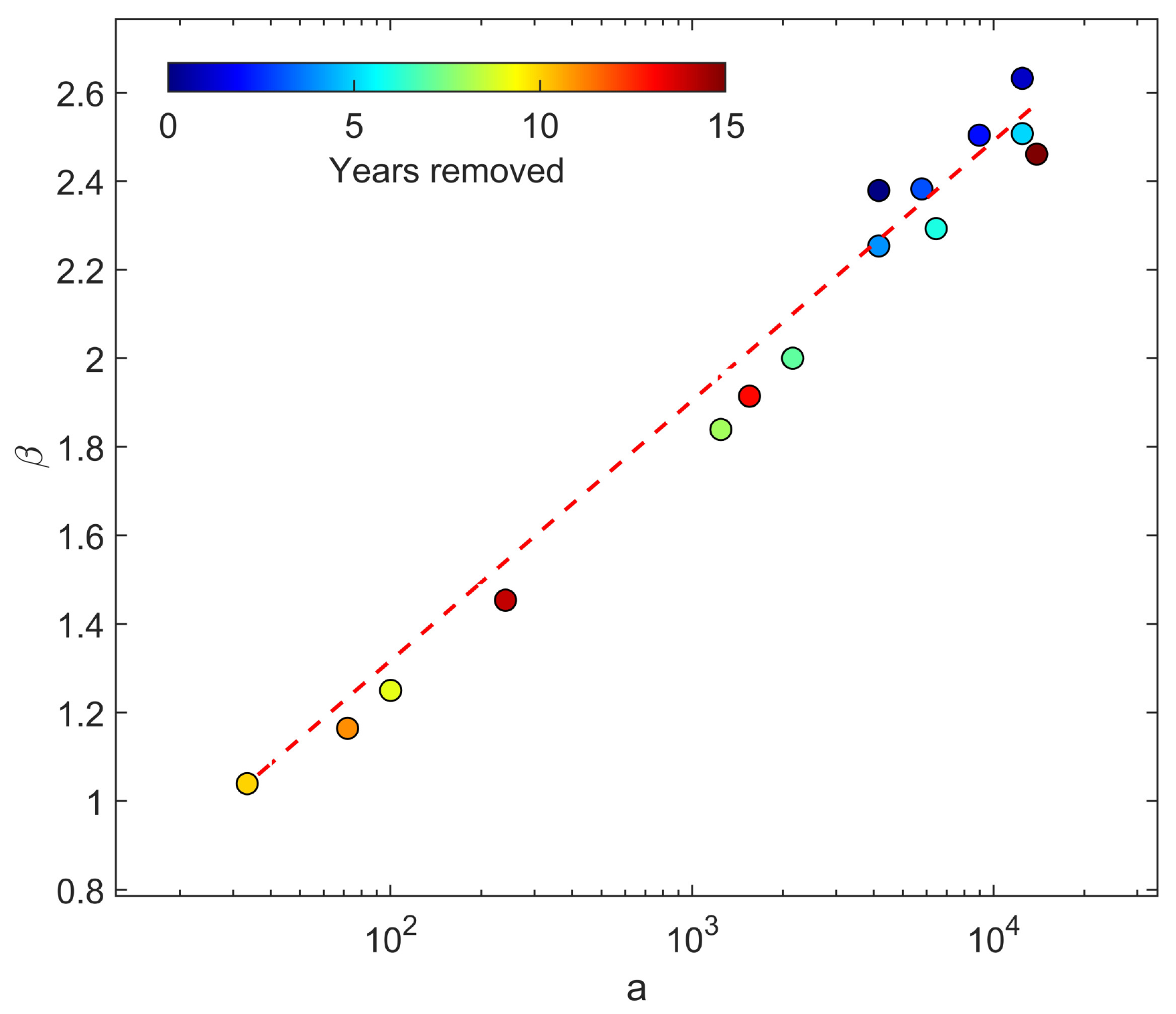}
	\caption{Correlation between the singularity exponent $\beta$ and the regularization parameter $a$ when both are fitted simultaneously for different analysis start dates (indicated by the colour bar). The parameters exhibit a strong positive linear relationship on a semi-logarithmic scale, demonstrating that they are highly covariant when left free to vary. This degeneracy motivates the sequential fitting strategy adopted in this work: $\beta$ is first estimated from the unregularized singularity model and then held fixed in the regularized model, allowing for a stable and physically interpretable estimate of the regularization amplitude $a$.}
	\label{fig11}
\end{figure}	

\begin{figure}[ht]
	\centering
	\includegraphics[width=\linewidth]{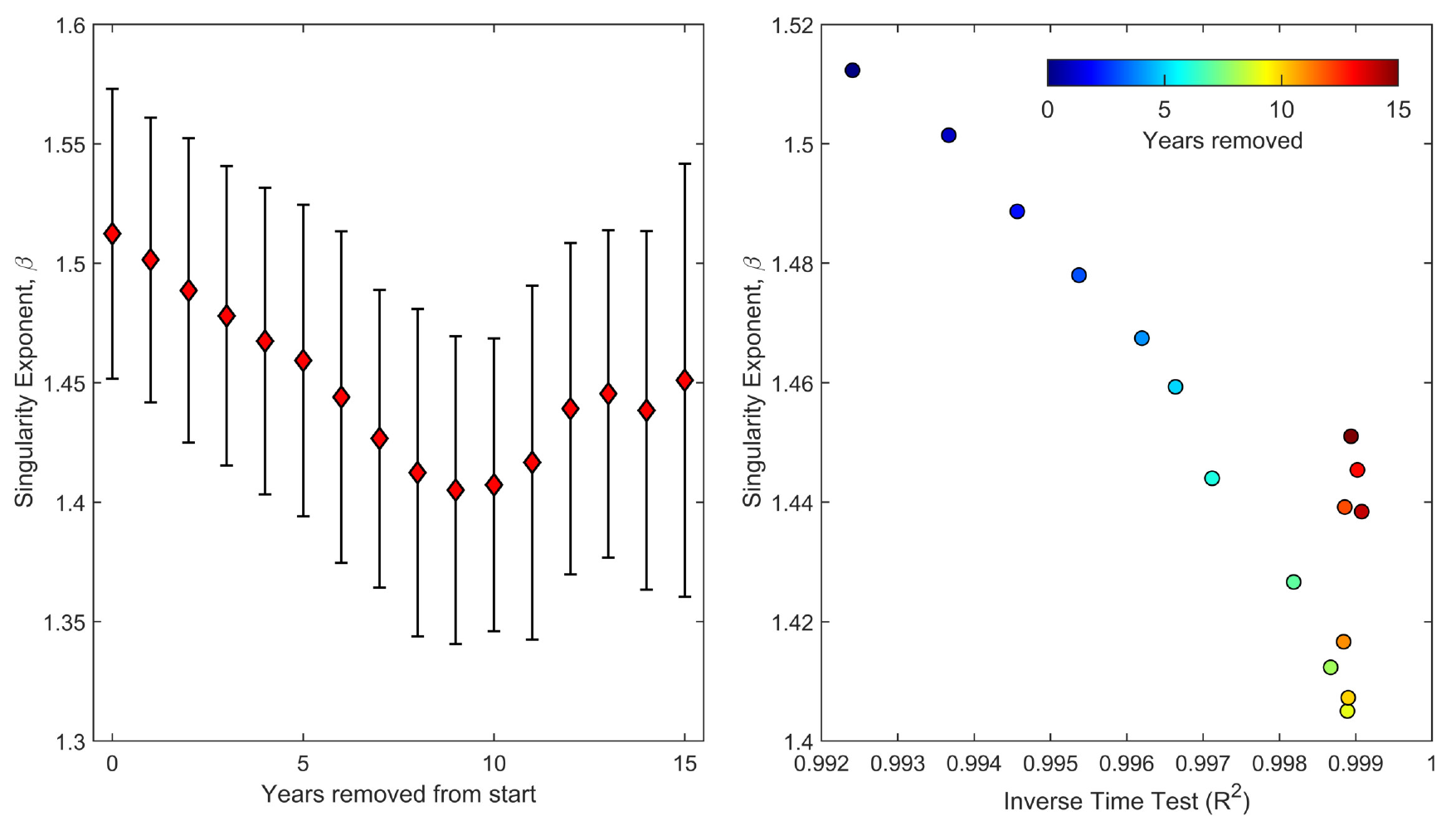}
	\caption{Evolution of the singularity exponent $\beta$ estimated from GNSS vertical displacement at station RITE. (a) $\beta$ as a function of the number of years removed from the start of the analysis window (beginning in 2000). (b) $\beta$ plotted against the inverse-time linearity diagnostic $R^2$ (see Figure~\ref{fig3}). Colours indicate the analysis start date. A systematic decreasing trend is observed, with $\beta$ stabilising around 1.4 for the most recent windows, coinciding with the emergence of a robust singular regime.}
	\label{fig12}
\end{figure}

\begin{figure}[ht]
	\centering
	\includegraphics[width=\linewidth]{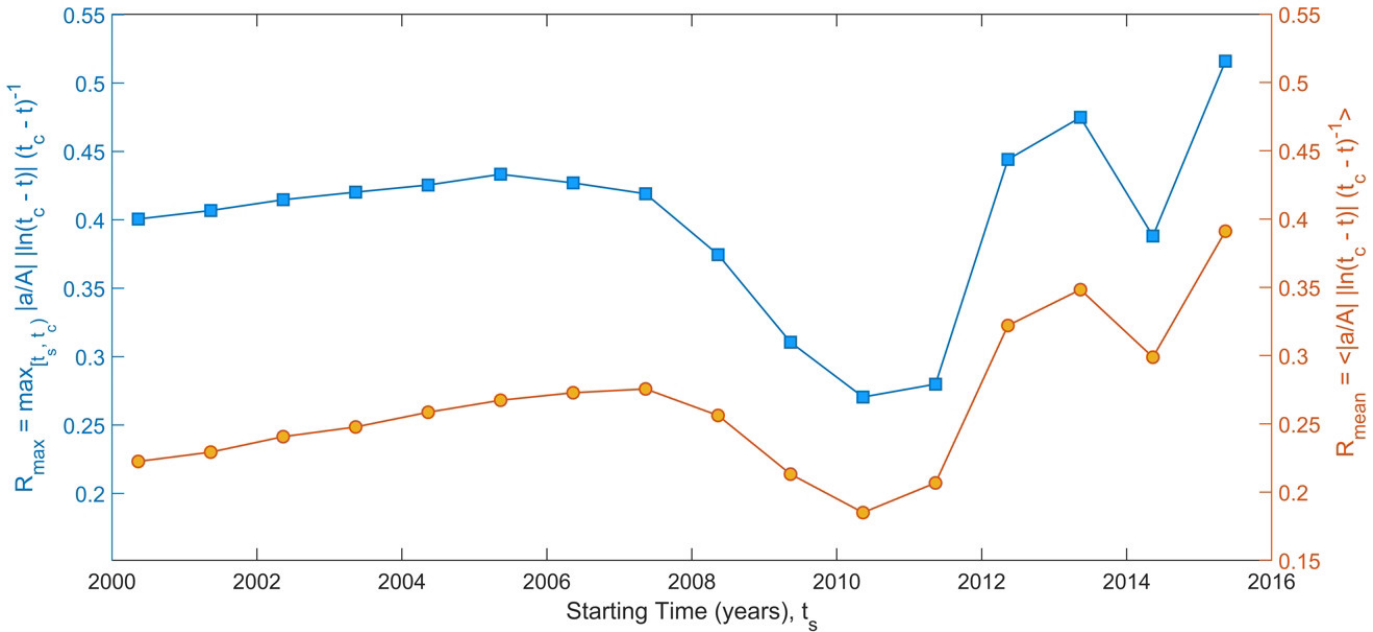}
	\caption{Relative importance of the regularization term compared to the leading singular term, quantified by the dimensionless ratio $R(t) = |a/A| \cdot |\ln(t_c - t)| \cdot (t_c - t)^{-1}$, as a function of analysis start date. The blue line (left axis) shows the maximum value of $R(t)$ over the interval $[t_{\text{start}}, t_c)$, while the orange line (right axis) shows the mean value over the same interval. Both metrics reach a minimum for windows starting between 2008 and 2011, corresponding to the optimal fitting interval identified in Supplementary Figure~\ref{fig7} and \ref{fig8}. The maximum ratio remains around 0.4 throughout, while the mean ratio increases from approximately 0.25 for earlier windows to 0.4 for the most recent interval (2016--2026). The slight linear increase over time is physically consistent with the growing influence of saturation effects as the system approaches the critical horizon.}
	\label{fig13}
\end{figure}

\begin{figure}[ht]
	\centering
	\includegraphics[width=\linewidth]{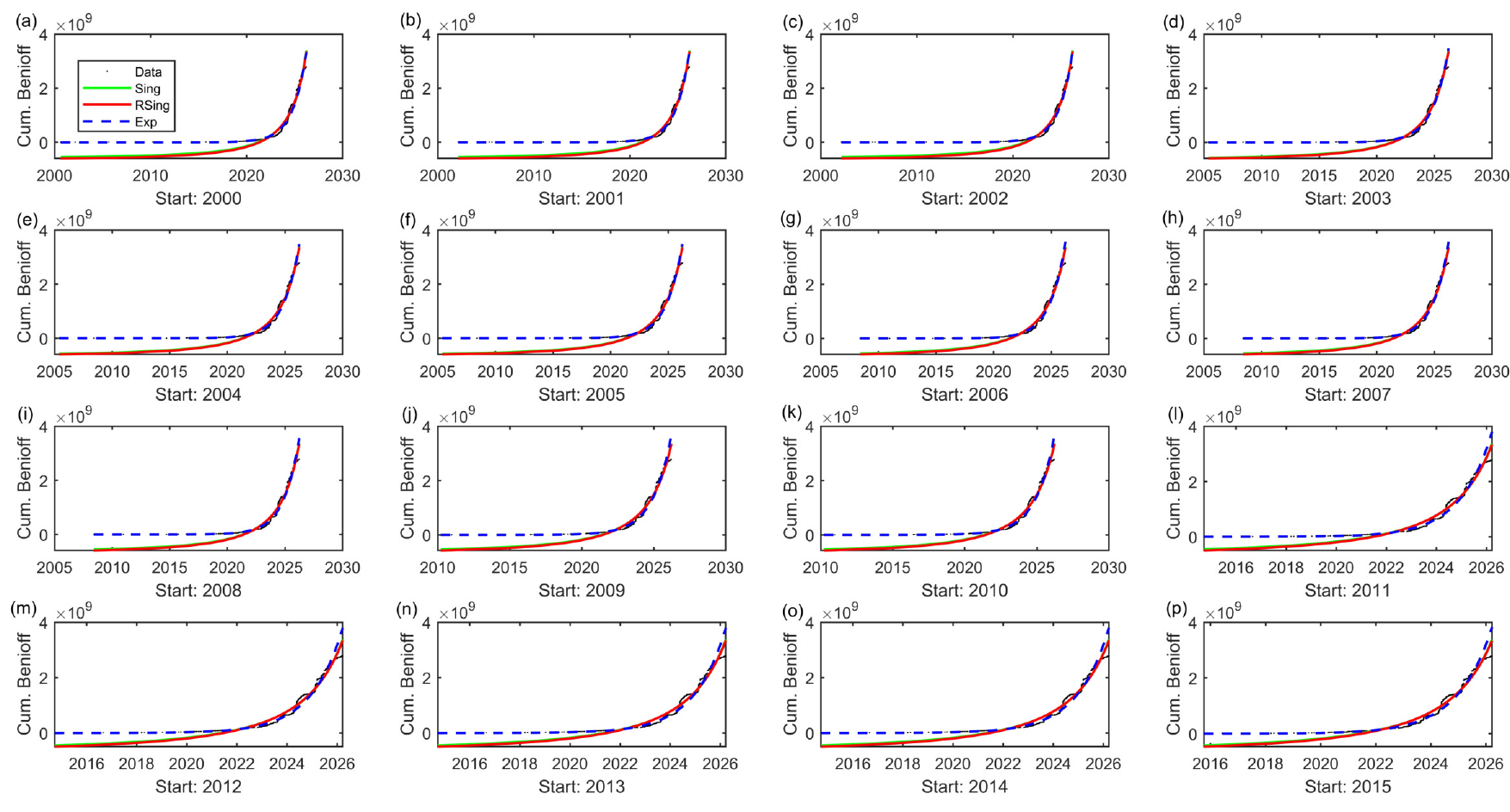}
	\caption{Benioff strain for analysis windows with start dates ranging from 2000 to 2015 (panels a-p). The dashed blue line represents the best-fitting exponential model, the solid green line the finite-time singularity model, and the solid red line the regularized singularity model.}
	\label{fig14}
\end{figure}

\begin{figure}[ht]
	\centering
	\includegraphics[width=\linewidth]{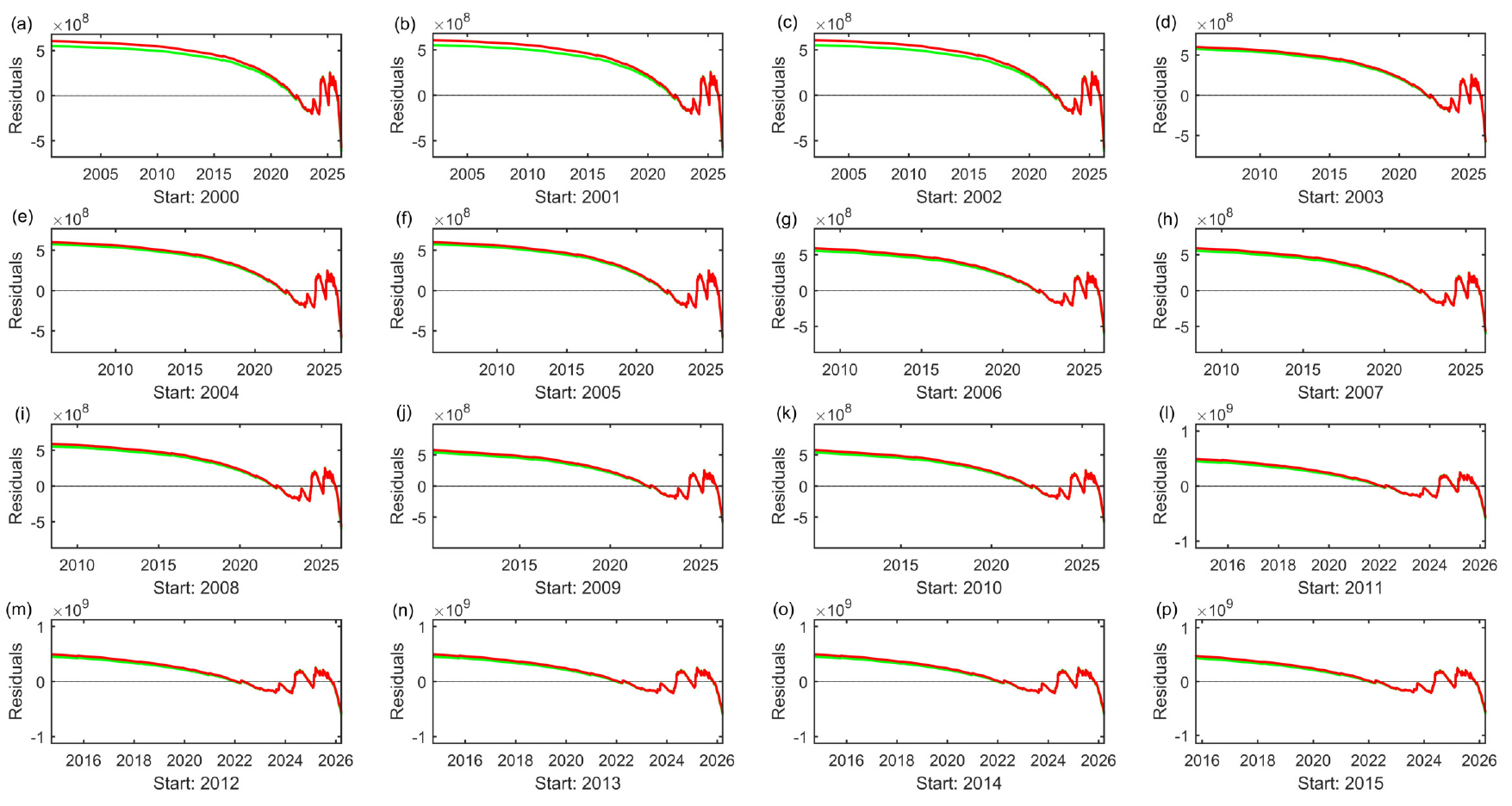}
	\caption{Residuals of the Benioff strain for different fitting functions and different analysis windows with start dates ranging from 2000 to 2015 (panels a-p). The solid green line the finite-time singularity residuals, and the solid red line the regularized singularity residuals.}
	\label{fig15}
\end{figure}	

\begin{figure}[ht]
	\centering
	\includegraphics[width=\linewidth]{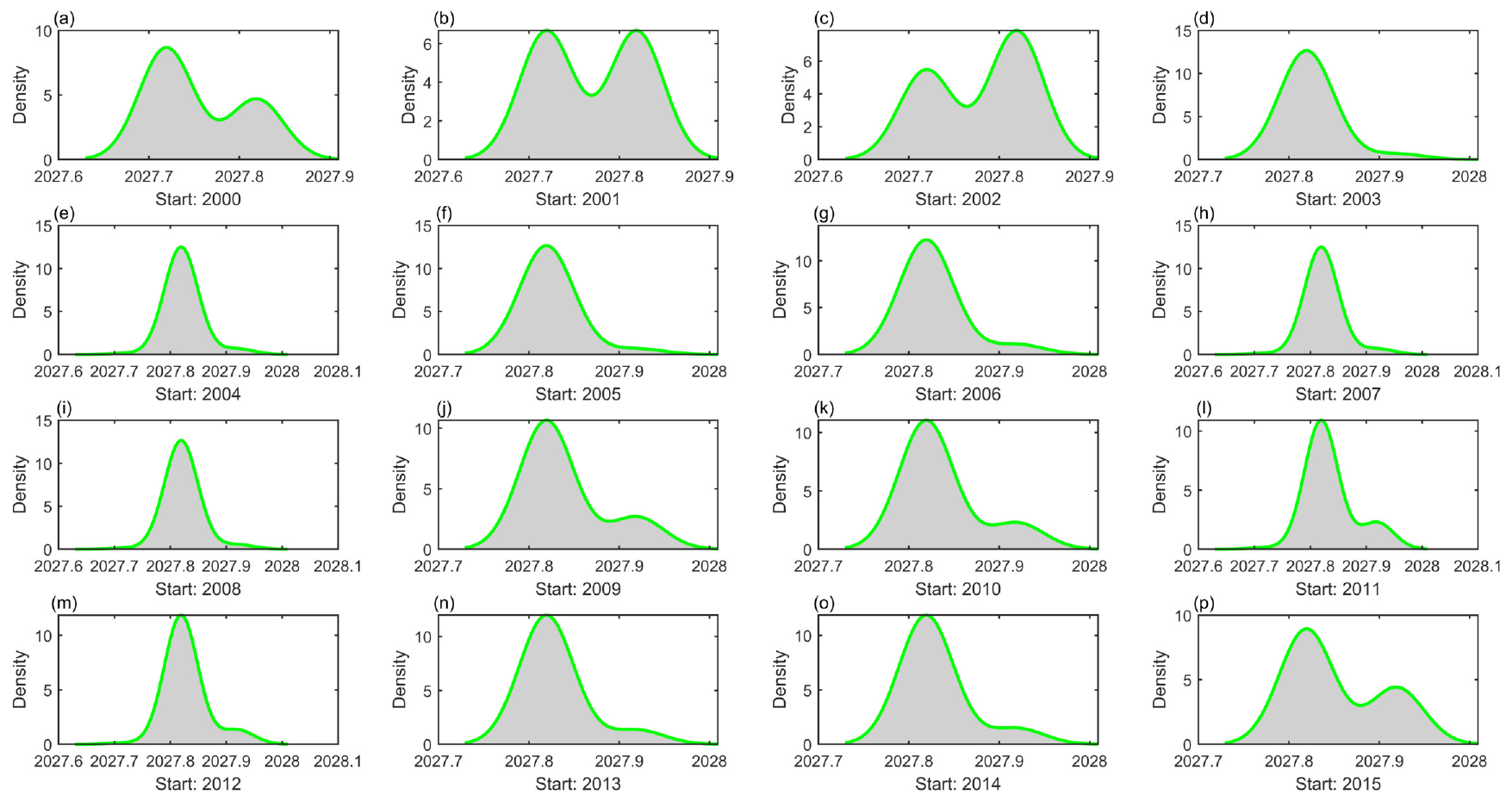}
	\caption{Probability density of the critical time $t_c$ estimated from the finite-time singularity model for the Benioff strain. Panels a-p correspond to analysis windows with start dates progressing from 2000 to 2015. For detailed average values with uncertainty for intervals starting beyond 2015, see Table 2.}
	\label{fig16}
\end{figure}

\begin{figure}[ht]
	\centering
	\includegraphics[width=\linewidth]{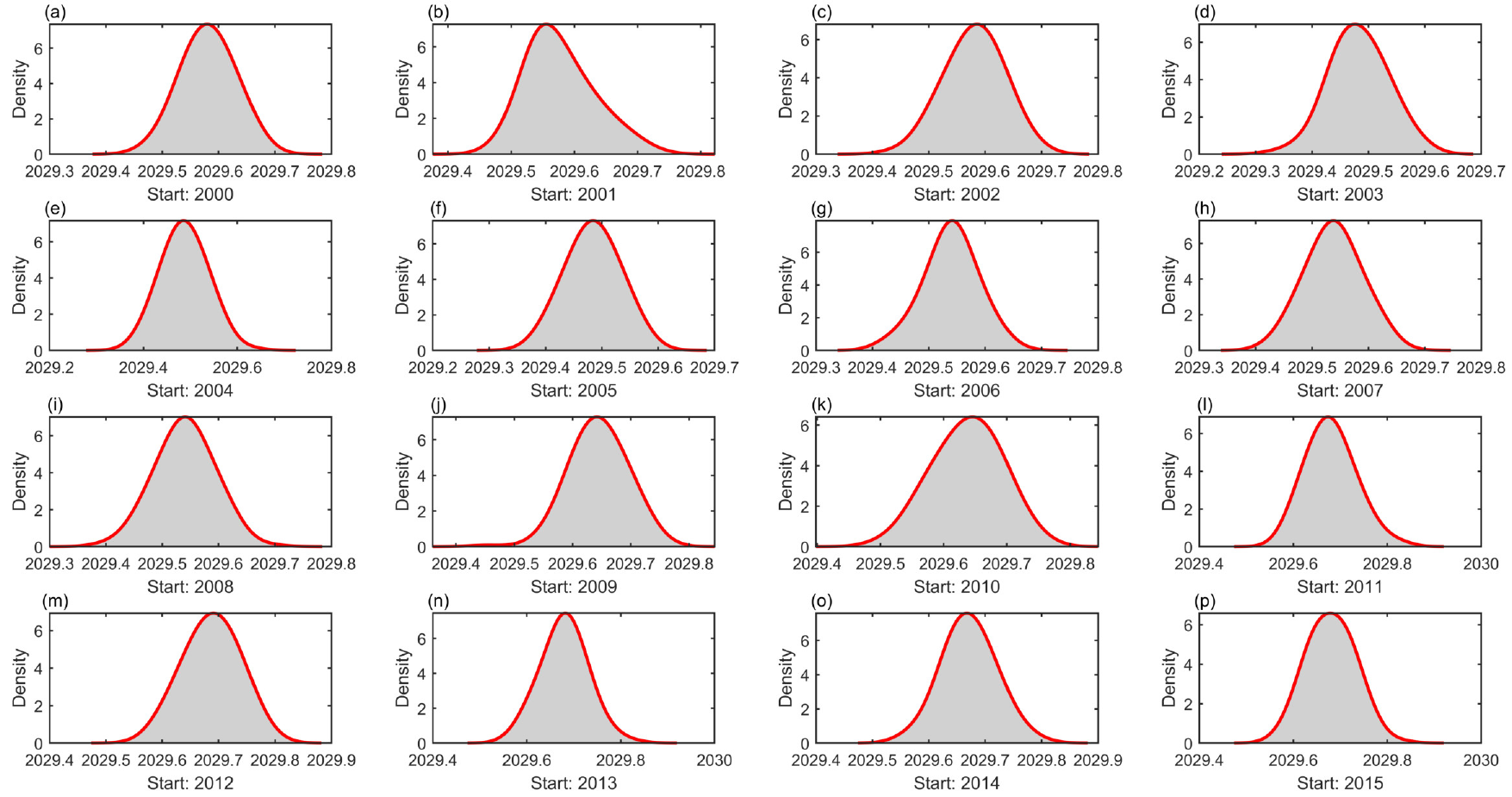}
	\caption{Probability density of the critical time $t_c$ estimated from the finite-time regularised singularity model for the Benioff strain. Panels a–p correspond to analysis windows with start dates progressing from 2000 to 2015. For detailed average values with uncertainty for intervals starting beyond 2015, see Table 4.}
	\label{fig17}
\end{figure}

\begin{figure}[ht]
	\centering
	\includegraphics[width=\linewidth]{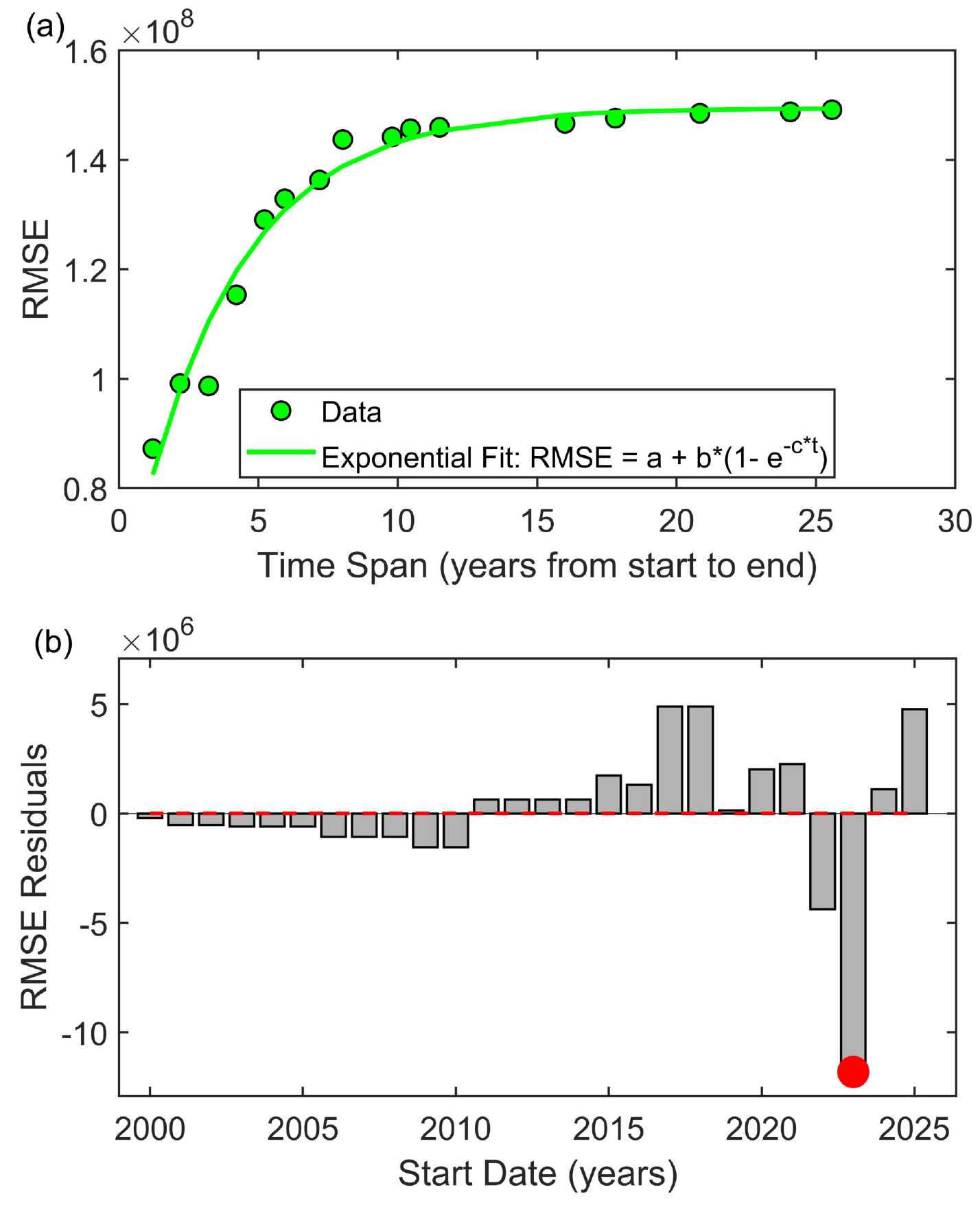}
	\caption{Identification of the optimal analysis window for the finite-time singularity model applied to the Benioff strain. (a) Root-mean-square error (RMSE) as a function of analysis start date, with an exponential fit capturing the systematic trend. (b) Residuals of the RMSE with respect to the linear fit. The minimum residual occurs at 2023 (red circle), indicating that the 2023--2026 window most faithfully reflects singular dynamics. This residual profile determines 75\% of the weighting scheme used to compute the final seismic critical time estimate $t_c^s$.}
	\label{fig18}
\end{figure}

\begin{figure}[ht]
	\centering
	\includegraphics[width=\linewidth]{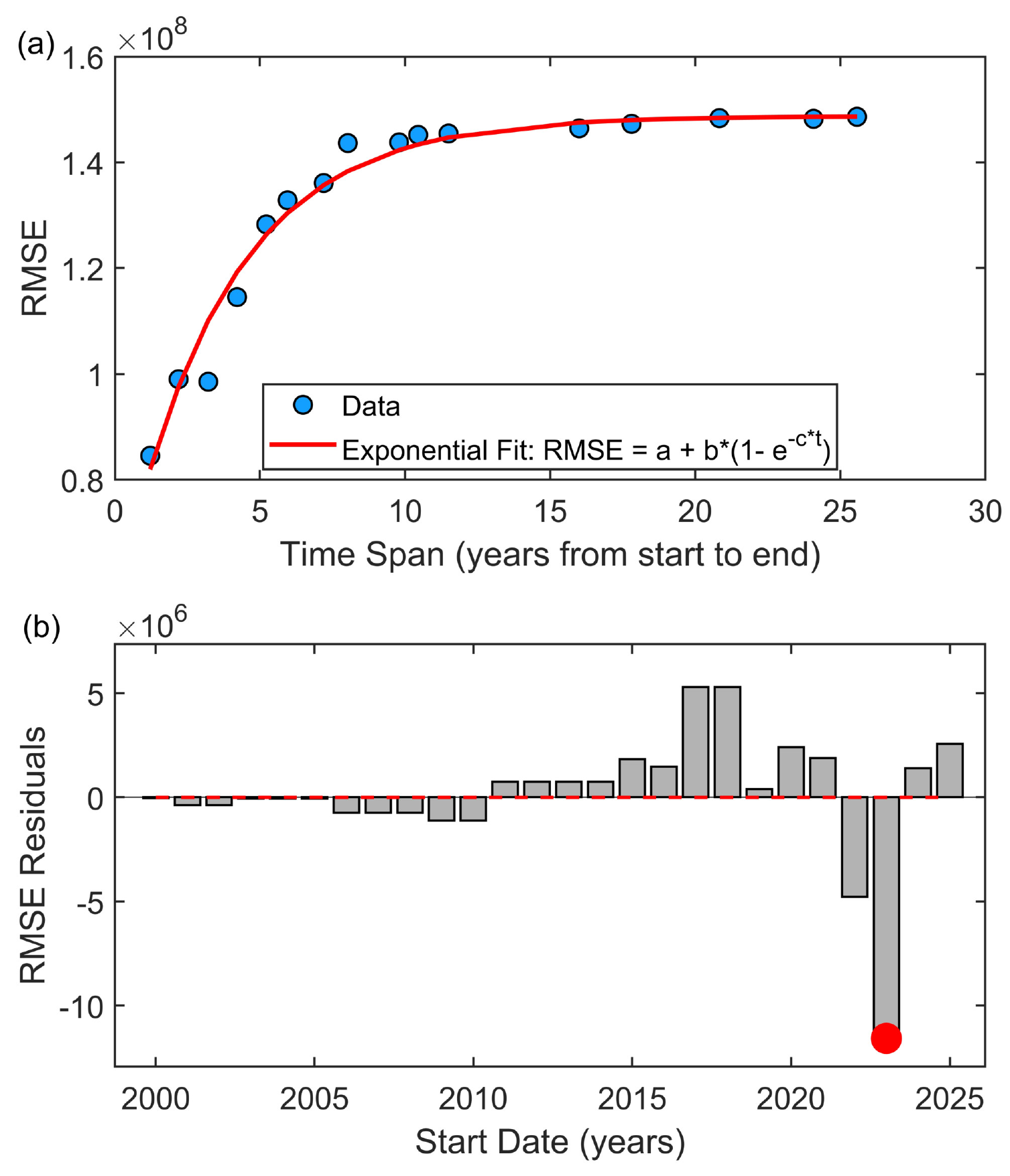}
	\caption{Identification of the optimal analysis window for the regularised finite-time singularity model applied to the Benioff strain. (a) Root-mean-square error (RMSE) as a function of analysis start date, with an exponential fit capturing the systematic trend. (b) Residuals of the RMSE with respect to the linear fit. The minimum residual occurs at 2023 (red circle), indicating that the 2023--2026 window most faithfully reflects singular dynamics. This residual profile determines 75\% of the weighting scheme used to compute the final seismic critical time estimate $t_c^s$.}
	\label{fig19}
\end{figure}	

\begin{figure}[ht]
	\centering
	\includegraphics[width=\linewidth]{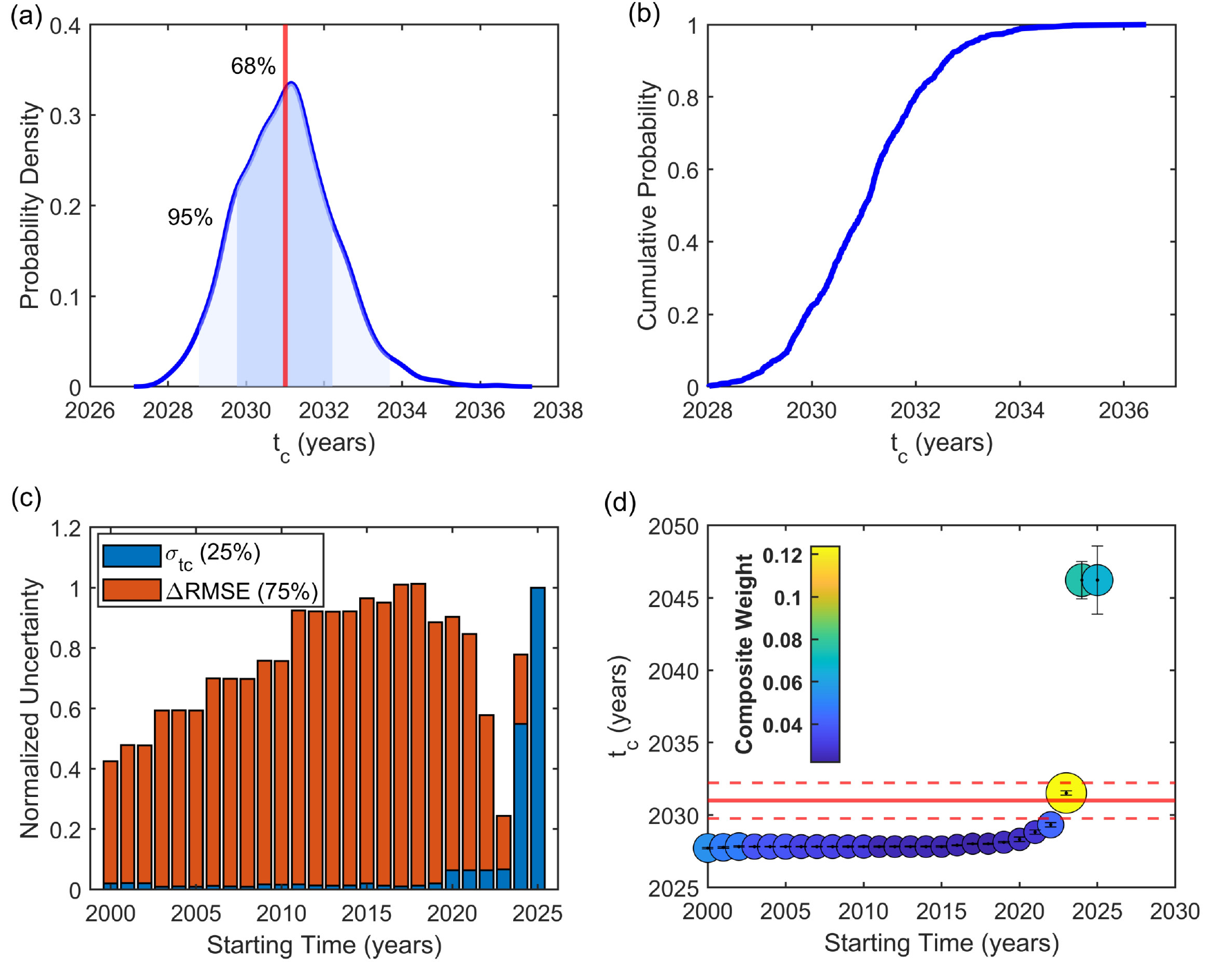}
	\caption{Weighting procedure for the combined estimation of the Benioff strain-based critical time $t_c$ from the finite-time singularity model. (a) Probability density function of $t_c$ obtained via weighted bootstrap resampling. (b) Cumulative distribution function corresponding to the density in panel a. The weights are defined as the inverse of the total normalized uncertainty, which combines the bootstrap uncertainty of $t_c$ (25\% contribution) and the RMSE residuals from Supplementary Figure~\ref{fig18}b (75\% contribution). (c) Individual $t_c$ estimates plotted against analysis start date. The size and colour of each point encode the composite weight, with the colour bar indicating the relative contribution of each window to the final probability density shown in panel a.}
	\label{fig20}
\end{figure}

\begin{figure}[ht]
	\centering
	\includegraphics[width=\linewidth]{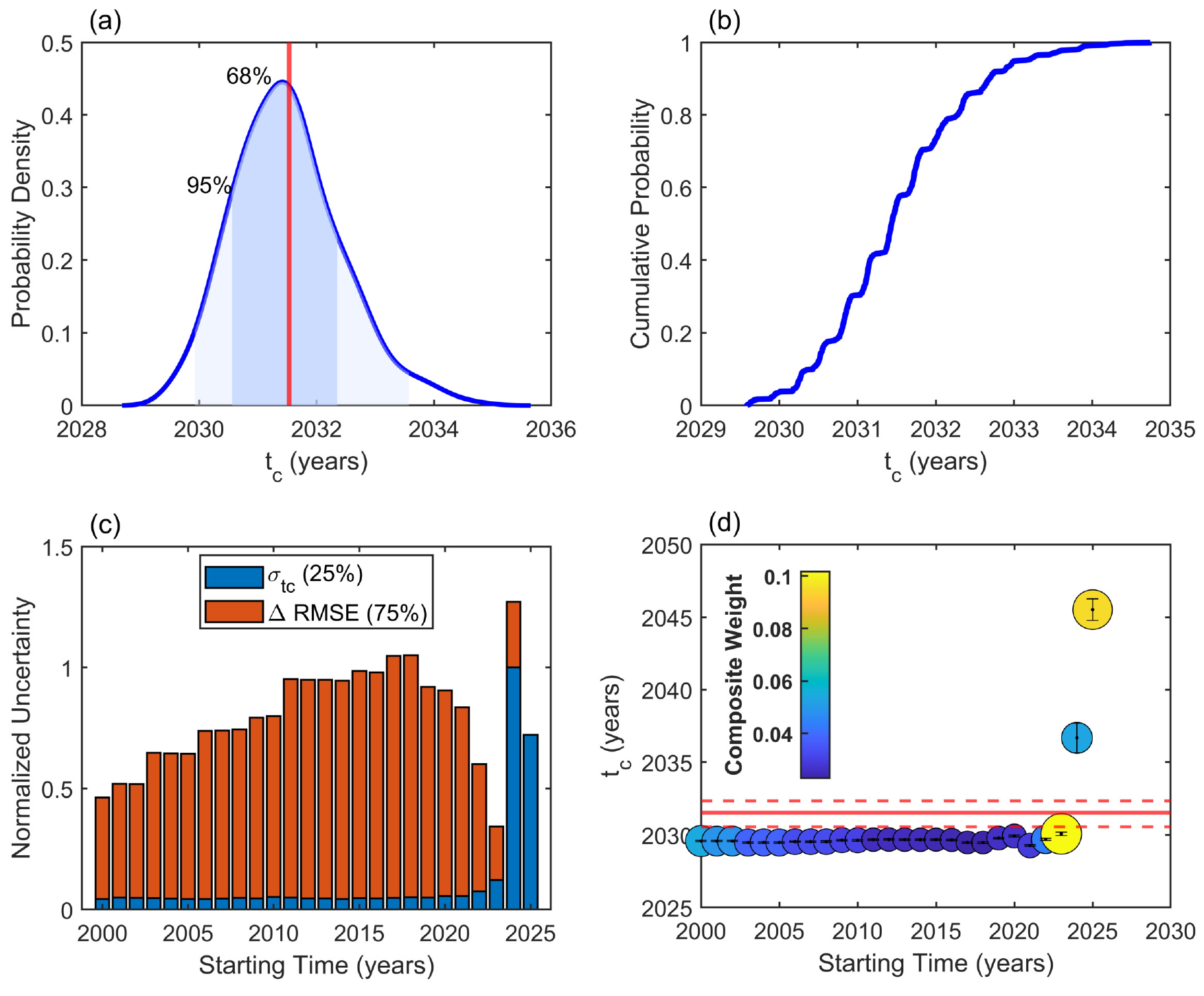}
	\caption{Weighting procedure for the combined estimation of the Benioff strain-based critical time $t_c$ from the regularised finite-time singularity model. (a) Probability density function of $t_c$ obtained via weighted bootstrap resampling. (b) Cumulative distribution function corresponding to the density in panel a. The weights are defined as the inverse of the total normalized uncertainty, which combines the bootstrap uncertainty of $t_c$ (25\% contribution) and the RMSE residuals from Supplementary Figure~\ref{fig19}b (75\% contribution). (c) Individual $t_c$ estimates plotted against analysis start date. The size and colour of each point encode the composite weight, with the colour bar indicating the relative contribution of each window to the final probability density shown in panel a.}
	\label{fig21}
\end{figure}

\begin{figure}[ht]
	\centering
	\includegraphics[width=\linewidth]{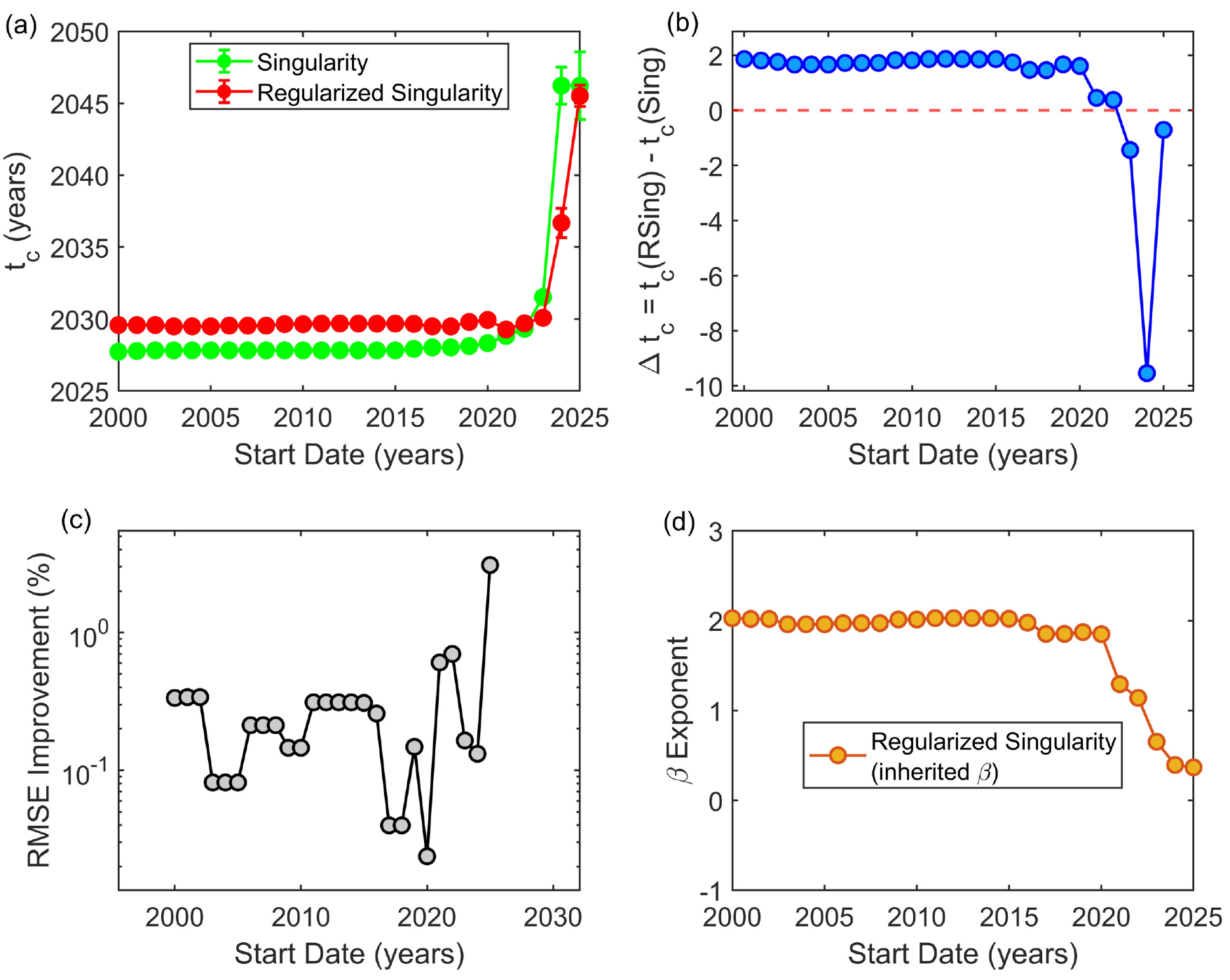}
	\caption{Comparison of the regularized and pure finite-time singularity models for cumulative Benioff strain at Campi Flegrei as a function of analysis start date. (a) Estimated critical time $t_c$ for both models. The estimate remains nearly constant for windows starting before 2020, reflecting the limited seismic energy release during the early phase of the unrest. A rapid forward shift occurs in the most recent windows, accompanied by substantially larger uncertainties, which contribute to the broader confidence intervals of the final seismic-based $t_c$ estimate. (b) Difference in $t_c$ between the two models ($\Delta t_c = t_c^{\text{RSing}} - t_c^{\text{Sing}}$). (c) Percentage improvement in RMSE achieved by the regularized model relative to the pure singularity model. (d) Singularity exponent $\beta$ as a function of start date, showing a marked decline in recent windows consistent with the transition to an accelerating regime.}
	\label{fig22}
\end{figure}

\begin{figure}[ht]
	\centering
	\includegraphics[width=\linewidth]{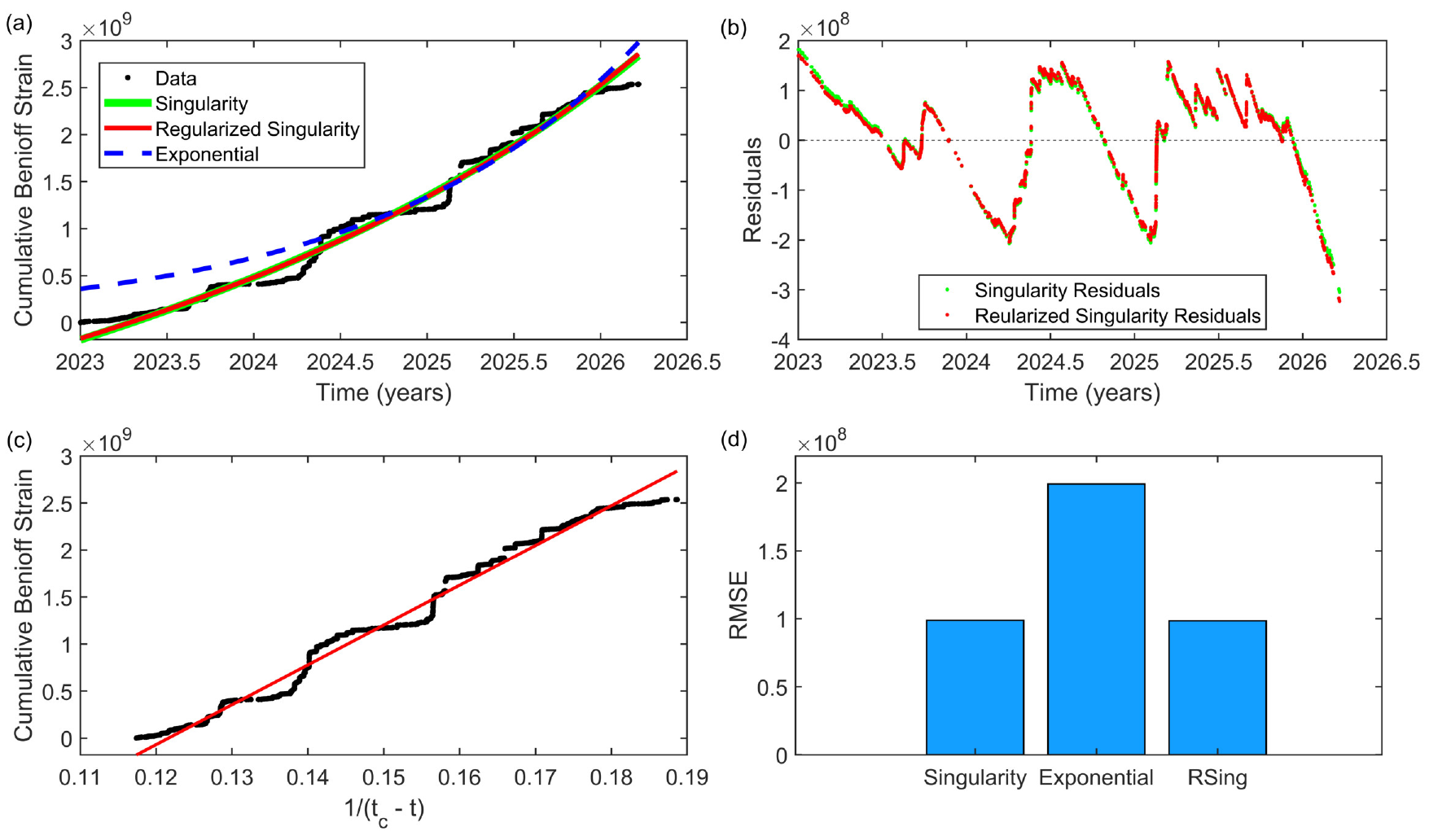}
	\caption{Benioff strain modelling for the optimal analysis window (2023--2026). (a) Cumulative Benioff strain with three competing model fits: exponential (dashed blue), finite-time singularity (solid green), and regularized singularity (solid red). (b) Residuals of the three fits relative to the observed data. (c) Regularized singularity fit plotted against the transformed time coordinate $1/(t_c - t)$, demonstrating the linear relationship characteristic of singular dynamics. (d) Root-mean-square error (RMSE) for the three models. The regularized singularity model achieves the lowest RMSE, outperforming both the pure singularity and exponential alternatives.}
	\label{fig23}
\end{figure}	

\begin{figure}[ht]
	\centering
	\includegraphics[width=\linewidth]{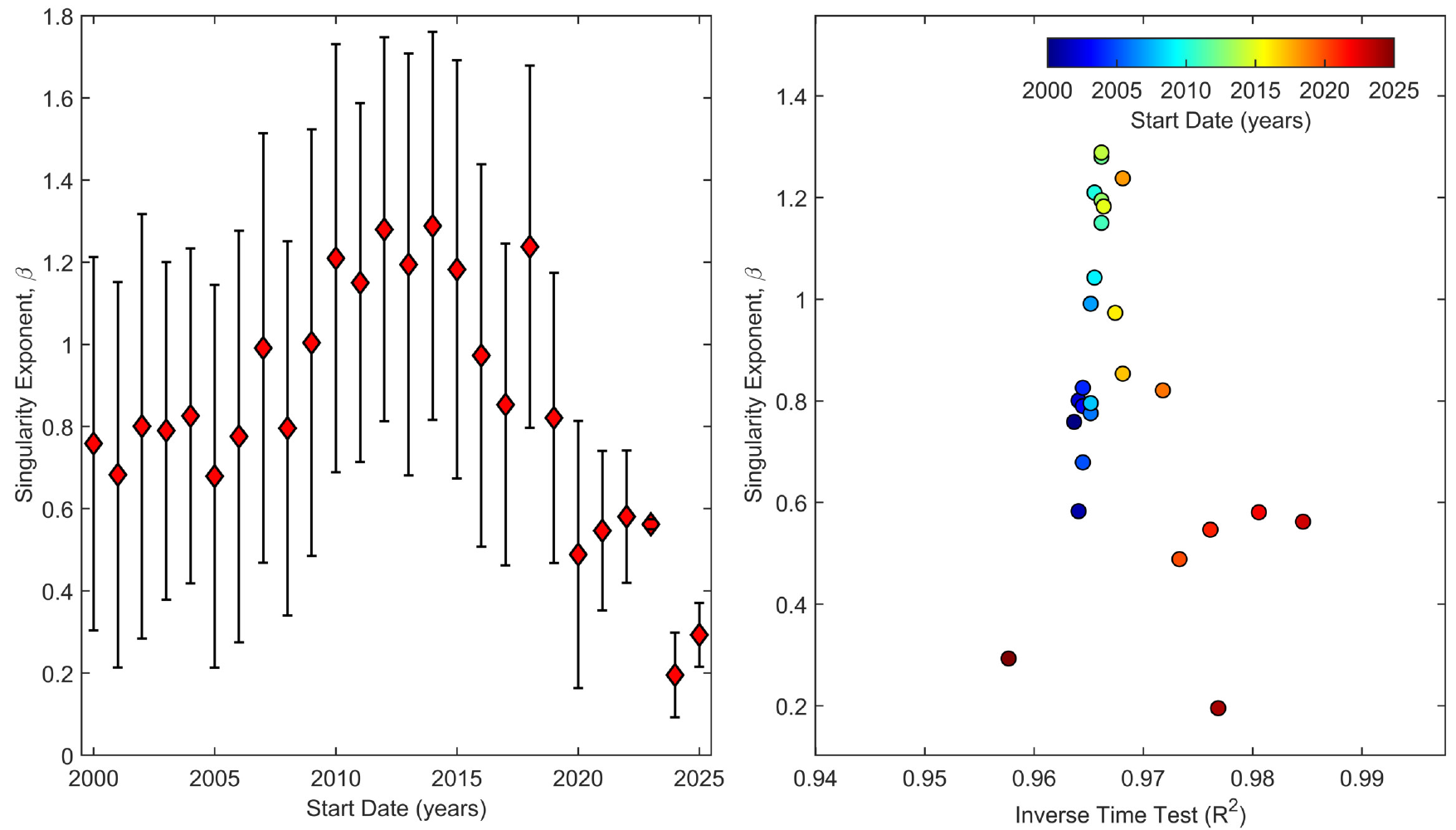}
	\caption{Evolution of the singularity exponent $\beta$ estimated from Benioff strain. (a) $\beta$ as a function of the number of the start of the analysis window (beginning in 2000). (b) $\beta$ plotted against the inverse-time linearity diagnostic $R^2$ (see Figure~\ref{fig3}). Colours indicate the analysis start date.}
	\label{fig24}
\end{figure}

\begin{figure}[ht]
	\centering
	\includegraphics[width=\linewidth]{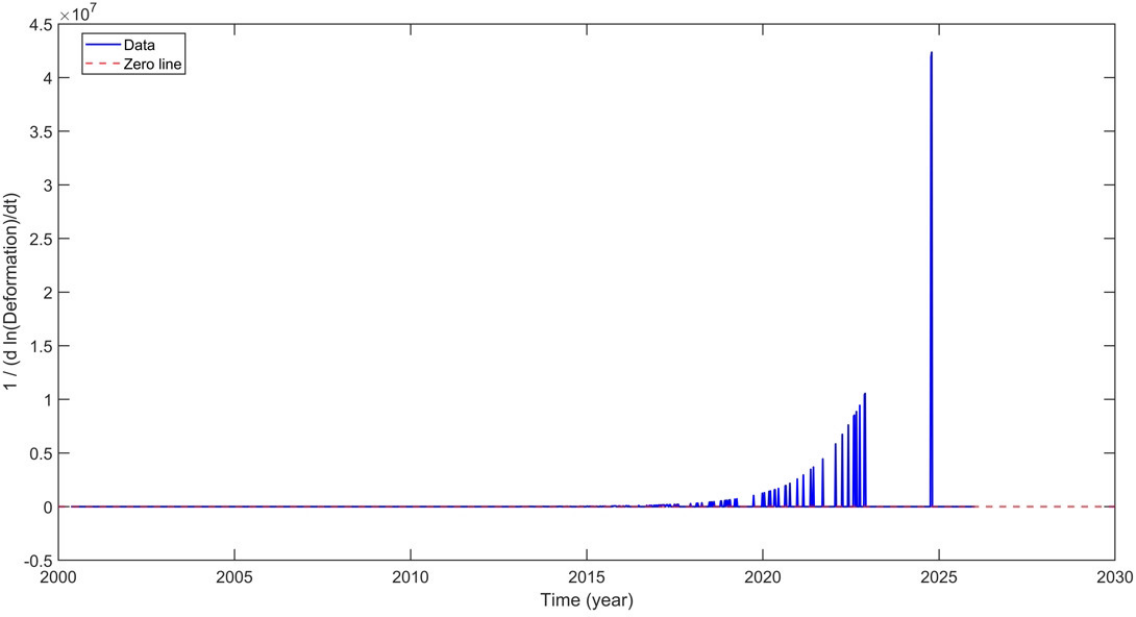}
	\caption{Temporal evolution of the inverse logarithmic derivative of GNSS vertical displacement, $1 / (d(\ln y)/dt)$, at station RITE. This quantity represents the characteristic timescale of deformation. Pronounced upward peaks correspond to rapid deformation bursts, each coinciding with episodes of heightened seismic activity. The overall trend reflects the interplay between transient accelerations and the secular evolution of the unrest.}
	\label{fig25}
\end{figure}

\begin{figure}[ht]
	\centering
	\includegraphics[width=\linewidth]{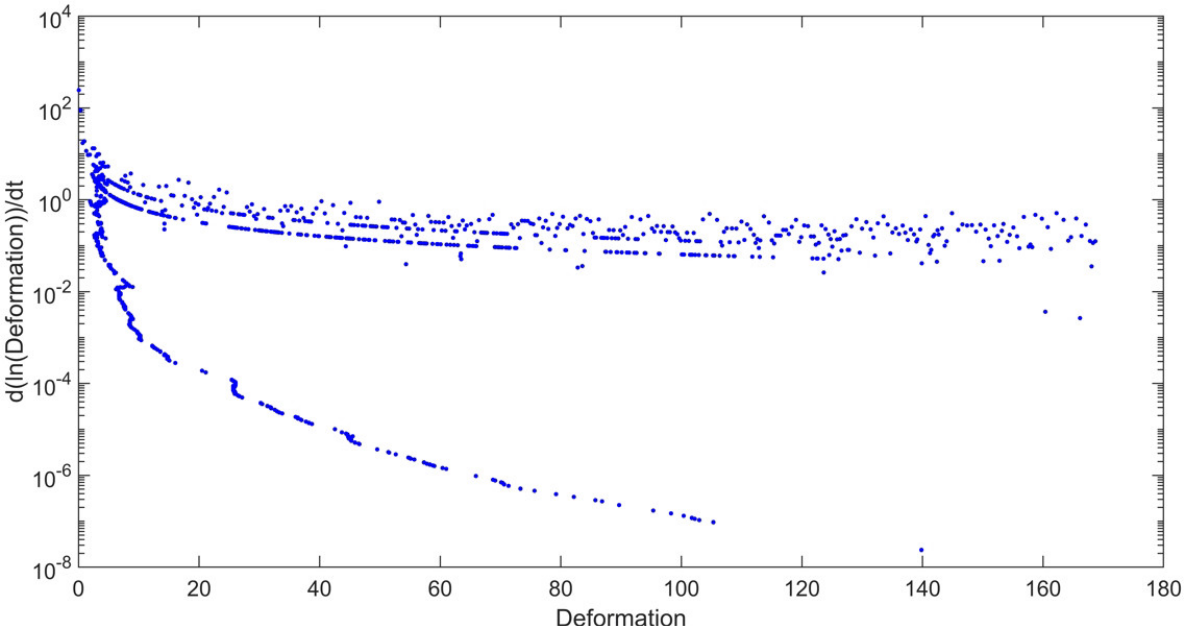}
	\caption{Logarithmic derivative of GNSS vertical displacement, $d(\ln y)/dt$, plotted against displacement on a semi-logarithmic scale. This representation separates the deformation dynamics into distinct declining trajectories, with upward concavity reflecting the interplay between fast and slow deformation modes. The clustering of points along different curves indicates the coexistence of multiple timescales in the destabilization process, consistent with a system driven by episodic forcing superimposed on a secular accelerating trend.}
	\label{fig26}
\end{figure}

\begin{figure}[ht]
	\centering
	\includegraphics[width=\linewidth]{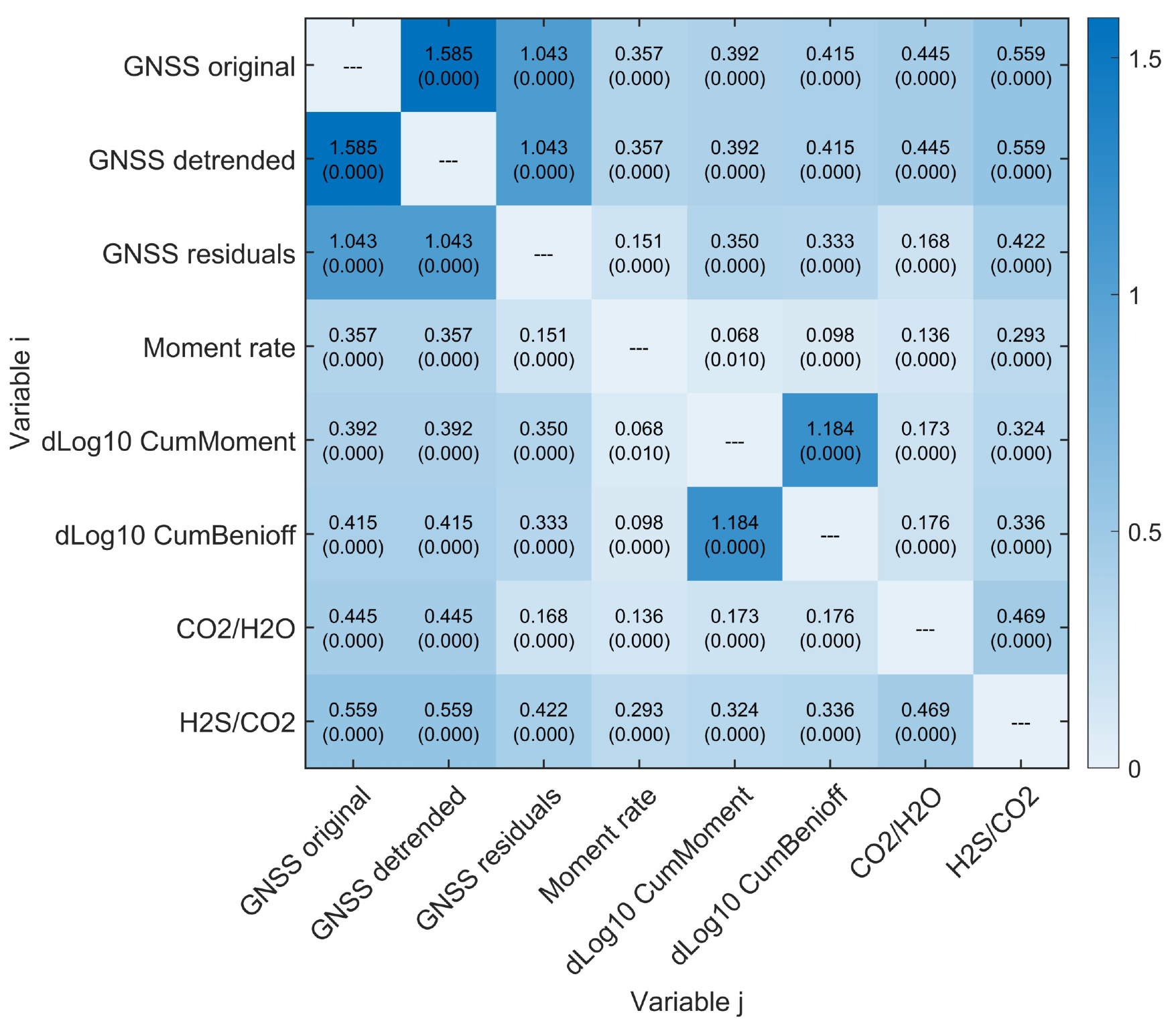}
	\caption{Mutual information matrix for the geophysical, geodetic and geochemical variables monitored at Campi Flegrei. Each cell quantifies the statistical dependence between variable pairs, capturing both linear and nonlinear relationships. The colour intensity (sky colormap) represents the mutual information magnitude, with brighter shades indicating stronger dependence. The matrix reveals a coupled system, with the strongest associations observed between GNSS deformation and seismic variables, and between relatively deep magmatic volatile tracer ($H_2S/CO_2$) and seismic energy release and geodetic deformation.}
	\label{fig27}
\end{figure}	

\begin{figure}[ht]
	\centering
	\includegraphics[width=\linewidth]{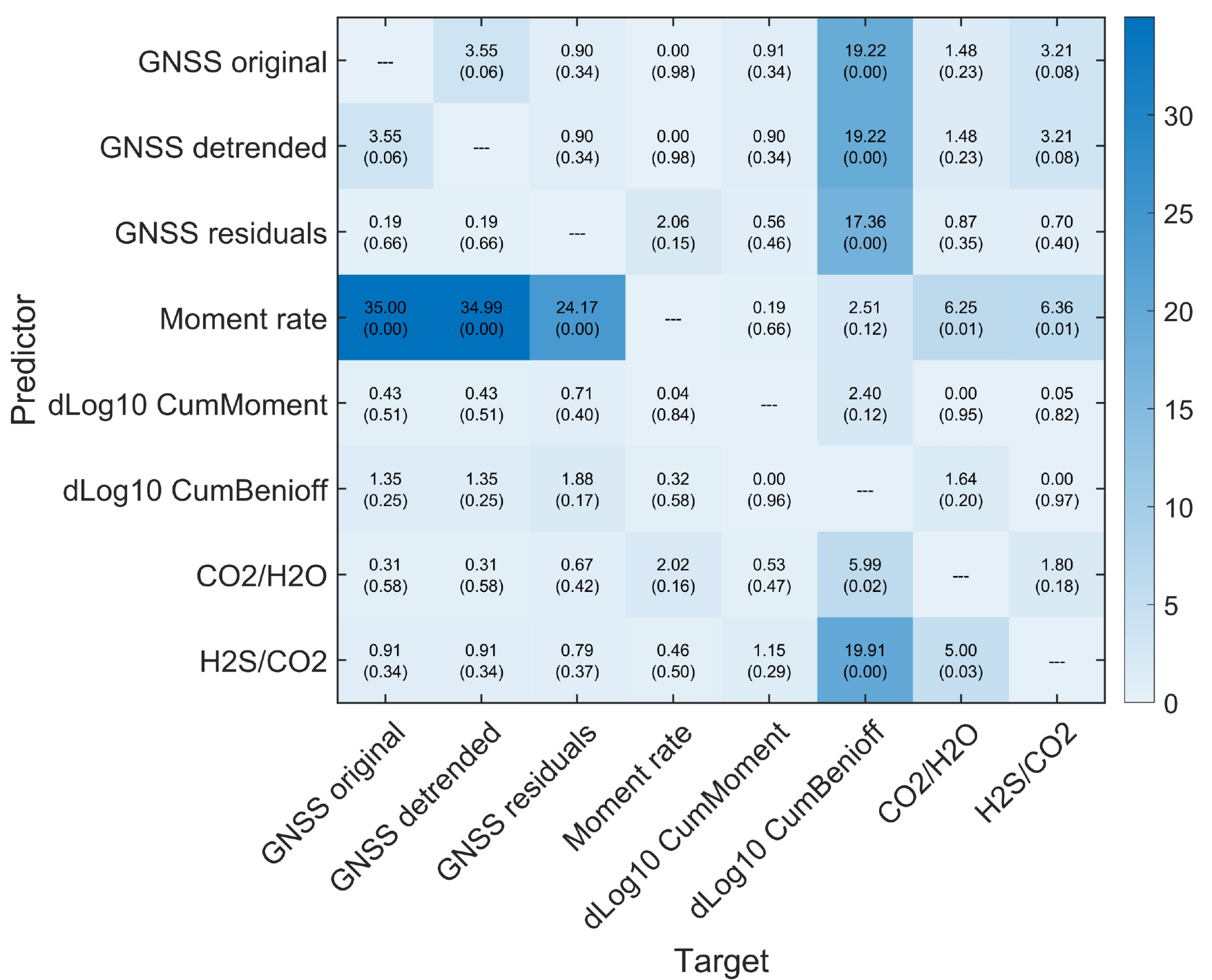}
	\caption{Granger causality matrix for the geophysical and geochemical variables monitored at Campi Flegrei. Each cell reports the $F$-statistic (and corresponding $p$-value) testing whether the past of the row variable significantly improves the prediction of the column variable beyond its own history (under the assumption of time series separability - for a more reliable assessment, see the matrix of transfer entropy in Figure 4b in the main text). The colour intensity (sky colormap) represents the strength of the causal link. Small values (e.g., $p < 0.01$) indicate directed predictive influence, with GNSS deformation emerging as the primary driver of seismicity and seismic moment rate of geodetic deformation and deep magmatic volatiles (traced by H$_2$S/CO$_2$) showing a possible causal control on seismic energy release represented by the Benioff strain.}
	\label{fig28}
\end{figure}

\begin{figure}[ht]
	\centering
	\includegraphics[width=\linewidth]{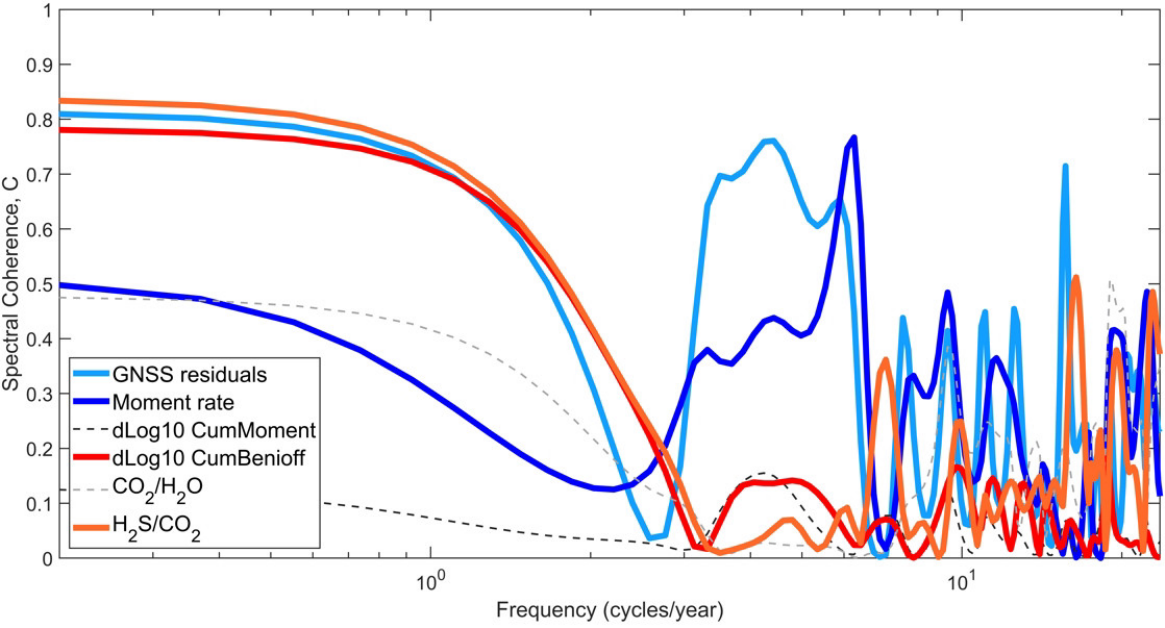}
	\caption{Spectral coherence between GNSS vertical displacement and other geophysical and geochemical variables at Campi Flegrei. Coherence quantifies the frequency-domain linear correlation, with values ranging from 0 (no relationship) to 1 (perfect linear coupling). Significant coherence at periods of 2--4 years between deformation and seismic variables indicates common forcing mechanisms operating at these timescales, likely associated with fluid pressure cycles. The coherence with geochemical ratios is weaker at higher frequency but strong for the deep marker - H$_2$S/CO$_2$, consistent with a mechanism of common slow driving dynamics and a possible indirect, time-lagged nature of fluid transport from depth to the surface.}
	\label{fig29}
\end{figure}

\newpage

\begin{landscape}
	\begin{table}[ht]
		\centering
		\caption{\textbf{Sensitivity analysis of the singularity model for GNSS vertical displacement at station RITE.} The analysis is performed by systematically advancing the starting date of the fitting window from 2000.4 to 2015.4 in six-month increments. For each window, we report the estimated critical time $t_c$ (years, with $2\sigma$ bootstrap uncertainty), the root-mean-square error of the singularity fit (RMSE$_{\text{Sing}}$) and exponential fit (RMSE$_{\text{Exp}}$), their ratio (Ratio = RMSE$_{\text{Exp}}$/RMSE$_{\text{Sing}}$), the Akaike Information Criterion difference ($\Delta$AIC = AIC$_{\text{Sing}} - \text{AIC}_{\text{Exp}}$), the Bayesian Information Criterion difference ($\Delta$BIC), the normalised BIC (BIC/$n$), the inverse-time linearity diagnostic ($R^2$), the stability index of $t_c$ estimates, the RMSE residual after linear trend removal (RMSE$_{\text{Res}}$), the fitted singularity exponent $\beta$, and the outcome of the model comparison. Negative $\Delta$AIC and $\Delta$BIC values favour the singularity model. The optimal window (minimum RMSE$_{\text{Res}}$) occurs at start date 2008.4, where $\beta$ stabilises near 1.41 and the singularity model decisively outperforms the exponential alternative.}
		\label{tab:sensitivity_gnss}
		\small
		\setlength{\tabcolsep}{3.5pt}
		\begin{tabular}{c c c c c c c c c c c c c c}
			\toprule
			\textbf{Start} & $\boldsymbol{t_c}$ & \textbf{RMSE$_{\textbf{Sing}}$} & \textbf{RMSE$_{\textbf{Exp}}$} & \textbf{Ratio} & $\boldsymbol{\Delta}$\textbf{AIC} & $\boldsymbol{\Delta}$\textbf{BIC} & \textbf{BIC/$\boldsymbol{n}$} & \textbf{Inv} $\boldsymbol{R^2}$ & \textbf{Stability} & \textbf{RMSE$_{\textbf{Res}}$} & $\boldsymbol{\beta}$ & \textbf{Best} \\
			\midrule
			2000.4 & 2032.24 $\pm$ 0.12 & 3.4873 & 3.2002 & 0.918 & 230.9 & 230.9 & 2.5143 & 0.9924 & 0.0000 & 0.2617 & 1.5123 & singularity weak \\
			2000.9 & 2032.37 $\pm$ 0.13 & 3.3657 & 4.8092 & 1.429 & -940.1 & -940.1 & 2.4436 & 0.9930 & 0.0000 & 0.2230 & 1.5054 & singularity \\
			2001.4 & 2032.51 $\pm$ 0.12 & 3.2108 & 4.5035 & 1.403 & -873.6 & -873.6 & 2.3497 & 0.9937 & 0.0001 & 0.1480 & 1.5014 & singularity \\
			2001.9 & 2032.67 $\pm$ 0.12 & 3.0895 & 4.3754 & 1.416 & -880.4 & -880.4 & 2.2730 & 0.9942 & 0.0001 & 0.1066 & 1.4933 & singularity \\
			2002.4 & 2032.79 $\pm$ 0.13 & 3.0027 & 4.0726 & 1.356 & -755.8 & -755.8 & 2.2162 & 0.9946 & 0.0001 & 0.0996 & 1.4887 & singularity \\
			2002.9 & 2032.93 $\pm$ 0.12 & 2.9059 & 4.1962 & 1.444 & -891.4 & -891.4 & 2.1511 & 0.9950 & 0.0001 & 0.0858 & 1.4811 & singularity \\
			2003.4 & 2033.08 $\pm$ 0.13 & 2.7954 & 4.6429 & 1.661 & -1204.4 & -1204.4 & 2.0739 & 0.9954 & 0.0001 & 0.0552 & 1.4780 & singularity \\
			2003.9 & 2033.25 $\pm$ 0.13 & 2.6666 & 4.7684 & 1.788 & -1349.5 & -1349.5 & 1.9798 & 0.9958 & 0.0002 & 0.0062 & 1.4722 & singularity \\
			2004.4 & 2033.42 $\pm$ 0.12 & 2.5545 & 4.4596 & 1.746 & -1264.8 & -1264.8 & 1.8943 & 0.9962 & 0.0002 & -0.0260 & 1.4675 & singularity \\
			2004.9 & 2033.55 $\pm$ 0.14 & 2.4900 & 4.5476 & 1.826 & -1336.0 & -1336.0 & 1.8435 & 0.9964 & 0.0002 & -0.0107 & 1.4613 & singularity \\
			2005.4 & 2033.71 $\pm$ 0.13 & 2.4233 & 4.6580 & 1.922 & -1414.1 & -1414.1 & 1.7896 & 0.9966 & 0.0002 & 0.0056 & 1.4593 & singularity \\
			2005.9 & 2033.85 $\pm$ 0.14 & 2.3622 & 4.8552 & 2.055 & -1521.5 & -1521.5 & 1.7390 & 0.9968 & 0.0003 & 0.0244 & 1.4517 & singularity \\
			2006.4 & 2034.04 $\pm$ 0.14 & 2.2646 & 5.4214 & 2.394 & -1798.2 & -1798.2 & 1.6550 & 0.9971 & 0.0003 & 0.0066 & 1.4440 & singularity \\
			2006.9 & 2034.28 $\pm$ 0.14 & 2.0885 & 5.9799 & 2.863 & -2112.3 & -2112.3 & 1.4935 & 0.9976 & 0.0003 & -0.0896 & 1.4380 & singularity \\
			2007.4 & 2034.58 $\pm$ 0.12 & 1.8125 & 5.6970 & 3.143 & -2237.8 & -2237.8 & 1.2105 & 0.9982 & 0.0004 & -0.2827 & 1.4267 & singularity \\
			2007.9 & 2034.81 $\pm$ 0.12 & 1.6774 & 5.3901 & 3.213 & -2220.3 & -2220.3 & 1.0561 & 0.9985 & 0.0004 & -0.3379 & 1.4182 & singularity \\
			2008.4 & 2035.03 $\pm$ 0.13 & 1.5724 & 5.3467 & 3.400 & -2264.1 & -2264.1 & 0.9274 & 0.9987 & 0.0004 & -0.3630 & 1.4124 & singularity \\
			2008.9 & 2035.22 $\pm$ 0.13 & 1.5043 & 5.3617 & 3.564 & -2285.2 & -2285.2 & 0.8393 & 0.9988 & 0.0005 & -0.3513 & 1.4050 & singularity \\
			2009.4 & 2035.40 $\pm$ 0.12 & 1.4556 & 5.4193 & 3.723 & -2295.2 & -2295.2 & 0.7741 & 0.9989 & 0.0005 & -0.3201 & 1.4050 & singularity \\
			2009.9 & 2035.56 $\pm$ 0.13 & 1.4258 & 5.2873 & 3.708 & -2217.5 & -2217.5 & 0.7333 & 0.9989 & 0.0005 & -0.2670 & 1.4084 & singularity \\
			2010.4 & 2035.60 $\pm$ 0.15 & 1.4426 & 4.9760 & 3.449 & -2030.6 & -2030.6 & 0.7575 & 0.9989 & 0.0005 & -0.1703 & 1.4273 & singularity \\
			2010.9 & 2035.57 $\pm$ 0.15 & 1.4620 & 4.9639 & 3.395 & -1941.1 & -1941.1 & 0.7849 & 0.9988 & 0.0006 & -0.0710 & 1.4295 & singularity \\
			2011.4 & 2035.44 $\pm$ 0.14 & 1.4368 & 4.8330 & 3.364 & -1868.1 & -1868.1 & 0.7507 & 0.9988 & 0.0006 & -0.0164 & 1.4296 & singularity \\
			2011.9 & 2035.31 $\pm$ 0.14 & 1.4161 & 4.9999 & 3.531 & -1874.6 & -1874.6 & 0.7225 & 0.9988 & 0.0006 & 0.0459 & 1.4324 & singularity \\
			2012.4 & 2035.11 $\pm$ 0.14 & 1.3528 & 4.6870 & 3.465 & -1781.9 & -1781.9 & 0.6319 & 0.9989 & 0.0006 & 0.0624 & 1.4392 & singularity \\
			2012.9 & 2035.00 $\pm$ 0.14 & 1.3195 & 5.8476 & 4.432 & -2057.5 & -2057.5 & 0.5829 & 0.9989 & 0.0006 & 0.1090 & 1.4503 & singularity \\
			2013.4 & 2035.27 $\pm$ 0.14 & 1.2196 & 6.8892 & 5.649 & -2302.8 & -2302.8 & 0.4263 & 0.9990 & 0.0006 & 0.0890 & 1.4454 & singularity \\
			2013.9 & 2035.54 $\pm$ 0.15 & 1.1479 & 6.7304 & 5.863 & -2260.4 & -2260.4 & 0.3062 & 0.9991 & 0.0006 & 0.0972 & 1.4361 & singularity \\
			2014.4 & 2035.55 $\pm$ 0.17 & 1.1647 & 6.7490 & 5.795 & -2150.5 & -2150.5 & 0.3364 & 0.9991 & 0.0006 & 0.1969 & 1.4384 & singularity \\
			2014.9 & 2035.51 $\pm$ 0.18 & 1.1765 & 7.1382 & 6.068 & -2113.1 & -2113.1 & 0.3576 & 0.9990 & 0.0006 & 0.2885 & 1.4473 & singularity \\
			2015.4 & 2035.53 $\pm$ 0.18 & 1.2024 & 7.1240 & 5.925 & -1992.7 & -1992.7 & 0.4025 & 0.9989 & 0.0006 & 0.3943 & 1.4510 & singularity \\
			\bottomrule
		\end{tabular}
	\end{table}
\end{landscape}

\begin{landscape}
	\small
	\setlength{\tabcolsep}{3.5pt}
	\begin{longtable}{c c c c c c c c c c c}
		\caption{\textbf{Sensitivity analysis of the singularity model for cumulative Benioff strain at Campi Flegrei.} The analysis is performed by systematically advancing the starting date of the fitting window from 2000.0 to 2025.0 in six-month increments. For each window, we report the number of events, the estimated critical time $t_c$ (years, with $2\sigma$ bootstrap uncertainty), the inverse-time linearity diagnostic ($R^2_{\text{lin}}$), the coefficient of determination of the singularity fit ($R^2_{\text{sing}}$), the root-mean-square error of the singularity fit (RMSE$_{\text{sing}}$), the ratio of exponential to singularity RMSE (RMSE$_{\text{ratio}} = \text{RMSE}_{\text{exp}}/\text{RMSE}_{\text{sing}}$), the Akaike Information Criterion difference ($\Delta$AIC = AIC$_{\text{sing}} - \text{AIC}_{\text{exp}}$), the RMSE residual after exponential trend removal (RMSE$_{\text{Res}}$), the fitted singularity exponent $\beta$, and the outcome of the model comparison. Negative $\Delta$AIC values favour the singularity model. The optimal window (minimum RMSE$_{\text{Res}}$) occurs at start date 2019.0, where $\beta$ stabilises near 0.74 and the singularity model decisively outperforms the exponential alternative.}
		\label{tab:sensitivity_benioff} \\
		\toprule
		\textbf{Start} & \textbf{Events} & $\boldsymbol{t_c}$ & $\boldsymbol{R^2_{\textbf{lin}}}$ & $\boldsymbol{R^2_{\textbf{sing}}}$ & \textbf{RMSE$_{\textbf{sing}}$} ($\times 10^8$) & \textbf{RMSE$_{\textbf{ratio}}$} & $\boldsymbol{\Delta}$\textbf{AIC} & \textbf{RMSE$_{\textbf{Res}}$} ($\times 10^4$) & $\boldsymbol{\beta}$ & \textbf{Best Model} \\
		\midrule
		\endfirsthead
		\multicolumn{11}{c}{\textit{Continued from previous page}} \\
		\toprule
		\textbf{Start} & \textbf{Events} & $\boldsymbol{t_c}$ & $\boldsymbol{R^2_{\textbf{lin}}}$ & $\boldsymbol{R^2_{\textbf{sing}}}$ & \textbf{RMSE$_{\textbf{sing}}$} ($\times 10^8$) & \textbf{RMSE$_{\textbf{ratio}}$} & $\boldsymbol{\Delta}$\textbf{AIC} & \textbf{RMSE$_{\textbf{Res}}$} ($\times 10^4$) & $\boldsymbol{\beta}$ & \textbf{Best Model} \\
		\midrule
		\endhead
		\bottomrule
		\endfoot
		2000.0 & 2483 & 2027.76 $\pm$ 0.05 & 0.9637 & 0.9705 & 1.4912 & 0.986 & 67.7 & 9.96 & 0.7584 & exponential \\
		2000.5 & 2483 & 2027.75 $\pm$ 0.05 & 0.9637 & 0.9705 & 1.4912 & 0.986 & 67.7 & 9.96 & 0.7198 & exponential \\
		2001.0 & 2482 & 2027.76 $\pm$ 0.05 & 0.9641 & 0.9707 & 1.4874 & 0.985 & 74.5 & -23.4 & 0.8299 & exponential \\
		2001.5 & 2482 & 2027.77 $\pm$ 0.05 & 0.9641 & 0.9707 & 1.4874 & 0.985 & 74.5 & -23.4 & 0.9592 & exponential \\
		2002.0 & 2482 & 2027.78 $\pm$ 0.05 & 0.9641 & 0.9707 & 1.4874 & 0.985 & 74.5 & -23.4 & 1.0085 & exponential \\
		2002.5 & 2481 & 2027.83 $\pm$ 0.03 & 0.9645 & 0.9708 & 1.4848 & 0.998 & 12.3 & -28.9 & 0.9311 & exponential \\
		2003.0 & 2481 & 2027.82 $\pm$ 0.03 & 0.9645 & 0.9708 & 1.4848 & 0.998 & 12.3 & -28.9 & 1.0558 & exponential \\
		2003.5 & 2481 & 2027.83 $\pm$ 0.04 & 0.9645 & 0.9708 & 1.4848 & 0.998 & 12.3 & -28.9 & 0.9664 & exponential \\
		2004.0 & 2481 & 2027.82 $\pm$ 0.02 & 0.9645 & 0.9708 & 1.4848 & 0.998 & 12.3 & -28.9 & 0.7851 & exponential \\
		2004.5 & 2481 & 2027.83 $\pm$ 0.03 & 0.9645 & 0.9708 & 1.4848 & 0.998 & 12.3 & -28.9 & 0.9896 & exponential \\
		2005.0 & 2481 & 2027.83 $\pm$ 0.02 & 0.9645 & 0.9708 & 1.4848 & 0.998 & 12.3 & -28.9 & 0.9565 & exponential \\
		2005.5 & 2480 & 2027.81 $\pm$ 0.03 & 0.9648 & 0.9709 & 1.4797 & 1.011 & -55.4 & -77.7 & 0.9366 & singularity \\
		2006.0 & 2479 & 2027.82 $\pm$ 0.02 & 0.9652 & 0.9711 & 1.4760 & 1.048 & -232.5 & -74.2 & 0.9884 & singularity \\
		2006.5 & 2479 & 2027.82 $\pm$ 0.01 & 0.9652 & 0.9711 & 1.4760 & 1.048 & -232.5 & -74.2 & 0.8682 & singularity \\
		2007.0 & 2479 & 2027.82 $\pm$ 0.02 & 0.9652 & 0.9711 & 1.4760 & 1.048 & -232.5 & -74.2 & 1.0676 & singularity \\
		2007.5 & 2479 & 2027.82 $\pm$ 0.02 & 0.9652 & 0.9711 & 1.4760 & 1.048 & -232.5 & -74.2 & 1.0963 & singularity \\
		2008.0 & 2479 & 2027.83 $\pm$ 0.02 & 0.9652 & 0.9711 & 1.4760 & 1.048 & -232.5 & -74.2 & 1.0759 & singularity \\
		2008.5 & 2478 & 2027.84 $\pm$ 0.04 & 0.9655 & 0.9714 & 1.4666 & 1.149 & -687.7 & -119.0 & 1.2009 & singularity \\
		2009.0 & 2478 & 2027.84 $\pm$ 0.04 & 0.9655 & 0.9714 & 1.4666 & 1.149 & -687.7 & -119.0 & 1.1921 & singularity \\
		2009.5 & 2478 & 2027.84 $\pm$ 0.04 & 0.9655 & 0.9714 & 1.4666 & 1.149 & -687.7 & -119.0 & 1.1229 & singularity \\
		2010.0 & 2478 & 2027.84 $\pm$ 0.04 & 0.9655 & 0.9714 & 1.4666 & 1.149 & -687.7 & -119.0 & 1.1857 & singularity \\
		2010.5 & 2477 & 2027.83 $\pm$ 0.03 & 0.9659 & 0.9716 & 1.4629 & 1.179 & -814.1 & -128.0 & 1.1925 & singularity \\
		2011.0 & 2476 & 2027.84 $\pm$ 0.04 & 0.9662 & 0.9717 & 1.4592 & 1.287 & -1250.9 & 103.0 & 0.9977 & singularity \\
		2011.5 & 2476 & 2027.83 $\pm$ 0.03 & 0.9662 & 0.9717 & 1.4592 & 1.287 & -1250.9 & 103.0 & 1.1957 & singularity \\
		2012.0 & 2476 & 2027.83 $\pm$ 0.03 & 0.9662 & 0.9717 & 1.4592 & 1.287 & -1250.9 & 103.0 & 1.2741 & singularity \\
		2012.5 & 2476 & 2027.84 $\pm$ 0.04 & 0.9662 & 0.9717 & 1.4592 & 1.287 & -1250.9 & 103.0 & 0.8065 & singularity \\
		2013.0 & 2476 & 2027.83 $\pm$ 0.03 & 0.9662 & 0.9717 & 1.4592 & 1.287 & -1250.9 & 103.0 & 0.8412 & singularity \\
		2013.5 & 2476 & 2027.83 $\pm$ 0.04 & 0.9662 & 0.9717 & 1.4592 & 1.287 & -1250.9 & 103.0 & 1.3890 & singularity \\
		2014.0 & 2476 & 2027.83 $\pm$ 0.03 & 0.9662 & 0.9717 & 1.4592 & 1.287 & -1250.9 & 103.0 & 1.1674 & singularity \\
		2014.5 & 2476 & 2027.84 $\pm$ 0.04 & 0.9662 & 0.9717 & 1.4592 & 1.287 & -1250.9 & 103.0 & 0.8956 & singularity \\
		2015.0 & 2475 & 2027.84 $\pm$ 0.04 & 0.9664 & 0.9718 & 1.4566 & 1.317 & -1363.7 & 213.0 & 1.3326 & singularity \\
		2015.5 & 2475 & 2027.84 $\pm$ 0.04 & 0.9664 & 0.9718 & 1.4566 & 1.317 & -1363.7 & 213.0 & 0.9476 & singularity \\
		2016.0 & 2469 & 2027.92 $\pm$ 0.02 & 0.9674 & 0.9723 & 1.4417 & 1.432 & -1773.9 & 172.0 & 0.9883 & singularity \\
		2016.5 & 2467 & 2027.92 $\pm$ 0.03 & 0.9678 & 0.9726 & 1.4320 & 1.456 & -1853.3 & 92.5 & 0.8944 & singularity \\
		2017.0 & 2465 & 2028.02 $\pm$ 0.01 & 0.9681 & 0.9724 & 1.4371 & 1.410 & -1692.2 & 529.0 & 1.3053 & singularity \\
		2017.5 & 2465 & 2028.02 $\pm$ 0.03 & 0.9681 & 0.9724 & 1.4371 & 1.410 & -1692.2 & 529.0 & 1.1040 & singularity \\
		2018.0 & 2465 & 2028.01 $\pm$ 0.02 & 0.9681 & 0.9724 & 1.4371 & 1.410 & -1692.2 & 529.0 & 1.1824 & singularity \\
		2018.5 & 2440 & 2028.12 $\pm$ 0.03 & 0.9712 & 0.9739 & 1.3888 & 1.506 & -1997.9 & 203.0 & 0.9308 & singularity \\
		2019.0 & 2436 & 2028.15 $\pm$ 0.05 & 0.9718 & 0.9748 & 1.3630 & 1.541 & -2108.0 & 54.0 & 0.7372 & singularity \\
		2019.5 & 2428 & 2028.34 $\pm$ 0.04 & 0.9726 & 0.9743 & 1.3726 & 1.517 & -2024.9 & 355.0 & 0.9621 & singularity \\
		2020.0 & 2421 & 2028.35 $\pm$ 0.05 & 0.9733 & 0.9758 & 1.3286 & 1.423 & -1709.3 & 238.0 & 0.9478 & singularity \\
		2020.5 & 2407 & 2028.35 $\pm$ 0.05 & 0.9743 & 0.9757 & 1.3278 & 1.362 & -1487.8 & 364.0 & 0.8179 & singularity \\
		2021.0 & 2375 & 2028.82 $\pm$ 0.07 & 0.9761 & 0.9766 & 1.2904 & 1.340 & -1390.5 & 260.0 & 0.7613 & singularity \\
		2021.5 & 2328 & 2029.28 $\pm$ 0.08 & 0.9793 & 0.9794 & 1.1938 & 1.290 & -1186.4 & -372.0 & 0.5955 & singularity \\
		2022.0 & 2301 & 2029.29 $\pm$ 0.08 & 0.9806 & 0.9805 & 1.1532 & 1.345 & -1363.0 & -412.0 & 0.5845 & singularity \\
		2022.5 & 2233 & 2030.50 $\pm$ 0.12 & 0.9839 & 0.9837 & 1.0310 & 1.891 & -2845.5 & -1210.0 & 0.5868 & singularity \\
		2023.0 & 2171 & 2031.54 $\pm$ 0.15 & 0.9846 & 0.9844 & 0.9869 & 2.017 & -3046.9 & -1270.0 & 0.5718 & singularity \\
		2023.5 & 1998 & 2033.76 $\pm$ 0.41 & 0.9826 & 0.9824 & 0.9844 & 1.904 & -2573.2 & -598.0 & 0.5002 & singularity \\
		2024.0 & 1716 & 2045.47 $\pm$ 1.34 & 0.9768 & 0.9768 & 0.9913 & 2.045 & -2455.1 & 97.4 & 0.1597 & singularity \\
		2024.5 & 1281 & 2036.91 $\pm$ 2.07 & 0.9591 & 0.9588 & 0.9778 & 2.376 & -2217.1 & 651.0 & 0.2033 & singularity \\
		2025.0 & 1087 & 2045.41 $\pm$ 1.95 & 0.9577 & 0.9563 & 0.8720 & 3.149 & -2493.7 & 428.0 & 0.3311 & singularity \\
	\end{longtable}
\end{landscape}

\begin{landscape}
	\begin{table}[ht]
		\centering
		\caption{\textbf{Comparison of singularity and regularized singularity models for GNSS vertical displacement at station RITE.} The analysis is performed for annual starting dates from 2000.4 to 2015.4. For each window, we report the estimated critical time $t_c$ (years, with $2\sigma$ bootstrap uncertainty) for both the pure singularity model ($t_c^{\text{Sing}}$) and the regularized singularity model ($t_c^{\text{RSing}}$), the root-mean-square error of each fit (RMSE$_{\text{Sing}}$ and RMSE$_{\text{RSing}}$), the Akaike Information Criterion for each model (AIC$_{\text{Sing}}$ and AIC$_{\text{RSing}}$), their difference ($\Delta$AIC = AIC$_{\text{RSing}} - \text{AIC}_{\text{Sing}}$), the regularization parameter $a$, and the singularity exponent $\beta$ (fixed at the value obtained from the unregularized fit). Negative $\Delta$AIC values favour the regularized model. The RSing model systematically achieves lower RMSE and AIC values across all windows.}
		\label{tab:gnss_rsing_comparison}
		\small
		\setlength{\tabcolsep}{4pt}
		\begin{tabular}{c c c c c c c c c c}
			\toprule
			\textbf{Start} & $\boldsymbol{t_c^{\textbf{Sing}}}$ & $\boldsymbol{t_c^{\textbf{RSing}}}$ & \textbf{RMSE$_{\textbf{Sing}}$} & \textbf{RMSE$_{\textbf{RSing}}$} & \textbf{AIC$_{\textbf{Sing}}$} & \textbf{AIC$_{\textbf{RSing}}$} & $\boldsymbol{\Delta}$\textbf{AIC} & $\boldsymbol{a}$ & $\boldsymbol{\beta}$ \\
			\midrule
			2000.4 & 2032.24 $\pm$ 0.12 & 2031.20 $\pm$ 0.08 & 3.4873 & 2.8634 & 3363.7 & 2833.8 & -529.9 & 5051.82 & 1.5123 \\
			2001.4 & 2032.51 $\pm$ 0.12 & 2031.20 $\pm$ 0.06 & 3.2108 & 2.6476 & 3018.0 & 2520.0 & -498.0 & 5051.82 & 1.5014 \\
			2002.4 & 2032.79 $\pm$ 0.14 & 2031.70 $\pm$ 0.09 & 3.0027 & 2.5084 & 2732.8 & 2286.8 & -446.0 & 5736.15 & 1.4887 \\
			2003.4 & 2033.09 $\pm$ 0.12 & 2031.71 $\pm$ 0.08 & 2.7954 & 2.3646 & 2446.5 & 2049.1 & -397.3 & 5736.15 & 1.4780 \\
			2004.4 & 2033.42 $\pm$ 0.14 & 2032.25 $\pm$ 0.11 & 2.5545 & 2.2058 & 2135.0 & 1801.8 & -333.2 & 6513.18 & 1.4675 \\
			2005.4 & 2033.71 $\pm$ 0.12 & 2032.23 $\pm$ 0.10 & 2.4233 & 2.1380 & 1921.4 & 1650.4 & -271.0 & 6513.18 & 1.4593 \\
			2006.4 & 2034.04 $\pm$ 0.14 & 2032.97 $\pm$ 0.18 & 2.2646 & 2.0577 & 1689.9 & 1492.4 & -197.4 & 7395.47 & 1.4440 \\
			2007.4 & 2034.59 $\pm$ 0.12 & 2033.50 $\pm$ 0.20 & 1.8125 & 1.7022 & 1168.0 & 1045.4 & -122.6 & 7880.46 & 1.4267 \\
			2008.4 & 2035.03 $\pm$ 0.13 & 2034.57 $\pm$ 0.48 & 1.5724 & 1.5341 & 843.3 & 797.7 & -45.7 & 8397.27 & 1.4124 \\
			2009.4 & 2035.39 $\pm$ 0.12 & 2036.51 $\pm$ 0.27 & 1.4556 & 1.4522 & 661.5 & 657.5 & -4.0 & 7395.47 & 1.4050 \\
			2010.4 & 2035.61 $\pm$ 0.14 & 2034.76 $\pm$ 0.22 & 1.4426 & 1.4411 & 607.0 & 605.3 & -1.7 & 2357.29 & 1.4273 \\
			2011.4 & 2035.45 $\pm$ 0.14 & 2036.25 $\pm$ 0.35 & 1.4368 & 1.4316 & 564.1 & 558.6 & -5.6 & 8397.27 & 1.4166 \\
			2012.4 & 2035.11 $\pm$ 0.14 & 2034.19 $\pm$ 0.31 & 1.3528 & 1.2870 & 439.3 & 367.8 & -71.5 & 9534.77 & 1.4392 \\
			2013.4 & 2035.27 $\pm$ 0.14 & 2033.83 $\pm$ 0.22 & 1.2196 & 1.1485 & 270.0 & 190.1 & -79.9 & 9534.77 & 1.4454 \\
			2014.4 & 2035.57 $\pm$ 0.15 & 2035.66 $\pm$ 0.43 & 1.1647 & 1.1329 & 192.6 & 158.8 & -33.8 & 10826.37 & 1.4384 \\
			2015.4 & 2035.53 $\pm$ 0.19 & 2033.49 $\pm$ 0.16 & 1.2024 & 1.1325 & 212.4 & 145.3 & -67.1 & 9534.77 & 1.4510 \\
			\bottomrule
		\end{tabular}
	\end{table}
\end{landscape}

\begin{landscape}
	\begin{table}[ht]
		\centering
		\caption{\textbf{Comparison of singularity and regularized singularity models for cumulative Benioff strain at Campi Flegrei.} The analysis is performed for annual starting dates from 2000.0 to 2025.0. For each window, we report the number of events, the estimated critical time $t_c$ (years, with $2\sigma$ bootstrap uncertainty) for both the pure singularity model ($t_c^{\text{Sing}}$) and the regularized singularity model ($t_c^{\text{RSing}}$), the root-mean-square error of each fit (RMSE$_{\text{Sing}}$ and RMSE$_{\text{RSing}}$), the coefficient of determination ($R^2_{\text{Sing}}$), the singularity exponent $\beta$ for both models (with $\beta_{\text{RSing}}$ fixed at the unregularized value), the regularization parameter $a$, and the Akaike Information Criterion difference ($\Delta$AIC = AIC$_{\text{RSing}} - \text{AIC}_{\text{Sing}}$). Negative $\Delta$AIC values favour the regularized model. The RSing model systematically achieves lower RMSE across all windows, with the greatest improvement occurring for recent starting dates.}
		\label{tab:benioff_rsing_comparison}
		\small
		\setlength{\tabcolsep}{3.5pt}
		\begin{tabular}{c c c c c c c c c c}
			\toprule
			\textbf{Start} & \textbf{Events} & $\boldsymbol{t_c^{\textbf{Sing}}}$ & $\boldsymbol{t_c^{\textbf{RSing}}}$ & \textbf{RMSE$_{\textbf{Sing}}$} & \textbf{RMSE$_{\textbf{RSing}}$} & $\boldsymbol{R^2_{\textbf{Sing}}}$ & $\boldsymbol{\beta_{\textbf{Sing}}}$ & $\boldsymbol{a}$ & $\boldsymbol{\Delta}$\textbf{AIC} \\
			& & & & \multicolumn{2}{c}{($\times 10^8$)} & & & ($\times 10^5$) & \\
			\midrule
			2000.0 & 2483 & 2027.72 $\pm$ 0.10 & 2029.58 $\pm$ 0.09 & 1.4912 & 1.4862 & 0.9705 & 2.0254 & 6.31 & -16.7 \\
			2001.0 & 2482 & 2027.77 $\pm$ 0.10 & 2029.58 $\pm$ 0.10 & 1.4874 & 1.4823 & 0.9707 & 2.0178 & 6.31 & -16.9 \\
			2002.0 & 2482 & 2027.82 $\pm$ 0.10 & 2029.58 $\pm$ 0.10 & 1.4874 & 1.4823 & 0.9707 & 2.0178 & 6.31 & -16.9 \\
			2003.0 & 2481 & 2027.82 $\pm$ 0.04 & 2029.48 $\pm$ 0.10 & 1.4848 & 1.4836 & 0.9708 & 1.9588 & 6.31 & -4.1 \\
			2004.0 & 2481 & 2027.82 $\pm$ 0.05 & 2029.49 $\pm$ 0.09 & 1.4848 & 1.4836 & 0.9708 & 1.9588 & 6.31 & -4.1 \\
			2005.0 & 2481 & 2027.82 $\pm$ 0.04 & 2029.48 $\pm$ 0.09 & 1.4848 & 1.4836 & 0.9708 & 1.9588 & 6.31 & -4.1 \\
			2006.0 & 2479 & 2027.82 $\pm$ 0.05 & 2029.54 $\pm$ 0.09 & 1.4760 & 1.4728 & 0.9711 & 1.9711 & 6.31 & -10.5 \\
			2007.0 & 2479 & 2027.82 $\pm$ 0.05 & 2029.54 $\pm$ 0.09 & 1.4760 & 1.4728 & 0.9711 & 1.9711 & 6.31 & -10.5 \\
			2008.0 & 2479 & 2027.82 $\pm$ 0.04 & 2029.54 $\pm$ 0.10 & 1.4760 & 1.4728 & 0.9711 & 1.9711 & 6.31 & -10.5 \\
			2009.0 & 2478 & 2027.82 $\pm$ 0.08 & 2029.64 $\pm$ 0.09 & 1.4666 & 1.4644 & 0.9714 & 2.0121 & 6.31 & -7.2 \\
			2010.0 & 2478 & 2027.82 $\pm$ 0.08 & 2029.64 $\pm$ 0.11 & 1.4666 & 1.4644 & 0.9714 & 2.0121 & 6.31 & -7.2 \\
			2011.0 & 2476 & 2027.82 $\pm$ 0.08 & 2029.68 $\pm$ 0.10 & 1.4592 & 1.4547 & 0.9717 & 2.0265 & 6.31 & -15.4 \\
			2012.0 & 2476 & 2027.82 $\pm$ 0.06 & 2029.69 $\pm$ 0.09 & 1.4592 & 1.4547 & 0.9717 & 2.0265 & 6.31 & -15.4 \\
			2013.0 & 2476 & 2027.82 $\pm$ 0.06 & 2029.68 $\pm$ 0.09 & 1.4592 & 1.4547 & 0.9717 & 2.0265 & 6.31 & -15.4 \\
			2014.0 & 2476 & 2027.82 $\pm$ 0.06 & 2029.67 $\pm$ 0.09 & 1.4592 & 1.4547 & 0.9717 & 2.0265 & 6.31 & -15.4 \\
			2015.0 & 2475 & 2027.82 $\pm$ 0.09 & 2029.68 $\pm$ 0.10 & 1.4566 & 1.4521 & 0.9718 & 2.0198 & 6.31 & -15.3 \\
			2016.0 & 2469 & 2027.92 $\pm$ 0.06 & 2029.66 $\pm$ 0.09 & 1.4417 & 1.4380 & 0.9723 & 1.9762 & 6.31 & -12.8 \\
			2017.0 & 2465 & 2028.02 $\pm$ 0.05 & 2029.49 $\pm$ 0.10 & 1.4371 & 1.4365 & 0.9724 & 1.8518 & 6.31 & -2.0 \\
			2018.0 & 2465 & 2028.02 $\pm$ 0.06 & 2029.48 $\pm$ 0.10 & 1.4371 & 1.4365 & 0.9724 & 1.8518 & 6.31 & -2.0 \\
			2019.0 & 2436 & 2028.12 $\pm$ 0.09 & 2029.79 $\pm$ 0.10 & 1.3630 & 1.3609 & 0.9748 & 1.8720 & 6.31 & -7.2 \\
			2020.0 & 2421 & 2028.32 $\pm$ 0.30 & 2029.93 $\pm$ 0.11 & 1.3286 & 1.3282 & 0.9758 & 1.8493 & 6.31 & -1.2 \\
			2021.0 & 2375 & 2028.82 $\pm$ 0.30 & 2029.27 $\pm$ 0.11 & 1.2904 & 1.2826 & 0.9766 & 1.2923 & 6.31 & -28.9 \\
			2022.0 & 2301 & 2029.32 $\pm$ 0.30 & 2029.70 $\pm$ 0.15 & 1.1532 & 1.1452 & 0.9805 & 1.1392 & 6.31 & -32.3 \\
			2023.0 & 2171 & 2031.52 $\pm$ 0.31 & 2030.08 $\pm$ 0.25 & 0.9869 & 0.9852 & 0.9844 & 0.6522 & 6.31 & -7.1 \\
			2024.0 & 1716 & 2046.22 $\pm$ 2.57 & 2036.68 $\pm$ 2.05 & 0.9913 & 0.9900 & 0.9768 & 0.3938 & 6.31 & -4.5 \\
			2025.0 & 1087 & 2046.22 $\pm$ 4.69 & 2045.51 $\pm$ 1.48 & 0.8720 & 0.8451 & 0.9563 & 0.3670 & 6.31 & -68.3 \\
			\bottomrule
		\end{tabular}
	\end{table}
\end{landscape}

\end{document}